\documentclass[a4paper, oneside, 11pt]{scrbook}
\usepackage[T1]{fontenc}
\usepackage[utf8]{inputenc}
\usepackage[english]{babel}
\usepackage{header}
\usepackage{comment}

\usepackage[backend=biber, style=ieee]{biblatex}
\thesistype{Master's Thesis} 
\title{Discriminating modelling approaches for Point in Time Economic Scenario Generation}

\author{Rui Wang}
\email{ruiwang16@ethz.ch}

\institute{Departement of Mathematics, D-MATH \\[2pt]
ETH Zürich}

\supervisors{Prof. Dr. Patrick Cheridito\\ Mr. Binghuan Lin (UBS)}

\date{\today}

\svgsetup{inkscapelatex=false}

\begin{document}

\frontmatter 
\maketitle

\cleardoublepage

\begin{acknowledgements}
The thesis topic about conditional economic scenario generation came up in the author's cooperation with UBS. The views expressed in this thesis are those of the author and do not necessarily reflect views of UBS AG, its subsidiaries or affiliates. The author declares no conflict of interest.
\end{acknowledgements}

\begin{abstract}
We introduce the notion of \textit{Point in Time Economic Scenario Generation} (PiT ESG) with a clear mathematical problem formulation to unify and compare economic scenario generation approaches conditional on forward looking market data. Such PiT ESGs should provide quicker and more flexible reactions to sudden economic changes than traditional ESGs calibrated solely to long periods of historical data. We specifically take as economic variable the S\&P500 Index with the VIX Index as forward looking market data to compare the nonparametric filtered historical simulation, GARCH model with joint likelihood estimation (parametric), Restricted Boltzmann Machine and the conditional Variational Autoencoder (Generative Networks) for their suitability as PiT ESG. Our evaluation consists of statistical tests for model fit and benchmarking the out of sample forecasting quality with  a strategy backtest using model output as stop loss criterion. We find that both Generative Networks outperform the nonparametric and classic parametric model in our tests, but that the CVAE seems to be particularly well suited for our purposes: yielding more robust performance and being computationally lighter. 
\end{abstract}

\tableofcontents
\mainmatter
\chapter{Introduction}
Regulatory requirements for financial institutions given e.g. by the Basel Framework (\cite{basel2}) or the Solvency II Directive (\cite{solv2}) require companies to be financially prepared for certain losses and demand appropriate internal models for the calculation and modelling of such risks.  Economic Scenario Generators (ESGs), models simulating realistic future paths of financial variables with unknown analytical formulation or distribution, are used to fulfil these regulatory guidelines, as shown in \cite{ESG}, \cite{10.2307/41300453}. 

As indicated in \cite{ESG}, Economic Scenario Generators can be any type of model, parametric or non parametric. They need to provide relevant and realistic views, permit reasonably easy recalibration to different attributes and be computationally efficient and numerically stable. Their modelling performance is often judged by their replication capabilities of financial time series properties, labelled stylized facts in \cite{stylizedfacts}, and are based on calibration to historical data. As mentioned in \cite{usehistory}, however, calibration solely on long periods of historical data might make ESGs inflexible to sudden changes in economic circumstances influencing variable development. The need thus arises to include information from both historical data and forward looking market data in the simulation process. Nevertheless, there is a lack of general consensus in academic literature on how to recalibrate models in such a way, resulting in inconsistencies in the mathematical problem setting and implementation. Approaches range from expert judgement on selecting scenario outputs to models calibration for fixed values at fixed future time points as described in \cite{cali}, \cite{MDS}. Encountering this problem, this thesis formally introduces the notion of \textit{Point in Time Economic Scenario Generation} with a clear mathematical problem formulation and compares different modelling approaches on the example of the S\&P500 and VIX indices.

Our work extends that of \cite{cali} and \cite{MDS}, which only introduce possible conditional ESGs without formal problem definition, to a consistent comparison between several recalibration approaches. We base our idea of taking information from the VIX index for a better modelling of the S\&P500 Index following the research of \cite{Giot92} and \cite{vixforecast}, who show the existence of forecasting quality in VIX values on the level and volatility of future S\&P500 Index returns. During the comparison we research ideas from nonparametric, parametric and data driven models, as they all pose different advantages and disadvantages.  Nonparametric models might be easy in implementation, but can require strict assumptions. Parametric models suffer from high model risk due to a priori assumptions or a trade off between usage simplicity and description power, but benefit from well known theoretical results. Data driven models, known in the context of ESGs as Market Generators after \cite{RBM}, are computationally intense and their calibration process might require a lot of expertise and effort. Due to the primarily industrial usage of ESGs, literature on this topic is scarce and thus any model recalibration proposal needs to be researched separately according to the model family. For the nonparametric model we choose a model free adaptation of the filtered historical simulation introduced in \cite{fhsori} with VIX Indices presented in \cite{fhs}. As parametric model we assume a GARCH process (\cite{garch}) with a joint likelihood estimation using an autoregressive model for the VIX as used in \cite{garchvixoption}, and we choose generative neural networks as data driven models, known in scenario generation as Market Generators following the notation in \cite{RBM}. More specifically, we extend the research of \cite{RBM} on using Bernoulli Restricted Boltzmann machines (\cite{RBMori}) as market generators with a conditioning factor and additionally include in the comparison a conditional Variational Autoencoder (\cite{cvaeori}), where we use labels for the forward looking VIX information similar to \cite{cvae}.

Extending above mentioned literature to the use and comparison against each other, our contribution towards the problem of conditionally modelling financial time series are the following: 
\begin{enumerate}
\item Formalizing the notion of Point in Time Economic Scenario Generation to mathematically clarify ESG recalibration with both historical and forward looking data;
\item Summarizing and comparing the chosen models using statistical tests for the in sample model fit evaluation;
\item Benchmarking the forecasting abilities of each model by using their generated outputs in a stop loss strategy backtest with a stop loss criterion reflecting important modelling aspects.
\end{enumerate}
This thesis is built up in the following way: chapter \ref{pb} and chapter \ref{related} describe the problem setting with a mathematical formulation and review in detail the existing literature on conditional economic scenario generation. Afterwards, chapter \ref{theory} introduces the classic theory behind each model approach that will be used. Then, chapter \ref{exp} reports experimental results starting first with the experiment setup (section \ref{setup}) containing the recalibration proposals for the chosen models to include conditional modelling; results on toy data (section \ref{toydata}) with and without time dependence as well as performance evaluation on real data in section \ref{realdatamodelling} with an evaluation of the model fits and the model forecasting abilities. Finally, we conclude in chapter \ref{conc} with a summary of the experiments and possible implications of their results.

\chapter{Problem setting} \label{pb}
Our central problem setting revolves around the modelling of a conditional \textit{point in time} distribution, which we, for a vector of $d$ risk factors $x \in \mathbb{R}^{d}$, define as modelling from its conditional distribution 
\begin{equation}
    f^{t, t+H}(x|\mathcal{F}_{t})
    \label{pbdist}
\end{equation} over the time period $[t, t+H]$ given the current economic situation at time $t$ denoted by the filtration $\mathcal{F}_{t}$.
Given this definition, we face the following problems: 
\subsubsection{Choice of filtration}
The choice of filtration depends on a priori assumptions about the importance and influence of certain economic indicators on the to be modelled risk factor and its mainly desired properties. A very detailed description might cover all necessary information, but contain redundant information that increases computational effort. A too narrow description however could miss out on information vital for proper modelling. 

A possible solution could be given by indices combining information from several economic areas, some of which have already been developed by governmental and private institutions. Research in this broad area is widespread and represents more the economic side of the main problem setting, we will thus not focus on all possible choices and their construction in this thesis, but will herefore refer to e.g. \cite{indexreview}, \cite{CISS}. 
\subsubsection{Choice of modelling}
(Unconditionally) modelling financial time series traditionally faces difficulties that arise from some inherent properties called stylized facts, examples of which include non-stationarity, their heavy tails or their volatility clustering as shown in \cite{stylizedfacts}. \cite{QRM}, \cite{greenspan}. Thus a reasonable and realistic ESG needs to be able to replicate chosen stylized facts. In order to achieve this, raw financial data needs to undergo transformations to be modelled without the hindrances of stylized facts and model assumptions need to be examined in order to justify their usage. Possible transformations include those described as quest for invariants in \cite{meucci} or the reformulation into their signatures as proposed in \cite{cvae}.

Additionally, there is a trade off between the (time) length of a modelling period and the amount of available modelling data for the chosen period length. A higher period length enables the model to learn more dependencies into the past, but given the limited length of available time series data ultimately yields fewer possible samples to learn from. A shorter period gives more samples, but learning from historical events might be compromised.
In order to include a conditioning factor, it is necessary to first find the explicit dependence of the modelled variables on the elements of the filtration and then to reasonably describe them within each modelling approach. In general, there are three types of modelling approaches: 
\begin{itemize}
    \item [] \textbf{Historical Approach.} Historical models based on the Bootstrap method (\cite{bsori}) in combination with the historical simulation first used in VaR estimations (e.g. \cite{varhs}) follow the central assumption of i.i.d. data samples.  Modelling consists solely of sampling from historical data with several variations of using sampling blocks or basing the sampling on intensity models. Its advantage lies in the simple computation and implementation. However, the i.i.d. assumption is a often too strong and violated in reality, thus possibly yielding unfitting model outputs. Additionally, independent of model assumption violations, model outputs are clearly solely limited to scenarios already observed in historical data. 
    \item [] \textbf{Parametric/ Distributional Approaches.} All models of this family require some a priori assumptions about the distribution of the data, as the base assumptions are a fixed constellation and the parameters usually only control the final combination of the single model assumptions. Exemplary models of this type include autoregressive models (e.g. \cite{garch}, \cite{arch}), copula models or models using stochastic differential equations (e.g. \cite{heston}, \cite{blackscholes}) and their variations. Parametric models benefit from a long period and large amount of research knowledge and often have analytical solutions. A clear disadvantage however is the high model risk and its inflexibility.
    \item[] \textbf{Data Driven Approaches.} Data Driven approaches follow the idea of extracting information from the given data to model its distribution without any a priori assumptions. Examples are generative networks based approaches like the VAEs or GANs (\cite{cvaeori}, \cite{GAN}), which approximate the underlying distribution by a distribution described by the network construction and its parameters with the goal of being similarity with respect to certain statistical measures. Its advantages lie in its flexibility to learn any kind of distribution due to its independence from any a priori distributional assumptions, its disadvantages in its computational heaviness, difficult training process and very often high demand of data.
\end{itemize}
\subsubsection{Choice of Evaluation}
Following the development and implementation of a fitting model, we need to find an appropriate evaluation measure for the comparison of the generated outputs. Given the very nature of an ESG, its main performance criterion is the model fit given by the ability of reproducing historically observed dynamics, the economic plausibility of its outputs as well as the effort for calibration as described in \cite{ESG}. We consider computational complexity as another criterion in our research, it should however be noted that it is not a performance measure per se, but rather considered as a limitation of available usable technology.

Classically, comparison between the similarity of distributional properties relies on statistical measures such as QQ Plots, Tail Behaviour measures, correlation measures etc as used in \cite{RBM} or \cite{roncalli}. These approaches however might face a variety of problems as described in \cite{cvae}, given by different usages of the scenario generation (see examples from introduction), be it optimization with respect to certain options or a portfolio performance. Other difficulties in performance evaluation arise for example also due to the non continuity of the given data or the intractability of the true underlying distribution. Approaches to counter these problems could be the usage of signature transforms and Maximum Mean Discrepancy as performance measure as given in \cite{cvae}.

Additionally, it might also be interesting to verify the forecasting quality of an ESG, possibly by using its outputs for a stop loss strategy backtest as indicated in \cite{roncalli}. Difficulties arise within the choice of the stop loss criterion, which should be fitted to the modelled condition to utilize sufficient information from the model output to be a valid indicator for the model performance.

In this thesis, we will focus on the last two problems regarding modelling and evaluation of the models and will leave the problem regarding a proper description of the filtration for further research.

\chapter{Related Works} \label{related}
Financial modelling approaches exist in various forms, from classical nonparamteric approaches (e.g. \cite{bsori}) over parametric approaches in discrete time (e.g. \cite{garch}, \cite{arch} or \cite{dccgarch}) and continuous time (e.g. \cite{heston}, \cite{SDESG}) to modern data driven approaches (e.g. \cite{RBM}, \cite{genmodels}). However, these approaches focus merely on the modelling of the unconditional distribution. 

Papers on approaches modelling distributions conditional on certain factors are rare, as pointed out in \cite{cali} and \cite{MDS}, and there is no common consensus on how such a conditional calibration could be done. \cite{cali} follows a similar idea as we, introducing a conditional ESG (CESG) in an attempt to unify calibration approaches to include forward looking views/information. He suggests an extension of the Black Litterman Model (\cite{BLM}), which forecasts returns densities conditional on views/ forward looking information for the expected return, into a multi period multi factor model to jointly model macroeconomic and financial variables. However, his approach is strictly parametric with a linear factor, asset and macro model as well as linear formulations for the conditioning views. Hence his CESG is is prone to inflexibility and model error. Similarly, \cite{timebl} propose a time dependent extension of the Black Litterman model, but restrict themselves to a specific factor model and only accept conditional views of the same time step as one simulation, hence ignoring views at further future horizons. \cite{condfacmod} suggest a conditional density forecast in discrete time, but their approach is also limited to a factor model assumption.  Other papers involving general conditional modelling in a scenario generation setting are given by \cite{combiprob}, who analyze ways to model expert judgement as combined probability distributions but provide no testing or inclusion in a forecasting framework; \cite{meanvol}, who extend the Black Litterman model by forecasting asset distributions conditionally on both views for its mean and variance but assume a linear Gaussian model for the asset returns; or \cite{MDS}, who propose a Market-Driven Scenario Approach based on \cite{MDSori} for forecasting financial assets conditional on fixed values for some of the considered assets and evaluate it on the P\&L distribution of a portfolio given forecasted Brexit scenarios. However, their approach does not consider a multi period time horizon. 

Seeing as information on general approaches to calibrate scenario generation on both historical values and forward looking information is scarce, research herefore is fragmented into works on the separate modelling approaches. Research to use forward looking information from the VIX index in GARCH models is conducted in \cite{garchvixformula}, who estimate model parameters jointly from the maximum likelihood of VIX and returns with an additive model for the VIX. Similarly, \cite{garchvixoption} compares the joint parameter estimation for several GARCH models with an auto regressive formulation for the VIX. A multivariate formulation is given in \cite{garchmulti}. However, these approaches are restricted to the models performance in option valuation and their performed comparison is limited only to other GARCH models.

In case of historical simulation, \cite{fhsori} introduces filtered historical simulation as a way to make historical simulations more flexible to changes in volatility by standardization with a volatility estimate of that day and scaling with a volatility forecast. This is a semi parametric approach as they suggest the usage of a parametric GARCH model for the volatility. Whereas \cite{bsvix} introduces a fully non parametric filtered historical simulation using values from the VIX for the filtering and compares classic historical simulation as well as a standard vanilla GARCH model. 

Papers for Market Generators are given by \cite{cvae}, who present a way to overcome the difficulties of small data sets by using Variational Autoencoders (VAE) and propose to condition the generation process on values for initial volatility, initial return level and the previous simulation output with the help of labels in a conditional VAE. Their main focus however lies on the comparison of the performance of two approaches for the data processing using signatures or the classical Log Returns, similarly, they do not conduct rigorous data backfilling, thus disadvantaging the Log Returns approach, and do not discuss the choice and setting of the conditioning factors. A comparison is done only with a stochastic rough path model. Similarly focusing solely on the use of generative models is \cite{cGANtrade}, who research the performance of a conditional GAN to produce realistic scenarios to use on trading strategy calibration and aggregation. Another approach is given in \cite{roncalli} to compare the scenario generating abilities between a conditional RBM and a conditional Wasserstein GAN for the joint unconditional distribution of the S\&P500 and VIX Indices, with conditional information consisting of 20 day previous data of both index values. The possibility of using the VIX Index as a forward looking indicating conditioning factor is only briefly touched upon with a recommendation for future research. 

Given the different goals of the papers for using the proposed models, the evaluation methodology also differs across all literature. We will focus on aforementioned (or new) literature that have the same objective as us and whose evaluation methods are thus relevant for review. Every evaluation should focus on the necessary requirements on an ESG, examples of which are detailed in \cite{ESG}. Following this, approaches contain the classical statistical measures such as QQ Plots, summary statistics as used in \cite{RBM} and \cite{roncalli}. Other ways to evaluate distributional approximation quality are given by distance measures like the KL-Divergence, Wasserstein distance or Maximum Mean Discrepancy, as used in \cite{cvae}. A economically more intuitive way of evaluation is hinted at in \cite{roncalli}, proposing the usage of ESG forecasts in a strategy indicator and then judging the strategy performance using this indicator. 

\chapter{Theory} \label{theory}
In this chapter we provide the theoretical background of the established non parametric, parametric and Machine Learning approaches to use for scenario generation.
\section{Bootstrap simulation}
Bootstrap simulation is an extension of historical simulation, a methodology assuming the distribution of historical data as a proxy for the future distribution and thus calculating future statistics based on historical observations. Financial applications involving historical simulation first appeared in the estimation of Value-at-Risk (e.g. \cite{varfed}, \cite{varhs}). On the other hand, bootstrapping methodology has its mathematical background in \cite{bsori}, who introduces the bootstrap method to approximate distributions of estimators purely by the sampling from observations. The bootstrap follows the assumption of i.i.d. observations. Combining bootstrap and historical simulation yields the bootstrap simulation, which, in its core, consists of sampling with replacement from historical observations with possible extra features about the sampling size or intensity.
\section{GARCH}
Generalized AutoRegressive Conditional Heteroscedasticity Models (GARCH) were first introduced by Bollerslev (\cite{garch}) as an extension to the ARCH models by Engle (\cite{arch}) to describe varying conditional variance in the modelling of time series to include a longer memory and more flexible lag structure. A one dimensional GARCH($p,q$) process $x_{t}$ is a real valued discrete time process defined by if it can be described as

\begin{equation} 
\begin{split}
x_{t} & = \mu_{t}+\epsilon_{t}, \\
\epsilon_{t}& = w_{t}\sigma_{t}, \\
\sigma_{t}^{2} &= \omega+\sum_{i=1}^{p}\alpha_{i}\epsilon_{t-i}^{2}+\sum_{j=1}^{q}\beta_{j}\sigma_{t-j}^{2} \label{eq1}
\end{split}
\end{equation}
with $\alpha_{i}, \beta_{j} \neq 0; w_{t} \text{ i.i.d. with } \mathbb{E}(w_{t})=0, \text{ Var}(w_{t})=1$, $\mu_{t}$ describing a time varying or constant conditional mean vector and $\sigma_{t}^{2}$ the conditional variance of $\epsilon_{t}$. A GARCH($p,q$) process is stationary if $\sum_{i=1}^{p}\alpha_{i}+\sum_{j=1}^{q}\beta_{j}<1$ \cite{garch}.
\subsubsection{Parameter estimation}
The parameters $\Theta_{G}=\{\omega, \alpha_{1},...,\alpha_{p}, \beta_{1},...,\beta_{q}\}$ can be estimated via Maximum Likelihood estimation. The log likelihood for $w_{t}\sim \mathcal{N}(0,1)$ is given by: 
\begin{equation}
   \text{log} \mathcal{L}_{x_{1},...,x_{T}}(\Theta_{G})= \sum_{t=1}^{T}\left(\text{log}(2\pi)+\text{log}(\sigma_t^{2})+\frac{x_{t}^{2}}{\sigma_t^{2}}\right)
   \label{llhnormret}
\end{equation}
In practice, higher lag orders do not necessarily yield better performance, as \cite{garchcomp} show in a comparison of several ARCH type models on the DM/\$ exchange rate and IBM returns.
\subsubsection{Sampling Process}
Forecasting $k$ time steps from a fitted standard GARCH(1,1) model consists of forecasting the conditional variance $\sigma_{t+k}$ by the recursive formula, which for $p=q=1$ reads:
\begin{equation}
    \begin{split}
         \mathbb{E}_{t}(\sigma_{t+k}^{2}) &= \mathbb{E}_{t}\left(\omega+\alpha_{1}\epsilon_{t+k-1}^{2}+\beta_{1}\sigma_{t+k-1}^{2}\right) \\
         &= \omega+ \left(\alpha_{1}+\beta_{1}\right)\mathbb{E}_{t}(\sigma_{t+k-1}^{2}) \\
         & \vdots \\
         &= \omega \sum_{i=0}^{k-1} \left(\alpha_{1}+\beta_{1}\right)^{i} +\left(\alpha_{1}+\beta_{1}\right)^{k-1} \sigma_{t+1}^{2}
    \end{split}
    \label{sgarchsim}
\end{equation}
as $\sigma_{t+1}^{2}$ is known at time $t$ due to the relation given in \eqref{eq1} and $\mathbb{E}_{t}$ denoting the conditional expectation at time $t$.

\section{Machine Learning Approaches}
The ML algorithms considered for our problem setting belong to the family of generative models, which approximate the unknown true distribution $p$ of a given data set $\textbf{x}$ with an approximation $p_{\theta}$ with parameter set $\theta$ containing the network weights and activation function parameters.\footnote{All graphical illustrations in this section come from \cite{d2020learning} with approval of the authors.} After sufficient training, generative networks are able to produce new samples from the approximate distribution $p_{\theta}(\textbf{x})$ as demonstrated in \cite{d2020learning}. We choose the Restricted Boltzmann Machine and the conditional Variational Autoencoder as generative models. Even though GANs are currently among the most popular and well known generative neural networks, we decide against using them, as their training requires high expertise and their stability and convergence are hard to guarantee, as mentioned in \cite{cvae}, and their learning objective is not based on a maximum likelihood estimation, which makes the comparison with our other chosen models difficult. 

\subsection{Bernoulli Restricted Boltzmann Machine} \label{rbm}
\begin{figure}[h]
    \centering
    \includegraphics[]{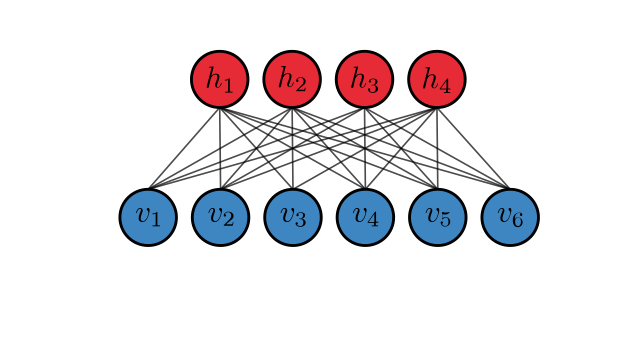}
    \vspace{-1.5cm}
    \caption[Restricted Boltzmann Machine network structure]{Underlying network structure of a RBM is an undirected graph connecting visible layers (blue) with hidden layers (red). Source: \cite{d2020learning}}
    \label{RBM}
\end{figure}

RBMs, first proposed in \cite{RBMori}, are a type of stochastic neural networks consisting of two layers: a visible layer $\textbf{v}=\{v_{1},..., v_{m}\}$ and a hidden layer $\textbf{h}=\{h_{1},..., h_{n}\}$. Usages of RBMs can be found in many areas, see for example \cite{d2020learning}, \cite{userbm1}.
All units in both layers of a Bernoulli RBM are binary, i.e. only take value 0 or 1, and there are no connection between units of each layer, i.e. the network structure of the RBM is a bipartite graph. A possible unique transformation to binary data (and back) is provided in \cite{RBM}. The visible layer is exposed to the given data set, hence each unit corresponds to one data point (after transformation into binary representation). The hidden layer provides additional degrees of freedom in learning the dynamics between the input variables. An illustration of the network structure is given in \ref{RBM}. The connections between the visible and hidden layer are undirected: Input from the visible layer is multiplied by its weights from the weights matrix $W \in \mathbb{R}^{m\text{x}n}$, added to the bias of the visible layer $\textbf{b} \in \mathbb{R}^{m}$ and passed into the sigmoid activation function $\sigma(x)=\frac{1}{1+e^{-x}}$ to generate a Bernoulli random variable. Similarly, the output of the hidden layers undergo the same transformation in the backward pass to visible layer with the hidden layer bias $\textbf{c} \in \mathbb{R}^{n}.$
Given this network construction, we can define the energy of a configuration ($\textbf{v}, \textbf{h}$) as shown in \cite{d2020learning}.\\
\begin{equation}
    E(\textbf{v}, \textbf{h})=-\textbf{b}^{T}\textbf{v}-\textbf{c}^{T}\textbf{h}-\textbf{h}^{T}W\textbf{v},
\end{equation}
and their joint probability given by the Boltzmann distribution
\begin{equation}
    p(\textbf{v}, \textbf{h})=\frac{1}{Z}e^{-E(\textbf{v}, \textbf{h})}
\end{equation}
with the partition function $Z=\sum_{(\textbf{v}, \textbf{h})}^{} e^{-E(\textbf{v}, \textbf{h})}$ summing over all combinations of $(\textbf{v}, \textbf{h}).$ The restricted property of RBM leading to unconnected units within each layer causes units of each layer to be mutually independent given the units of the other layer and the conditional probabilities of each unit is given by
\begin{equation}
\begin{split}
    p(h_{j}=1|\textbf{v})&=\sigma\left(c_{j}+\sum_{i=1}^{m} w_{ij}v_{i}\right), \\
    p(v_{i}=1|\textbf{h})&=\sigma\left(b_{i}+\sum_{j=1}^{n} w_{ij}h_{j}\right).
\end{split}
\label{condind}
\end{equation}
with $\sigma(x)$ the sigmoid function and 
\begin{equation}
    p(\textbf{v}|\textbf{h}) = \prod_{i=1}^m p(v_i|\textbf{h}) \quad \text{and} \quad p(\textbf{h}|\textbf{v}) = \prod_{j=1}^n p(h_j|\textbf{v})\,,
\label{eq:product_cond}
\end{equation}
\subsubsection{Parameter Estimation}
\begin{figure}[h]
    \centering
    \includegraphics[width=\textwidth]{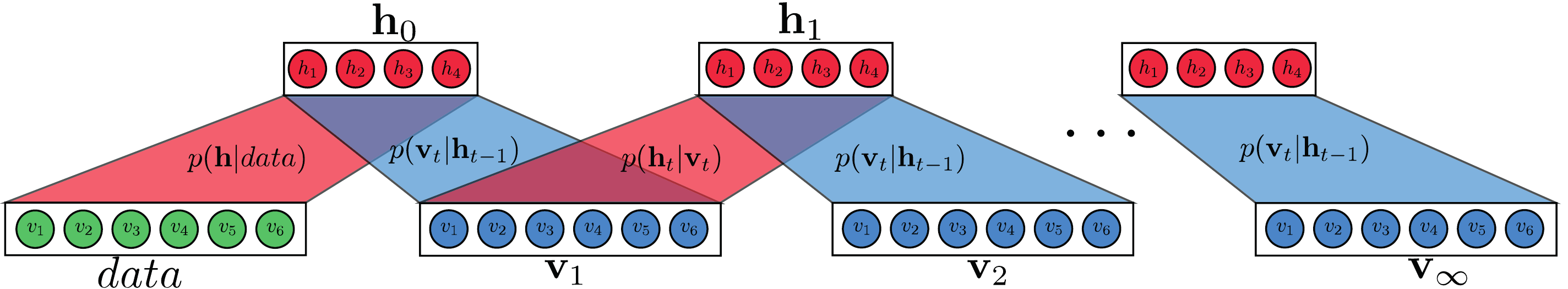}
    \caption[Illustration Block Gibbs Sampling]{Block Gibbs Sampling alternatively updates all units in each layer simultaneously. The initial input data can be chosen from real data to speed up the training process. Source: \cite{d2020learning}}
    \label{blockgibbs}
\end{figure}

The parameters $\Theta_{RBM}=\{\textbf{b},\textbf{c},W\}$ for a binary data set  $\textbf{x} \in \mathbb{R}^{N \text{x} m} $  of N i.i.d realization of dimension m can be found by performing a stochastic gradient descent with learning rate $\eta$ that minimizes a loss function $L(\textbf{x}, \Theta_{RBM})$ and iteratively updating the parameter estimates with
\begin{equation}
    \Theta_{RBM}^{t+1}=\Theta_{RBM}^{t}-\eta \frac{\partial L(\textbf{x}, \Theta_{RBM})}{\partial \Theta_{RBM}^{t}}
\end{equation}
The loss function $L(\textbf{x}, \Theta_{RBM})$ can be chosen as either the negative log likelihood of $\textbf{x}$
\begin{equation}
\begin{split}
    \text{log} \mathcal{L}_{\textbf{x}}(\Theta_{RBM})&= N \log (Z) - \sum_{k=1}^N \textbf{b}^T \textbf{x}_k \\
    &- \sum_{k=1}^N\sum_{j=1}^n \log \left(1 -  e^{(c_j + \sum_{i=1}^{m} w_{ij} x_{k i})}\right),
\end{split}
\end{equation}
with partition function 
\begin{equation}
Z  = \sum_{\{\textbf{h}\}} e^{\textbf{c}^T \textbf{h}} \prod_{i=1}^m \left(1 + e^{b_i + \sum_{j=1}^{n} w_{ij} h_j } \right),
\end{equation}
or equivalently (as shown in \cite{rbmcond}) as the Kullback-Leibler divergence measuring the similarity between the real underlying distribution $p(\textbf{x})$ and the learned approximation $p_{\theta}(\textbf{x})$ (with $\theta$ denoting $\Theta_{RBM}$)
\begin{equation}
D_{\text{KL}}\big(p(\textbf{x}) ||p_{\theta}(\textbf{x}) \big) = \sum_{\textbf{x}}p(\textbf{x})\log\frac{p(\textbf{x})}{p_{\theta}(\textbf{x})} \geq 0.
\end{equation}
In case of the RBM, the parameters are updated with: 
\begin{equation}
\begin{split}
\frac{\partial L(\textbf{x}, \Theta_{RBM})}{\partial \Theta_{RBM}} &= \mathbb{E}_{\textbf{x} \sim p(\textbf{x})} \left[ \frac{\partial E}{\partial \Theta_{RBM}}  \right]- \mathbb{E}_{\textbf{x}\sim p_{\theta}(\textbf{x})} \left[  \frac{\partial E}{\partial \Theta_{RBM}} \right],  \\
	\frac{\partial L(\textbf{x}, \Theta_{RBM})}{\partial w_{ij}} &= \big\langle h_j x_i  \big\rangle_{\textbf{x} \sim p(\textbf{x})} - \big\langle  h_j x_i \big\rangle_{\textbf{x} \sim p_{\theta}(\textbf{x})}, \\
	\frac{\partial L(\textbf{x}, \Theta_{RBM})}{\partial b_j} &= \big\langle x_i  \big\rangle_{\textbf{x} \sim p(\textbf{x})} - \big\langle  v_i \big\rangle_{\textbf{x} \sim p_{\theta}(\textbf{x})}, \\
	\frac{\partial L(\textbf{x}, \Theta_{RBM})}{\partial c_{i}} &= \big\langle h_j \big\rangle_{\textbf{x} \sim p(\textbf{x})} - \big\langle  h_j   \big\rangle_{\textbf{x} \sim p_{\theta}(\textbf{x})} \label{up}, 
\end{split}
\end{equation}
with $\langle \cdot \rangle$ indicates the expectation under the distribution given in the subscript. Due to the bipartite graph structure and the resulting conditional independence from \eqref{eq:product_cond}, updating units in each layer can be done in parallel according to the \textit{block Gibbs Sampling} as illustrated in \ref{blockgibbs}. Despite this simplification, training process can still be computationally expensive, as many forward and backward passes between visible and hidden layer might be necessary until the approximative distribution $p_{\theta}$ reaches equilibrium. \cite{ hinton2002training} addresses this problem by introducing the \textit{k-step contrastive divergence} algorithm, which starts the block Gibbs Sampling process by passing a real data sample into the visible layer instead of random noise and thus needs fewer updating steps (k) to properly train the RBM. 
\subsubsection{Sampling Process }
Generating samples from the RBM consists of a performing a possibly large number of backward and forward passes from a random input until the model reaches equilibrium state. Since all units of the Bernoulli RBM are binary, the finally sampled output possibly has to transformed back into a non binary representation.

\subsection{Variational Autoencoder} \label{vae}
Variational Autoencoders were first introduced by \cite{Kingma2014} and are latent variable models for variational interference consisting of an encoder and decoder network that aim to maximize the lower bound of the log likelihood of the data. The assumption behind the VAE framework is that the unknown true underlying distribution is generated involving some unobserved continuous latent variables $\textbf{z}$ with prior distribution $p(\textbf{z})$ which is usually chosen to be the multivariate standard normal Gaussian $\mathcal{N}(0, \textbf{I})$.  The encoder part takes in the input and tries to approximate the usually intractable true posterior $p(\textbf{z}|\textbf{x})$ by an approximation $q_{\phi}(\textbf{z}|\textbf{x})$ with neural network parameters $\phi$ of, while the decoder generates values of $\textbf{x}$ given a latent variable $\textbf{z}$ according to $p_{\theta}(\textbf{x} | \textbf{z})$ with neural network parameters $\theta$. The goal is to infer the generative model $p_{\theta}(\textbf{x},\textbf{z})$ needed for the data generation process by choosing parameters that maximize the marginal log likelihood $\log p_\theta(\textbf{x})$ in the following way \cite{d2020learning}: 
\begin{equation}
\begin{split}
	\log p_\theta(\textbf{x}) &= \log \int p_\theta(\textbf{x},\textbf{z}) \, \mathrm{d}\textbf{z} = \log \int p_\theta(\textbf{x}| \textbf{z}) p(\textbf{z}) \, \mathrm{d}\mathbf{z}\\ 
	&= \log \int p_\theta(\textbf{x}| \textbf{z}) p(\textbf{z}) \frac{q_\phi(\textbf{z}|\textbf{x})}{q_\phi(\textbf{z}|\textbf{x})}\, \mathrm{d}\mathbf{z} \\ 
	&\geq \mathbb{E}_{\textbf{z} \sim q_\phi(\textbf{z}|\textbf{x})} \left[ \log p_\theta(\textbf{x}|\textbf{z}) \right] -  D_{\text{KL}} \left(q_\phi(\textbf{z}|\textbf{x}) || p (\textbf{z}) \right) \\
	&=: -L\left(\textbf{x},\Theta_{VAE}\right)\, 
\end{split}
\label{eq:ELBO}
\end{equation}
\begin{figure}[]
    \centering
    \includegraphics[]{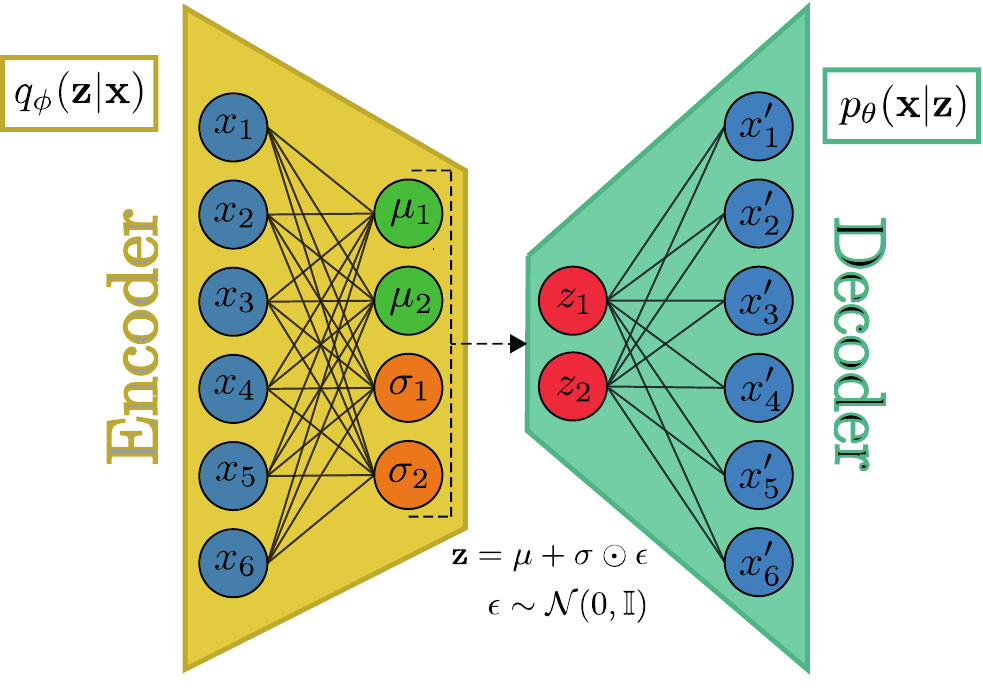}
    \caption[Variational Autoencoder network structure]{A VAE consists of an encoder network and a decoder network, each with possibly multiple layers in addition to input and output layer. Source: \cite{d2020learning}}
    \label{vae}
\end{figure}
with the fourth step given by Jensen's inequality and $\Theta_{VAE}:=\{\theta,\phi\}$.
\subsubsection{Parameter Estimation}
Since integration over $\textbf{z}$ is intractable, we include in the second step the approximate posterior distribution $q_{\phi}(\textbf{z}|\textbf{x})$ to get a tractable bound, the \textit{evidence lower bound} (ELBO), which thus becomes our to be minimized loss function $L\left(\textbf{x},\Theta_{VAE}\right)$. The terms of $L\left(\textbf{x},\Theta_{VAE}\right)$ are a combination of a reconstruction error $\mathbb{E}_{\textbf{z} \sim q_\phi(\textbf{z}|\textbf{x})} \left[ \log p_\theta(\textbf{x}|\textbf{z}) \right]$ enforcing more similarity between generated output and input, as well as a KL-Divergence term encouraging similarity between the approximate posterior and the chosen prior to ensure a smooth encoding in the latent representation.

Often, the approximate posterior is chosen to be a multivariate Gaussian with diagonal covariance structure, i.e. $ q_{\phi}(\textbf{z}|\textbf{x}) = \mathcal{N}(\textbf{z}|\mathbb{\mu}, \mathbb{\sigma^{2} I})$. The encoder thus maps inputs to the parameters of the encoder, i.e. $\mathbf{x}_{i} \xrightarrow{} \{\mathbb{\mu}_{i}(\mathbf{x}), \mathbf{\sigma}_{i}(\mathbf{x})\}$. Combining this with the assumption of a standard Gaussian prior, the KL-divergence in the loss function can explicitly rewritten into (\cite{KL}):
\begin{equation}
    D_{\text{KL}}\left(q_\phi(z_{i}|\textbf{x}) || p (z_{i}) \right)= -\frac{1}{2}\text{ln}(\sigma_{i})+\frac{1}{2}\sigma_{i}^{2}+\frac{1}{2}\text{ln}(\mu_{i})-\frac{1}{2}
\end{equation}
For non binary input data the model likelihood is often chosen as a Gaussian $p_\theta(\textbf{x}|\textbf{z})=\mathcal{N}(x|\mu_{\theta}(\textbf{z}),\sigma_{\theta}(\textbf{z})\textbf{I)})$ with $\mu_{\theta},\sigma_{\theta}$ deterministic functions and the reconstruction term in the loss function is calculated as average over $l$ Monte Carlo estimates. In practice however, often an identity matrix is chosen as variance of the model likelihood (i.e. $\sigma_{\theta}$=1 ) and the Monte Carlo estimate is calculated only over one sample for computational efficiency \cite{Ghosh2020From}. Given these assumptions, the reconstruction error becomes a simple mean squared difference between the input $\textbf{x}$ and the generated sample $\hat{\textbf{x}}$ \cite{mse}.  

In order to backpropagate through the network and to calculate the gradients of the reconstruction error w.r.t. its parameters $\phi=\{\mathbb{\mu}_(\textbf{x}),\mathbb{\sigma}_(\textbf{x})\}$, a so called \textit{reparameterization trick} is needed, as backpropagation through a stochastic node from the encoder output is not possible \cite{Kingma2014}. Hence we rewrite a vector $\textbf{z}$ from the latent space as $\textbf{z}=\mathbb{\mu}_(\textbf{x})+\mathbb{\sigma}_(\textbf{x}) \odot \epsilon$ where $\epsilon \sim \mathcal{N}(0,1)$ and $\odot$ denotes element wise multiplication. This way we can rewrite elements from the latent space into a differentiable and invertible transformation of another auxiliary random variable. From a practical point of view, often the $\log(\sigma_{\theta}(\textbf{x}))$ is modelled, as it is followed by a transformation with the exponential function which guarantees the positivity of the final variance. 
\subsubsection{Sampling Process}
Sampling from a trained VAE consists of passing a sample from the latent space distributed according to the chosen prior through the decoder. The output is thus a sample from the approximating model distribution $p_{\theta}$. 

\chapter{Experiments} \label{exp}
In this chapter we focus on the practical implementation and results of the described models. We begin with a proposal of how to recalibrate the previously introduced models to incorporate the volatility conditioning factor and afterwards conduct validity checks with synthetic data sets from both a known distribution and a known time dependent process to assess modelling quality for pure distributions as well as time dependent properties. In the end we report our main results on real historical financial data.

\section{Experiment Setup} \label{setup}
Similar to the problem setting we first relate to the modelling goal and specify the distribution to be modelled as
\begin{equation}
    f^{t,t+H}(RF|\mathcal{F}_{t}) 
    \label{conddist}
\end{equation}
where we choose $\left(RF_{k}\right)_{k \in [t,t+H]}$ to be the daily log return of the S\&P500 Index in the simulation period, i.e. $RF_{k}=\text{log}\left(\frac{\text{SPX}_{k}}{\text{SPX}_{k-1}}\right)$, and the use as proxy for the filtration historical values of the VIX Index as well as past daily Log Returns of the S\&P500 Index up to time $t$.
\begin{figure}[h]
    \centering
    \includegraphics[width=0.8\textwidth]{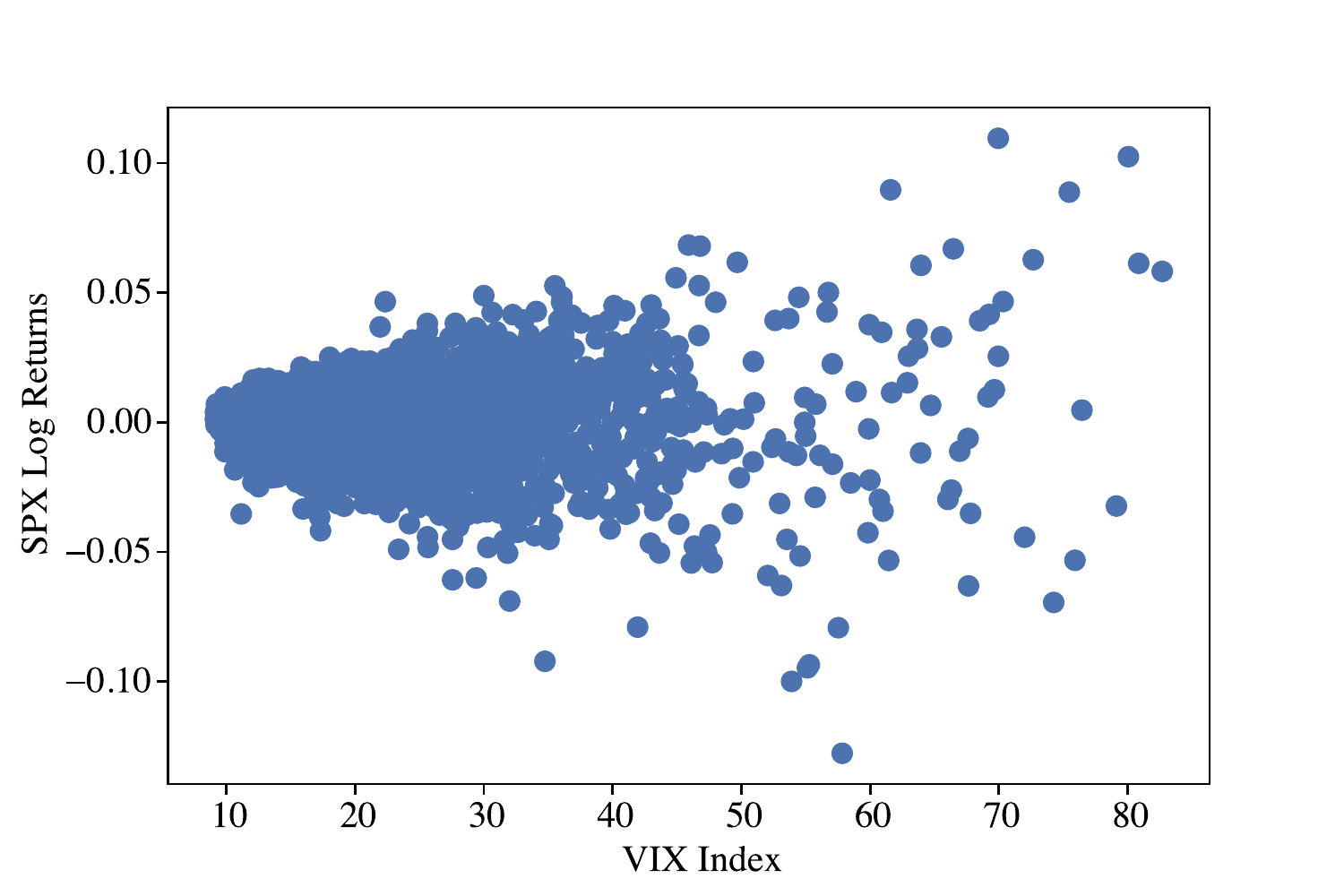}
    \caption[Scatterplot historical S\&P500 and VIX indices]{Historical S\&P500 Log-Returns with one day ahead VIX Index level from 03.01.2000 to 18.12.2020.}
    \label{spxvix}
\end{figure}\\
Keeping in mind our goal of properly replicating the distribution and its special properties such as the stylized facts low/no autocorrelation, fat tails, volatility clustering etc, we first show in \ref{spxvix} the relationship between the one day lagged VIX Index vs. SPX daily Log-Returns. Judging by the scatter plot, there seems to be a  slightly positive correlation between very high values of the VIX Index and positive SPX Log-Returns, also high (low) values of the VIX (above/below 40) seem to coincide with wider (tighter) spreads of the SPX index. Even though our analysis of the above mentioned relationship is not statistically conclusive, the observations visible with the naked eye do coincide with findings from the literature such as \cite{Giot92}, who shows the existence of some evidence of high (low) VIX levels indicating higher (lower) future expected returns in the S\&P100 Index. Additionally, \cite{vixforecast} and \cite{jrfm12040156} show the predictive power of the VIX Index for the second moment of future SPX Log-Returns.

In summary, given our modelling goal for the S\&P500 Index, our motivation for our choice of filtration proxy are the following: 
\begin{enumerate}
    \item The VIX by construction has market implied information on the S\&P500 Index \cite{Whaley12}; 
    \item It contains some predictive information about future SPX returns, as observed in \cite{Giot92};
    \item It contains some predictive information about future SPX volatility, as observed in \cite{vixforecast}.
\end{enumerate}
After clarifying the choice of modelled and conditional variables, we now introduce suggestions on how to include the conditioning factor into the proposed models.

\subsection{Filtered Historical Simulation} 
To include information from the VIX we consider the filtered historical simulation (FHS) (from \cite{fhsori}, \cite{fhs}), which tries to overcome the drawbacks of the standard bootstrap simulation by transforming the historical daily log-return sample $r_{t}$ of time $t$ into 
\begin{equation}
r_{t}^{*}= \sigma_{N}\frac{r_{t}}{\sigma_{t}}    
\label{fhs}
\end{equation}
with $\sigma_{k}being \text{One day volatility forecast for $r_{k}$ made at the end of day $k-1$}$, and $N$ being the day for which we wish to obtain a simulation. The motivation behind this transformation is twofold (\cite{fhs}): 
\begin{enumerate}
    \item Market variables scaled by their volatility estimate have an approximately stationary distribution;
    \item Multiplying by the most current volatility forecast allows an adjustment to current market situations.
\end{enumerate}
Thus, the FHS data samples are more likely to fulfil the i.i.d. assumption necessary for the bootstrap simulation and allows for samples different than historically observed ones. As volatility forecast, \cite{fhsori} proposes to use a GARCH/EWMA model, thus turning the bootstrap simulation from non parametric into a semi parametric approach. We however follow the non parametric filtered historical simulation given in \cite{bsvix} by transforming the 1 day log-returns of the SPX $r_{t}$ into 
\begin{equation}
r_{t}^{*}= \text{VIX}_{N}\frac{r_{t}}{\text{VIX}_{t}}    
\label{filtered}
\end{equation}
where $\text{VIX}_{t}$ denotes the closing value of the VIX index on day $t$. This approach benefits from a higher robustness of being non parametric, but has limitations as the VIX is only defined for the SPX and is a 22-day forecast. However, volatility indices for other major stock indices exists and \cite{bsvix} shows the superiority of this nonparametric approach also for time steps $\delta \neq 22$, hence we do not consider these drawbacks.\\

\subsection{GARCH}
We deviate in two ways from the previously described standard GARCH model in order to conditionally model the S\&P500 Index on the VIX 
\begin{enumerate}
    \item The distribution of the innovation term is chosen to follow a student-t distribution, as it is known that real equity returns are not normal \cite{stylizedfacts};
    \item In order to include information from the VIX, we perform parameter estimation under a joint likelihood of returns and VIX Index as in \cite{garchvixformula}.
\end{enumerate}
\textbf{Changing the innovation distribution}\\
Following the research of \cite{stdt1} and \cite{stdt2}, who identify the best fitting distribution from the Pearson family for daily log-returns on the S\&P500 Index to be a student-t distribution with 3.0-4.5 and 4.5 degrees of freedom, respectively, we change the distribution of the innovation term in the GARCH model to be a student-t distribution with 4 degrees of freedom. The log likelihood function from \eqref{llhnormret} then changes into
\begin{equation}
     \text{log} \mathcal{L}_{x_{1},...,x_{T}}(\Theta_{G})= \sum_{t=1}^{T}\left(\text{log}(\Gamma\left(\frac{5}{2})\right)-\text{log}\left(\Gamma\left(2\right)\right)-\frac{1}{2}\text{log}\left(\sigma^{2}2\pi\right)-\frac{5}{2}\text{log}\left(1+\frac{x_{t}^{2}}{2\sigma^{2}}\right)\right)
    \label{llht}
\end{equation}
with $\Gamma$ being the gamma function. \\
\textbf{Including the VIX likelihood for parameter estimation}\\
In order to use information from the VIX in our GARCH estimation, we follow the approach of \cite{garchvixformula} but assume an autocorrelated process for the VIX as in \cite{garchvixoption}. Hence we interpret VIX as measure of the risk neutral expectation of integrated variance of the S\&P500 \cite{rewritevix} and use this reformulation for the case of discrete time and no jumps
\begin{equation}
    \frac{1}{\tau}\left(\frac{\text{VIX}_t}{100}\right)^{2}=\frac{1}{T}\mathbb{E}_{t}^{\mathbb{Q}}\sum_{k=1}^{T} \sigma_{t+k}^{2}
\end{equation}
where $\mathbb{E}_{t}^{\mathbb{Q}}$ denoted the conditional expectation at time $t$ under the risk neutral expectation derived as in \cite{garchQ}. The annualizing parameter $\tau$ as well as $T$ depend on the day counting convention, \cite{rewritevix} use an actual/365 convention, hence choosing $\tau=365$ and $T=30$. We however choose the convention trading day count convention of $\tau=252, T=22$ with the same reasoning as \cite{garchvixoption} that the return likelihood will be estimated on trading days. Thus yielding the reformulation 
\begin{equation}
    \text{VIX}_t=100 \left[\frac{\tau}{T}\sum_{k=1}^{T} \mathbb{E}_{t}^{\mathbb{Q}}(\sigma_{t+k}^{2})\right]^{\frac{1}{2}}. \label{vix}
\end{equation}
To write out the explicit formulation, we recall the risk neutral univariate GARCH specification for the S\&P500 Index, which for a standard GARCH(1,1) formulation under $\mathbb{P}$ has the same formulation under $\mathbb{Q}$ with different parameters \cite{garchvixformula}. We make the additional change of using the Standardized errors $w_{t}$ instead of the residuals $\epsilon_{t}$ in the variance regression of \eqref{eq1}, yielding the specifications under $\mathbb{Q}$:
\begin{equation}
\begin{split}
x_{t} & = \mu_{t}+\epsilon_{t}, \\
\epsilon_{t}& = \tilde{w_{t}}\sigma_{t}, \\
\sigma_{t}^{2} &= \tilde{\omega}+\tilde{\alpha_{0}}w_{t-i}^{2}+\tilde{\beta_{0}}\sigma_{t-j}^{2} \label{eq3}
\end{split}   
\end{equation}
with $\tilde{w} \sim \mathcal{N}(0,1)$ under $\mathbb{Q}$ and $\tilde{\beta_{0}},\tilde{\alpha_{0}},\tilde{\omega}>0$ and $ \tilde{\beta_{0}}+\tilde{\alpha_{0}}<1$ for stationarity. Given this formulation, it follows directly
\begin{equation}
    \mathbb{E}_{t}\left(\sigma_{t+k}^{2}\right)=\bar{\omega}\sum_{j=0}^{k-2} \tilde{\beta_{0}}^{j}+\tilde{\beta_{0}}^{k-1}\sigma_{t+1}^2
\end{equation}
with the notation $\bar{\omega}:=\tilde{\omega}+\tilde{\alpha_{0}}$, as from \eqref{eq3} $\sigma_{t+1}^{2}$ is known at time $t$. Thus \eqref{vix} becomes
\begin{equation}
    \text{VIX}_{t}=100\left[\left( \bar{\sigma} \left(1-\gamma\right) + \gamma \sigma_{t+1}^{2} \right)\tau \right]^{\frac{1}{2}} \label{vix2}
\end{equation}
with $\bar{\sigma}:=\frac{\bar{\omega}}{1-\beta_{0}}$, which is exactly the unconditional variance under $\mathbb{Q}$, and $\gamma:=\frac{1-\tilde{\beta_{0}}^{T}}{T(1-\tilde{\beta_{0}})}.$ For the derivation of the VIX log likelihood we, as in \cite{garchvixoption}, assume the following autoregressive model for $d_{t}=\text{VIX}^{O}- \text{VIX}_{t}$, the difference of observed real market $\text{VIX}^{O}$ and the $ \text{VIX}_{t}$ from \eqref{vix2} 
\begin{equation}
\begin{split}
    d_{t}&=\rho d_{t-1}+\epsilon_{t} \\
    \epsilon_{t} & \sim NID(0, \sigma)
\end{split}
\end{equation}
We assume $d_{t} \sim \mathcal{N}(0, \Sigma)$ and get, following \cite{automle}, the log likelihood function for VIX
\begin{equation}
\begin{split}
    \text{log} \mathcal{L}_\text{VIX}(\Theta_{G})=&-\frac{T}{2}\left(\text{ln}\left(2\pi\right)+\text{ln}\left(\Sigma\left(1-\rho^{2}\right)\right)\right)+\frac{1}{2}\left(\text{ln}(\Sigma\left(1-\rho^{2}\right)\right)-\text{ln}\left(\Sigma)\right)\\
    &-\frac{1}{2}\left(d_{1}^{2}+\sum_{z=2}^{T}\frac{\left(d_t-\rho d_{t-1}\right)^{2}}{1-\rho^{2}}\right)
\end{split}
\end{equation}
Hence the optimal final parameters $ \Theta_{G}^{*}$ are determined by the sum of the two likelihood functions:
\begin{equation}
 \Theta_{G}^{*} = \underset{\omega,\alpha_{0}, \beta_{0}}{\operatorname{argmax}} \hspace{0.2cm} \text{log} \mathcal{L}_\text{VIX} +  \text{log}\mathcal{L}_{x_{1},...,x_{T}}.
\end{equation}
The optimization procedure is implemented using the derivative free unconstrained method of "Nelder-Mead" with random initial starting points. The constraints are included by forcefully returning an artificially unlikely likelihood value in case of constraint breach, thus discouraging the algorithm from choosing parameters outside the optimal set. 

\subsection{RBM} \label{theorbmcond}
Since the RBM is designed to learn the dynamics between all input variables, the inclusion of additional information is straightforward: It is enough to simply extend the amount of visible units by the size of binary formulation for the VIX Index. Hence we feed into the visible layers all the historical data of length $H$ together with the conditioning values to train parameters that describe the joint distribution and use the transformation algorithm described in \cite{RBM} and \ref{appendix}. After the training process, we fix the visible units for the conditioning variables during the entire sampling process, i.e. do not choose random noise as input to these units but rather reset the values for these units as described in \cite{RBM} or \cite{rbmcond}. This way we infer the conditional distribution from the learned joint distribution, as in every sampling step, the values of the hidden layer calculated with \eqref{condind} are conditioned on the same condition. The chosen to be modelled joint distribution consists of daily data from Monday to Friday, hence choosing $H=5$, conditioned on the previous week Friday VIX value.

To encrypt the information from the VIX index in an efficient but still informative way, we round all VIX values to the next smallest integer and lift levels below 10 to be 10, as well as cut levels above 40 to be 40, with the chosen boundaries determined by the spread behaviour of the S\&P500 Log Returns in \ref{spxvix}. This approach allows us to discretize the conditioning space while keeping the most relevant forecasting information: It is clear that a VIX index below 10 clearly indicates a non stressed economy, while a VIX level above 40 shows heightened economic stress. The transformed rounded/ cut-off VIX value is then passed to the visible units corresponding to the VIX Index during training and sampling. 
\subsection{CVAE} \label{cvae}
In order to forecast a conditional distribution, we use a conditional VAE \cite{cvaeori} which encodes and decodes data based an additional labels/conditional values $\textbf{c}$. We then have the following changes in \eqref{eq:ELBO}: the approximate posterior becomes $q_\phi(\textbf{z}|\textbf{x}, \textbf{c})$ and the model likelihood is then  $p_\theta(\textbf{x}|\textbf{z}, \textbf{c})$. This also enables the decoder to generate outputs conditioned on previous levels of the VIX. Additionally, to avoid possible problems with the data range, we scale all values into [0,1]. The modelled multivariate distribution is the same as in the RBM, consisting of a 5-dimensional weekly distribution.

Practically speaking, it is only necessary to increase the number of neurons in the input layers of decoder and encoder by the amount of conditions. To sample from the conditional distribution we sample from the trained decoder as described in \ref{vae}, but pass additionally the conditioning values. Same as with the RBM, the conditioning VIX value is first rounded/cut-off, then transformed and passed to the CVAE encoder/decoder during training/sampling, respectively.
\section{Toy Example} \label{toydata}
In this section we cross check the validity of our models for some toy data. Due to the nature of our model usage, we differentiate between data with and without time dependency. 

\subsection{Gaussian Mixture}
\begin{figure} [h]
    \centering
    \includegraphics[width=0.5\textwidth]{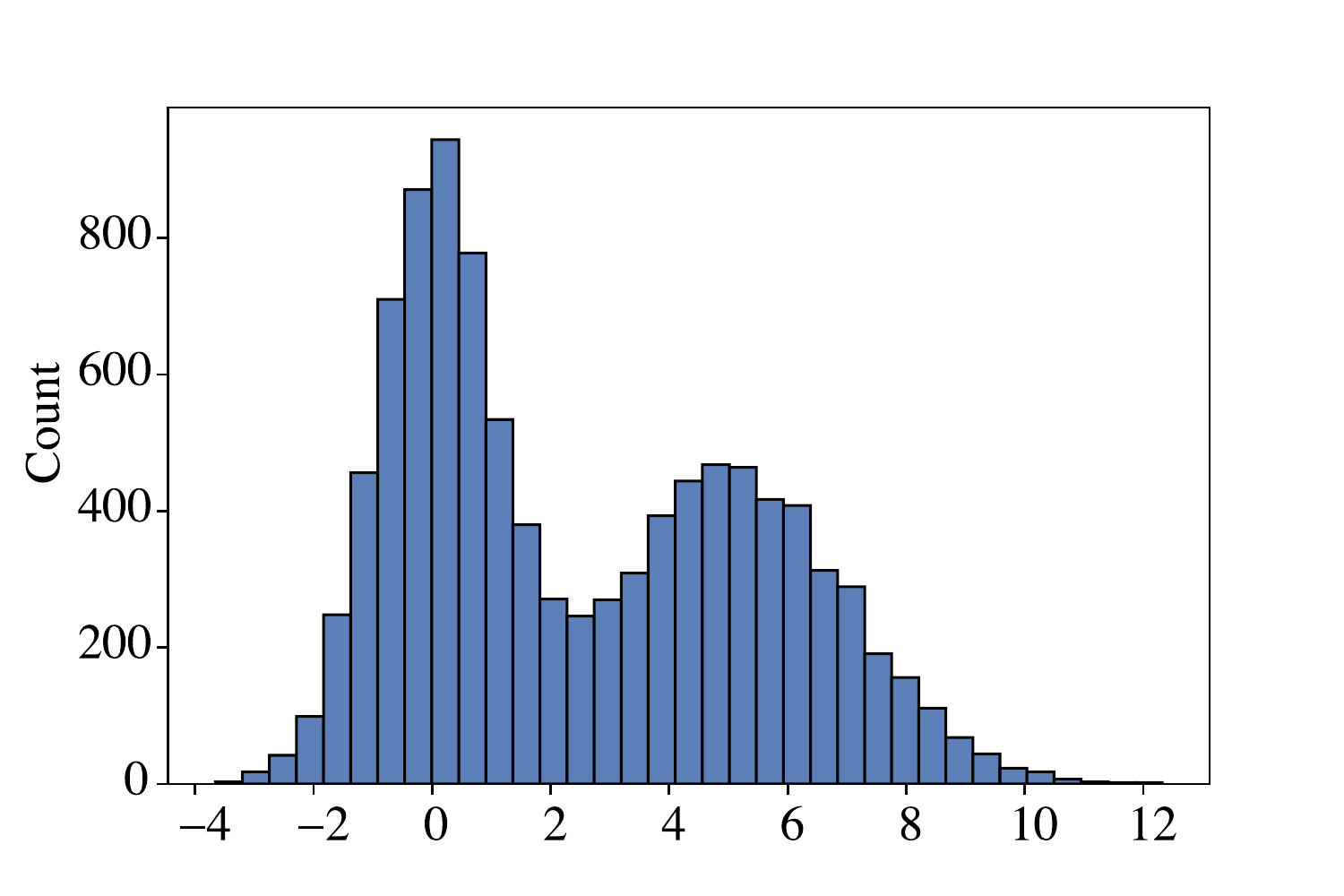}
    \caption[Histogram Toy Gaussian Mixture]{Histogram of toy data from equally weighted Gaussian Mixture of $\mathcal{N}(0,1)$, $\mathcal{N}(5,2)$.}
    \label{GMtoy}
\end{figure}

\begin{figure}
    \centering
    \subfigure[Histogram of bootstrapped samples]{\includegraphics[width=0.45\textwidth]{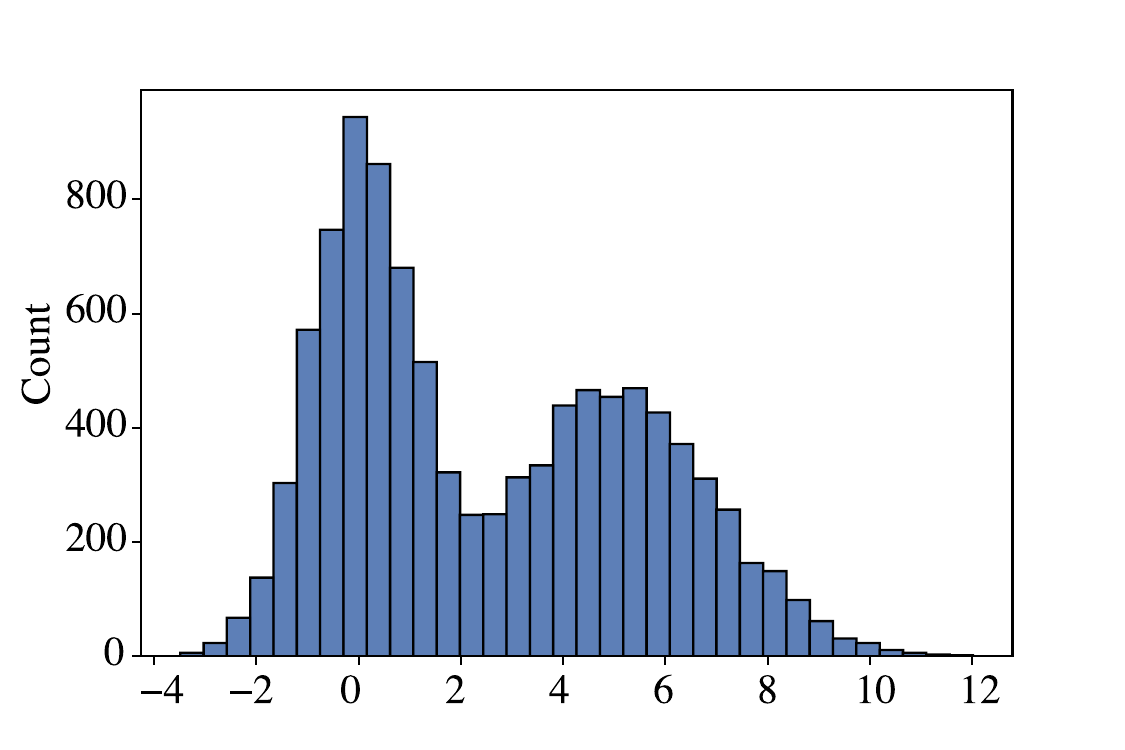}}
    \subfigure[QQ Plot real data vs. bootstrapped samples]{\includegraphics[width=0.45\textwidth]{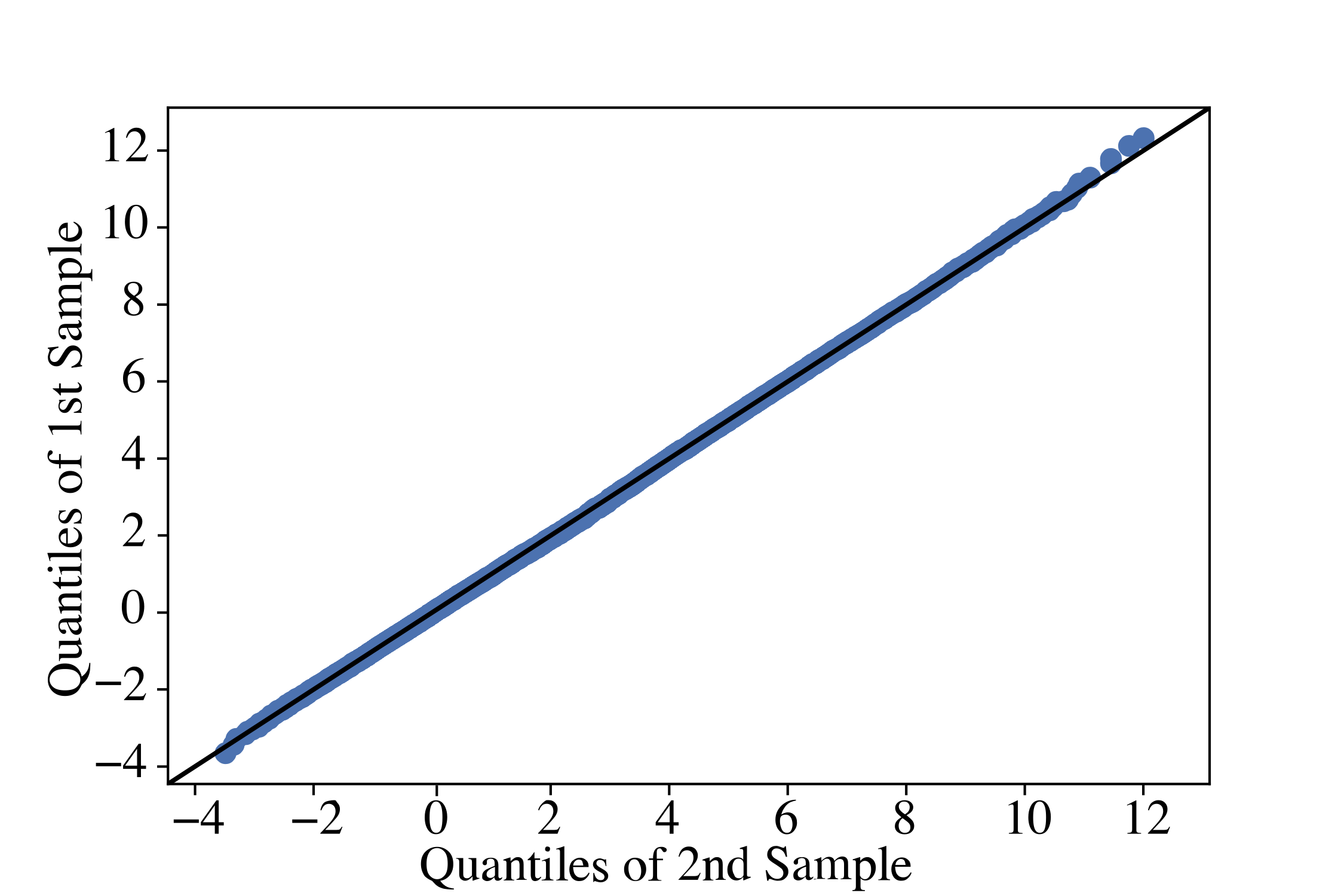}} 
    \subfigure[Histogram of RBM samples]{\includegraphics[width=0.45\textwidth]{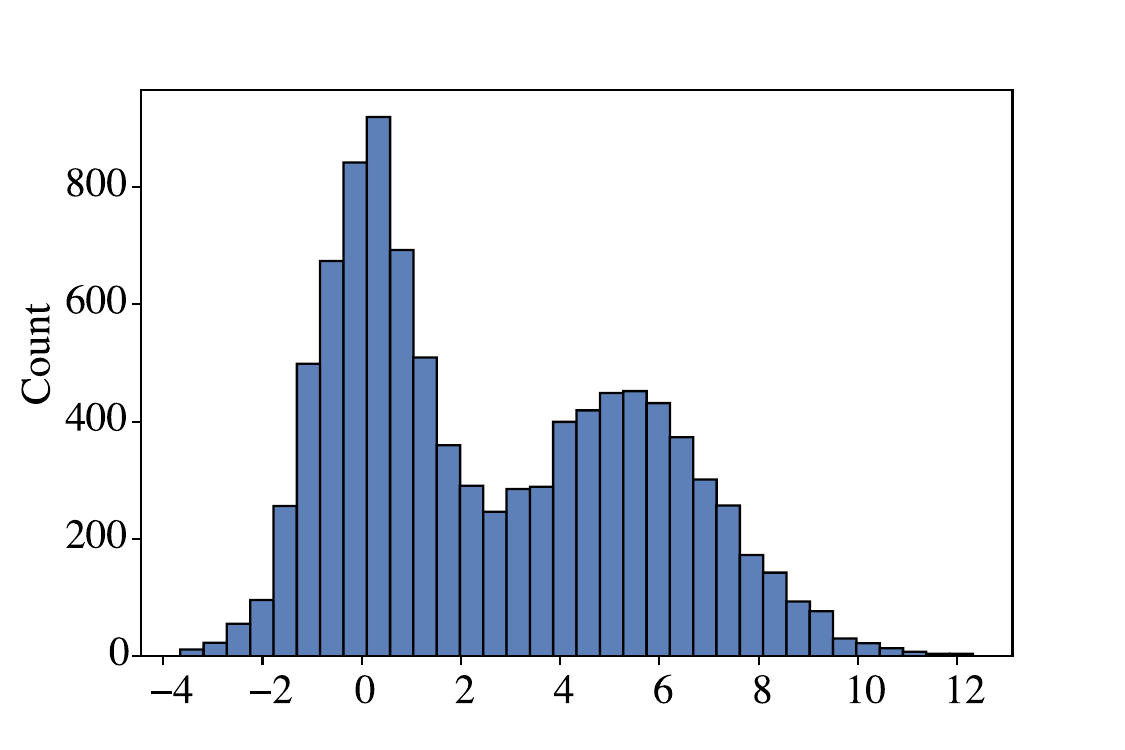}}
        \subfigure[QQ Plot real data vs. RBM samples]{\includegraphics[width=0.45\textwidth]{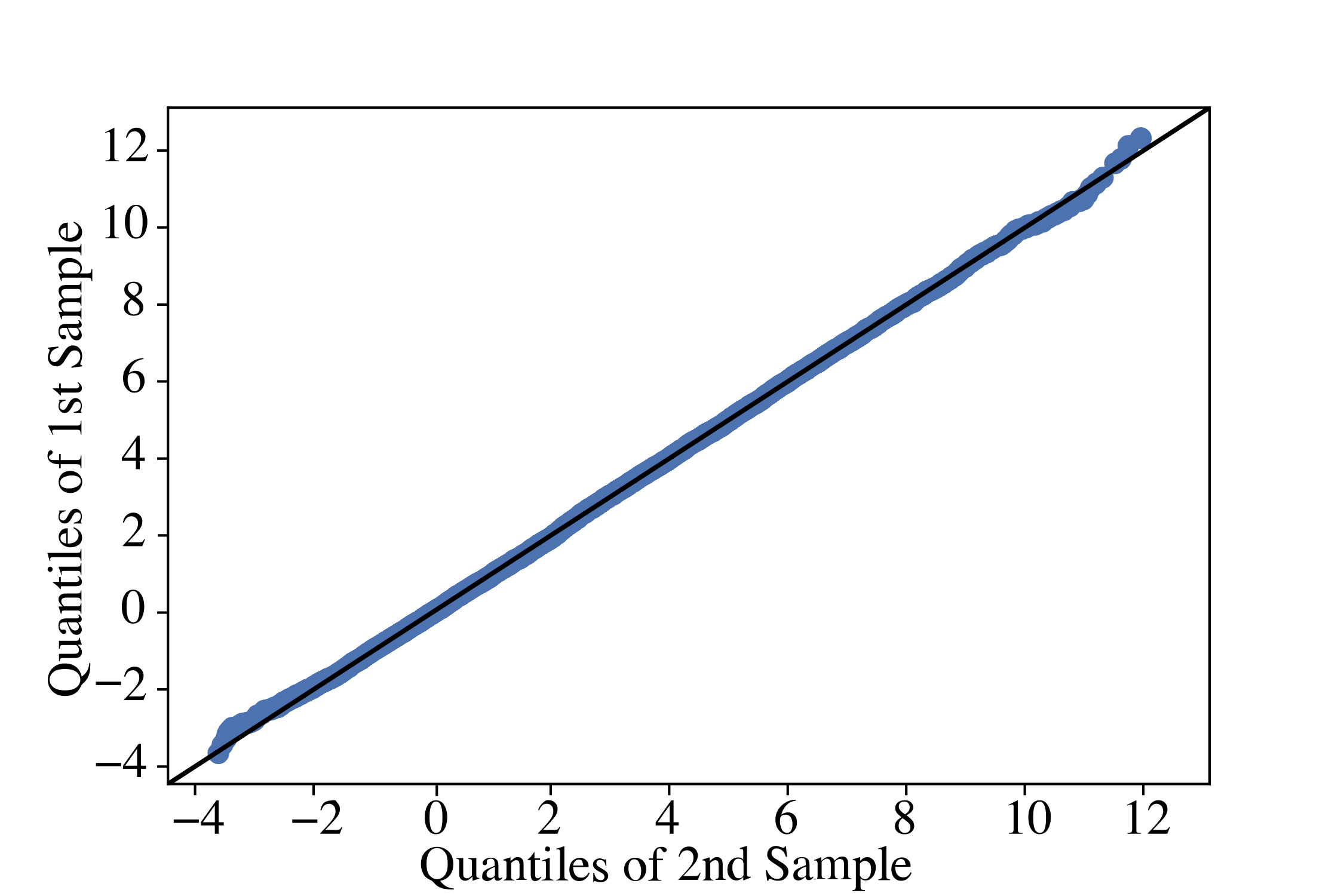}} 
     \subfigure[Histogram of VAE samples]{\includegraphics[width=0.45\textwidth]{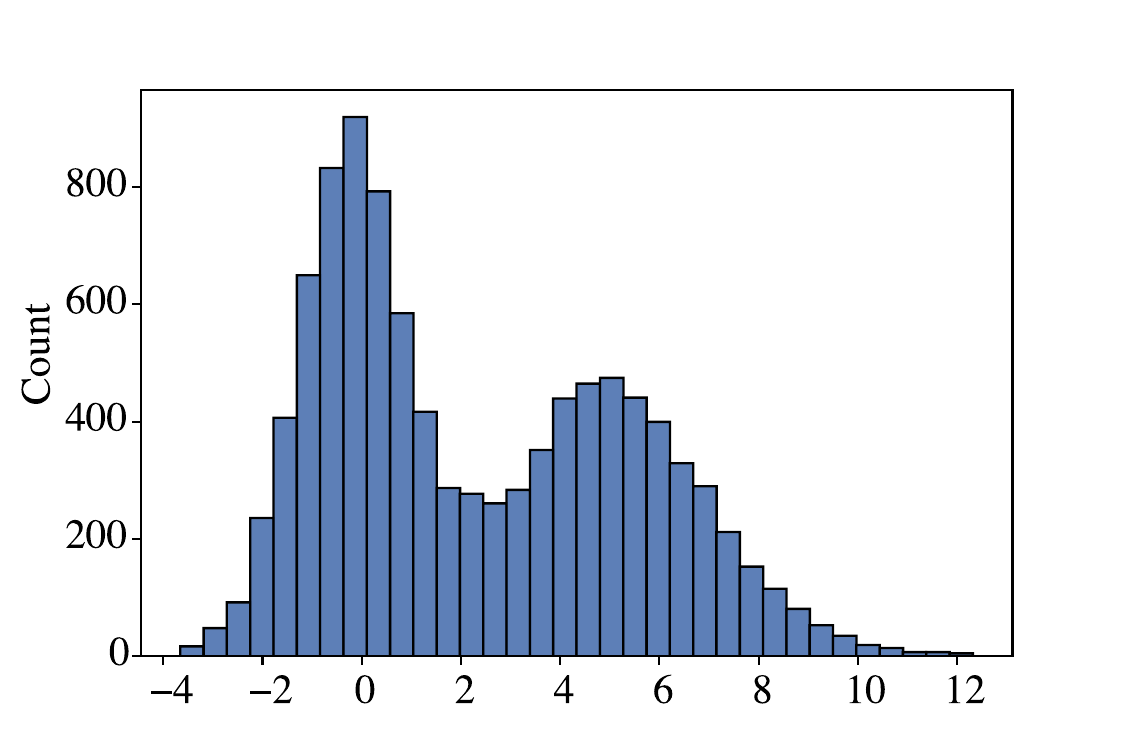}}
    \subfigure[QQ Plot real data vs. VAE samples]{\includegraphics[width=0.45\textwidth]{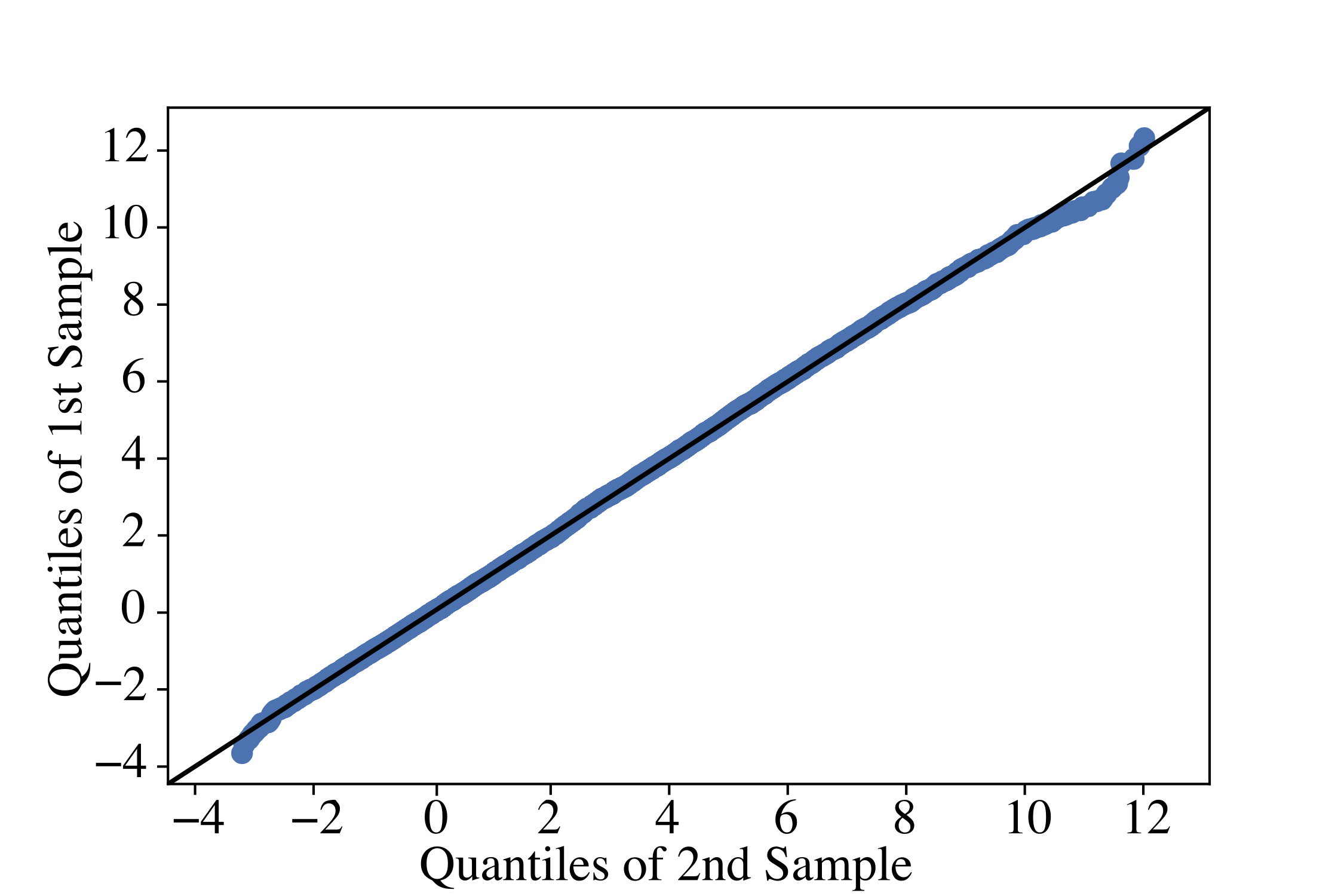}} 
    \caption[Results toy example Gaussian Mixture]{Histograms and QQ Plots of different modelling samples to compare with real data from an equally weighted Gaussian consisting of $\mathcal{N}(0,1)$, $\mathcal{N}(5,2)$.}
    \label{GM}
\end{figure}

For the first part of the validity check we create 10,000 samples of a univariate mixture Normal distribution from two equally weighted Gaussian distributions $\mathcal{N}(0,1)$, $\mathcal{N}(5,2)$ with its histogram in \ref{GMtoy}. Generated samples as histograms and QQ Plots to compare with real data are in reported in \ref{GM}.
In the following, we describe the sampling process for each considered approach. \\
\paragraph{Bootstrap Simulation} We generate 5 samples of 10,000 data points each and average over all paths to report the sample histogram and QQ Plot of the average simulations. \\
\paragraph{GARCH} As the GARCH process by definition describes time series, this toy data is unsuitable for checking the model's validity. Thus, no results are reported for the GARCH in this subsection. \\
\paragraph{RBM} We first convert all values into 16 digit binary numbers according to the algorithm described in \cite{RBM} and thus get an RBM with 16 visible units. We construct the hidden layer with 10 hidden units to work as a information bottleneck. The RBM is trained with the contrastive divergence algorithm for 50,000 epochs and a learning rate of 0.001. Afterwards we sample 5 times from the trained model with 10,000 samples each time and calculate the average values over the 5 sampling times. \\
\paragraph{VAE} For the VAE we design an encoder with input dimension 1, two hidden layers of dimensions 30 and 15 as well as a latent space of dimension 1. Similarly, the decoder has two hidden layers of dimension 15 and 30 each. We train the VAE for 50,000 epochs with a learning rate 0.005. Afterwards we sample 5 times with 10,000 samples each time to recreate the learned distribution.
\subsection{Time dependent data}
\begin{figure}[h]
    \centering
   \subfigure[Squared GARCH samples]{\includegraphics[width=0.45\textwidth]{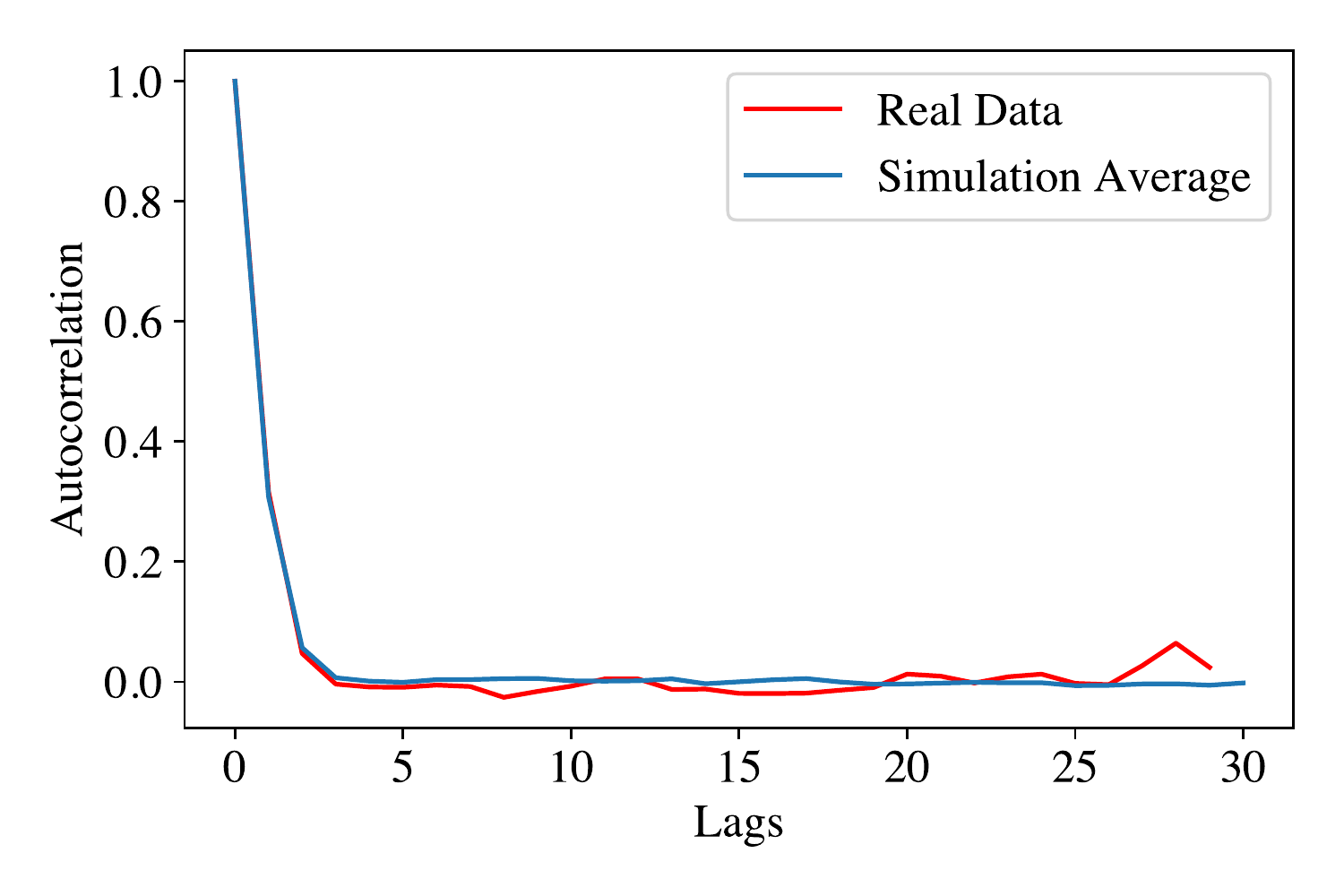}}
    \subfigure[Squared RBM samples]{\includegraphics[width=0.45\textwidth]{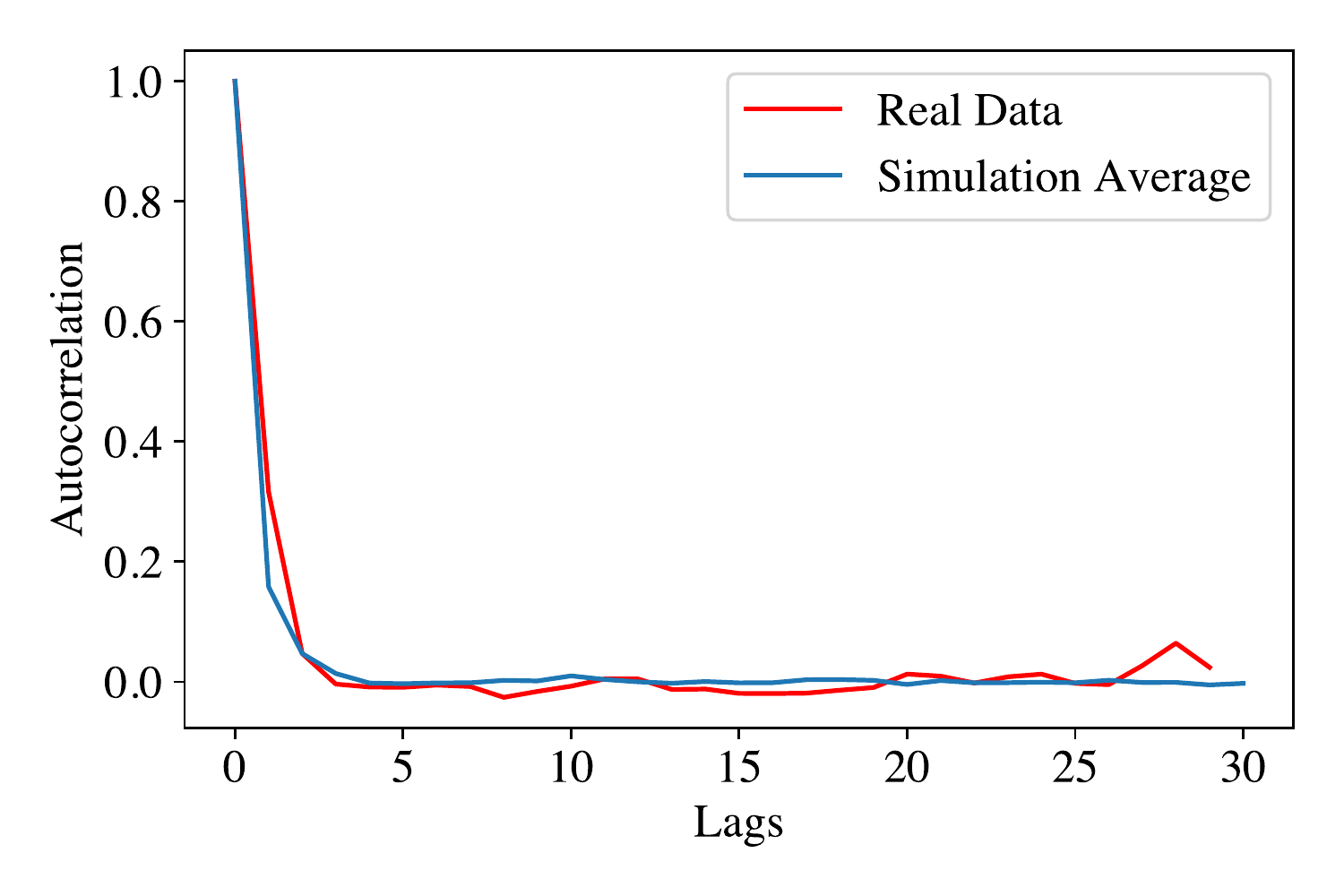}}
    \subfigure[Squared CVAE samples]{\includegraphics[width=0.45\textwidth]{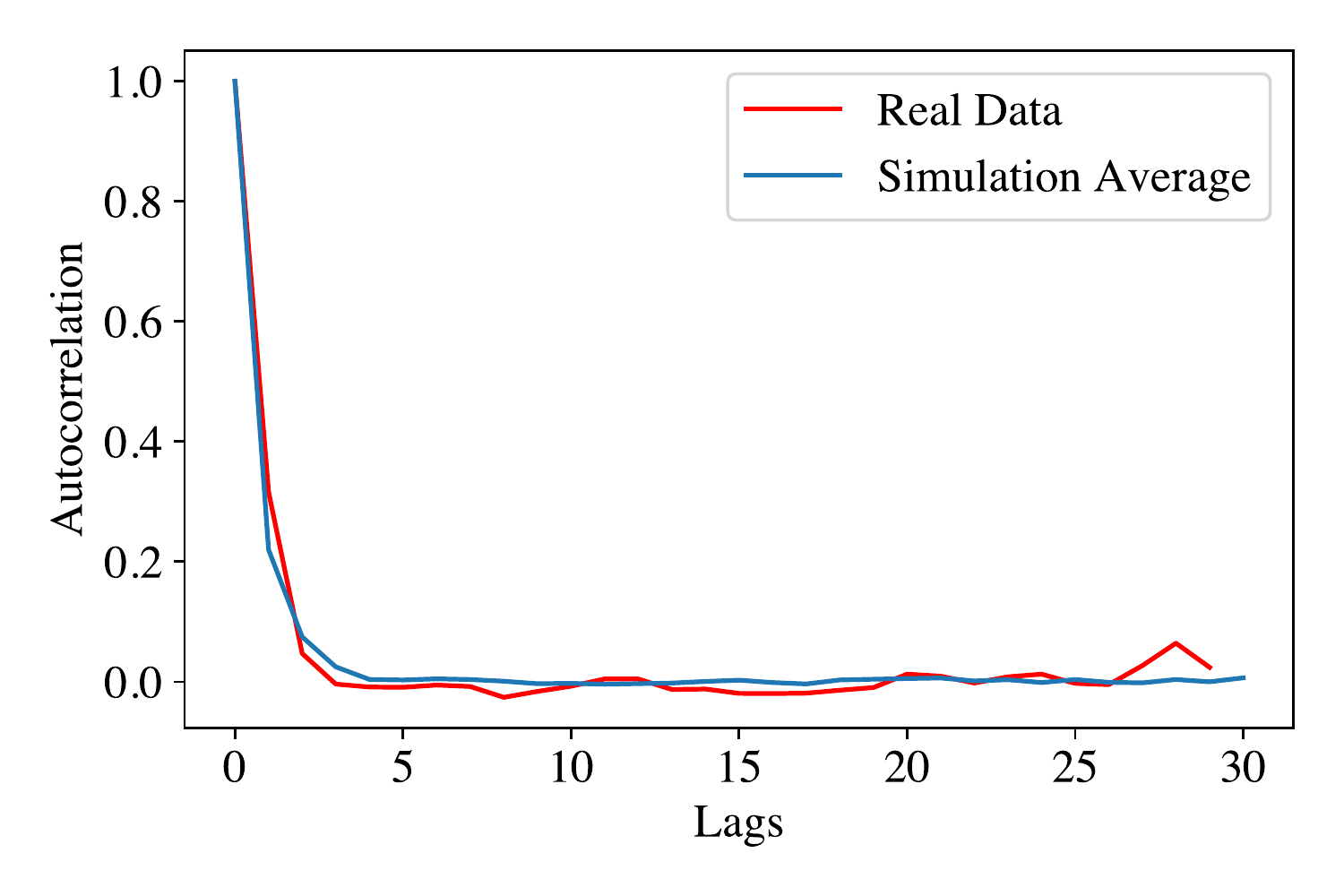}}
    \caption[Autocorrelation of squared samples plot for synthetic time series]{Average autocorrelation of squared generated returns vs. autocorrelation of squared synthetic time series.}
    \label{sqauto}
\end{figure}

In a second step, we generate 5,000 data points following a GARCH process with Gaussian innovations, parameters $\omega=0.7,\alpha_{0}=0.4, \beta_{0}=0.3$ with initial return and volatility equal zero,  to see whether our model specifications for time dependent data are able to capture the trademark process specific properties such as its distribution and volatility clustering. From each approach we generate 5000 simulation steps and compare their statistical properties to that of the real data in form of some summary statistics \ref{statstssims} and plots to verify the capturing of volatility clustering \ref{sqauto}.\\
\paragraph{Bootstrap Simulation}As the historical bootstrap assumes all samples to be i.i.d., it does not make sense to test it on data which is known to violate this assumption, as the GARCH process does. Therefore we do not test the historical bootstrap for this data set.\\
\paragraph{GARCH}We fit a GARCH model with Gaussian innovations on the data points and then simulate conditionally on the real data from time step $t=0$ the next 5000 time steps from the fitted model. We simulate 20 of such time series and use their averages to conduct the statistical tests.\\
\paragraph{RBM}Also here we first transform all data points with the unique transformation from \cite{RBM} into 16-digit binary values and pass as training data for each time step tuple $(r_{t}, r_{t-1})$ as input to the RBM. We thus have 32 units in the visible layer and choose 16 units in the hidden layer, so that the hidden layer creates an information bottleneck. Again, the RBM was trained with the contrastive divergence algorithm. After training, we sample 5000 consecutive time steps of the time series by fixing the last 16 visible units (corresponding to the values from the previous day) first on the known value of $t=0$ and afterwards on the generated next-day-value from the first 16 visible units. We generate each sample following 1000 Gibbs Sampling steps. We simulate 20 times series and average over their values to get the reported results.\\
\paragraph{CVAE} In order to use a VAE to learn time dependence between data, we come back to the conditional VAE (CVAE) as described in \ref{cvae}. We transform the data first by scaling it to $[0,1]$ and, as with the RBM, pass each time step value conditioned on the value of the day before to the CVAE. We thus have an input dimension of 2 for the encoder, add one hidden layers of dimension 20 to the encoder and decoder and choose a latent space of dimension 1. We train the CVAE with a learning rate of 0.005 and a weight of 0.003 for the KL-divergence part of the loss function for 50,000 epochs. Afterwards we generate a time series of length 5000 by consecutively conditioning on the previous day value, starting from the value at $t=0$, and updating the conditioning value in each step with the generated output of the previous step. Also here we generate 20 time series and average over their sorted values for the result evaluation. \\
\\
A detailed comparison of properties of generated vs. real time series with summary statistics over the model generations is given in \ref{statstssims}.
\begin{table*}[]
\centering
\begin{adjustbox}{max width=\textwidth}
\begin{tabular}{lrrrr}
\toprule
                                        &  \multicolumn{1}{c}{\textbf{Synthetic Data}}     & \multicolumn{1}{c}{\textbf{GARCH}}        & \multicolumn{1}{c}{\textbf{RBM}} & \multicolumn{1}{c}{\textbf{CVAE}}  \\ \midrule \midrule
\textbf{Mean}                                    & 0.027         & 0.028 ($\pm$ 0.009)   & 0.055 ($\pm$ 0.014)  & 0.025 ($\pm$ 0.007) \\
\textbf{Standard Deviation}                      & 1.018         & 0.983  ($\pm$ 0.020)  & 1.035 ($\pm$ 0.022)   & 0.997 ($\pm$ 0.013)  \\
\textbf{$\text{1}^{\text{st}}$ percentile }      & -2.677        & -2.674 ($\pm$ 0.131)  & -2.808 ($\pm$ 0.130)  & -2.594 ($\pm$ 0.074) \\
\textbf{$\text{99}^{\text{th}}$ percentile}      & 2.865         & 2.708 ($\pm$ 0.091)   & 2.902 ($\pm$ 0.138)   & 2.750 ($\pm$ 0.105) \\
\bottomrule
\end{tabular}
\end{adjustbox}
\caption[Sample statistics on synthetic time series]{Summary Statistics on synthetic time series data from a GARCH process with Gaussian innovations vs. generated sample paths over 20 simulations. Format: average ($\pm$ 1 Standard deviation).}
\label{statstssims}
\end{table*}
Additionally, we check whether all models can learn the volatility clustering exhibited in a GARCH process. We do so by looking at the autocorrelation functions of some squared generated samples and the squared real data in \ref{sqauto}.
\\
\subsubsection{Summary of Toy Examples}
We have shown in the above experiment on the Gaussian Mixture how the bootstrap simulation is a powerful tool to model the distribution of time independent data. In the case of time dependent data, such as financial time series, the data needs to undergo transformations to fulfil the i.i.d. assumptions of the bootstrap simulation and only afterwards can this approach be checked for suitability as ESG. On the other hand, the GARCH model is by design a model for time dependent processes. It performs well in our toy data, as the real statistical values all lie within a range of two standard deviations from the sample mean. It is however clear that this performance is highly dependent on the correctness of the choice of the underlying distribution, as here the innovation distribution was chosen to be same as in the data generation process. Thus analysis needs to be undertaken on the best fitting parametric distribution for the data before a GARCH model can be properly fitted as an ESG. 

The machine learning approaches RBM and (C)VAE can by construction be used for both time dependent and time independent data. The RBM shows very good results in case of the time independent Gaussian Mixture toy data, clearly showing its distribution modelling capabilities. However, it does not seem to capture time dependence, as seen in the too low autocorrelation in the squared generated samples for the synthetic GARCH process. Also, the data statistics in \ref{statstssims} are the most far off among all models especially in mean and standard deviation, possibly due to not enough training epochs and thus unfit parameter choice. Possibly, more careful and extensive training could lead to better results, although the uncertainty about capturing time dependent properties still remains. We note that training, calibration and samlping from the RBM were the most time consuming.

The (C)VAE also shows good results for the Gaussian Mixture experiment. Its sample statistics for the synthetic GARCH process perform similarly to those of the GARCH model and the standard deviation of mean and standard deviation throughout all 20 samples are the lowest among all compared models, possibly indicating a better convergence of the model. We see that on average the captured volatility clustering is better than with the RBM, as the average sampled autocorrelation in the squares is higher in the first lag (though still lower than the real data) and also that the shown autocorrelation of squares functions are in general better fitting to that of the real data than the function of the RBM. An additional advantage of the CVAE over the RBM is its faster and easier calibration, training and sampling. Based on all these results, the CVAE appears to be a promising candidate for an application as PiT ESG.

\section{Real data modelling} \label{realdatamodelling}
We now use the methods\footnote{Details on the parameters and network architecture of RBM and CVAE used in this section are reported in \ref{paramgrid}.} described in section \ref{setup} on real data and generate 500 paths of the conditional distribution \ref{conddist}. First we test the quality of model fits, for which we, as with the toy examples, present some statistical analyses for the sample paths. Afterwards we additionally evaluate the forecasting abilities of each model by using the model projections on a stop loss strategy and compare the performances of the thus simulated portfolios, simultaneously checking for model robustness against changes in the input data.

Our total data set consists of daily Log Returns on the closing prices on the S\&P500 as well as the VIX Index from Yahoo Finance. The entire data set dates from 03 January 2000 to 18 December 2020, with 194 days missing due to national American holidays without trading. These missing days were backfilled with the last existing value for the levels, thus leading to backfilled log-returns of 0 - which is economically plausible with the fact that these days do not affect the returns of a portfolio containing futures on the indices. 
\subsection{Statistical Analysis}\label{realstats}
After training/fitting all the models to the data set, we choose the date 2018-01-01 as starting date for our simulation process to generate, following the findings in \cite{voladuration} about the duration of volatility regimes to be between 50 and 80 days, the future paths of the next three months (65 days) to check the model fitting qualities. Results on the comparison of statistical properties between generated samples and real time series measuring the modelling capabilities on the risk factor distribution in general are reported in \ref{statstsreal}. Afterwards we examine in detail the replication quality of known stylized facts of the S\&P500 Index like volatility clustering, heavy tails, no to low autocorrelation, as detailed in \cite{equitystylizedfacts}.

\begin{figure}
\centering  
\subfigure[Data vs. Normal]{\includegraphics[width=0.32\linewidth, trim=2.7cm 0.3cm 3.9cm 1.2cm, clip]{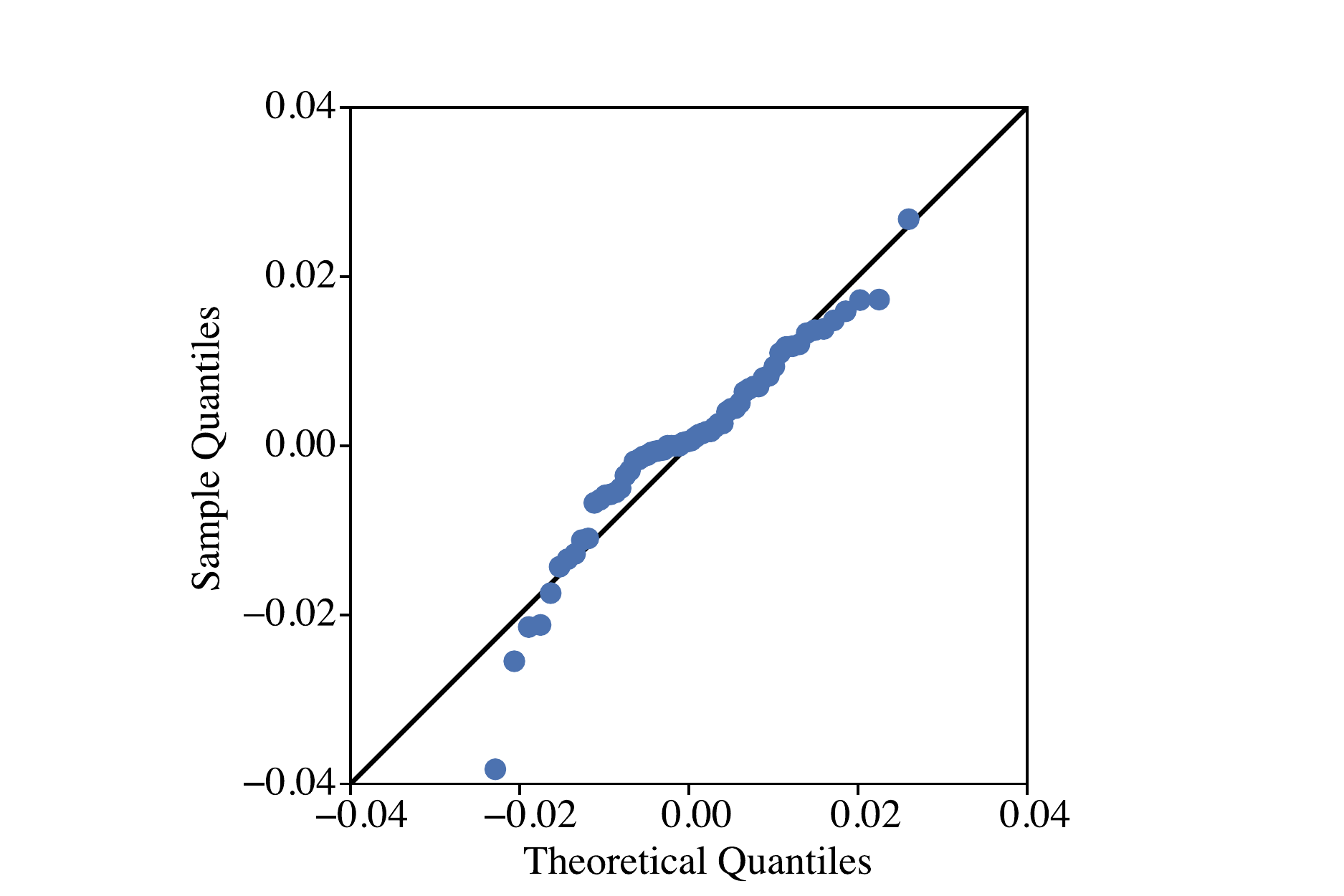}}
\subfigure[FHS vs. Normal]{\includegraphics[width=0.32\linewidth, trim=2.7cm 0.3cm 3.9cm 1.2cm, clip]{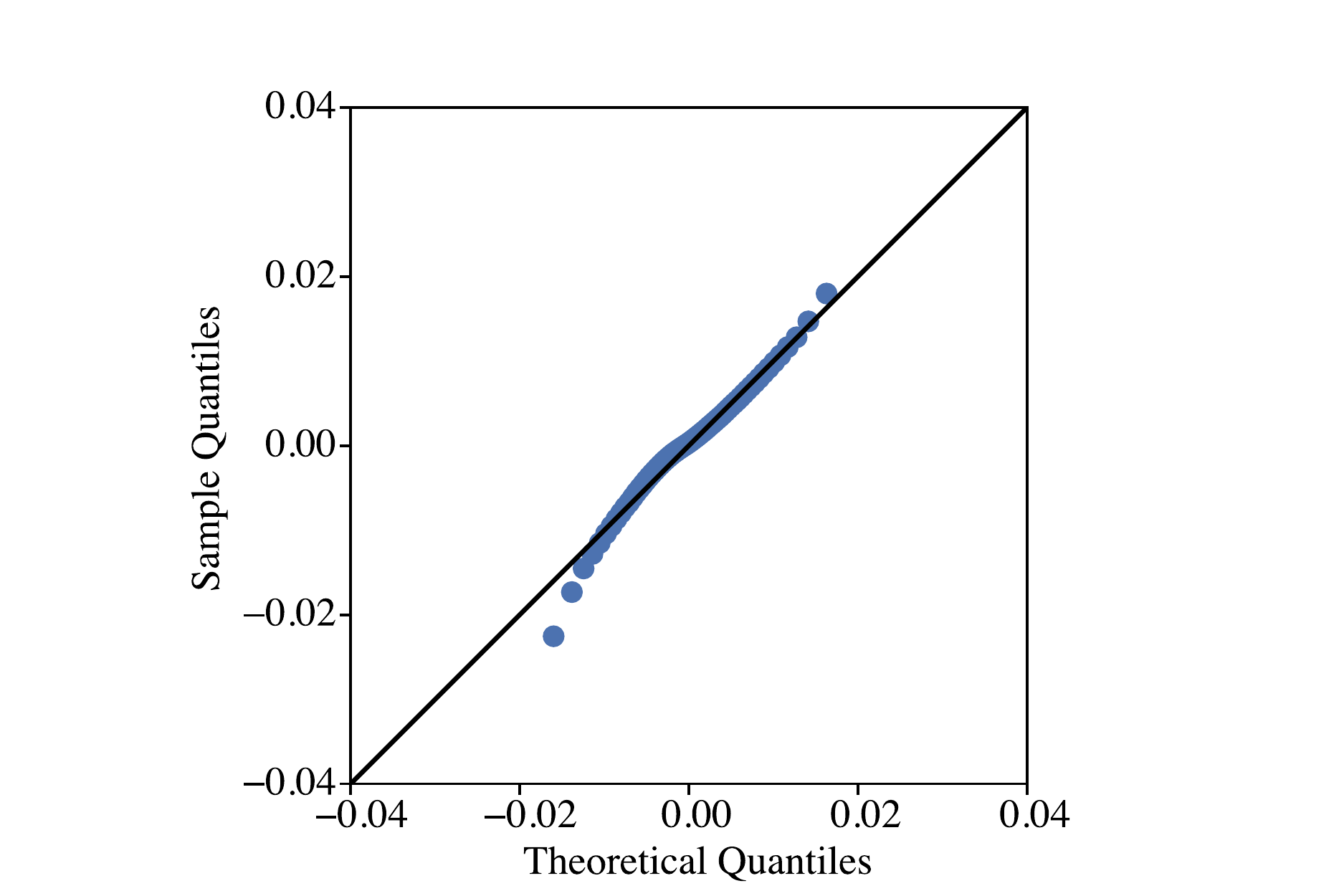}}
\subfigure[Data vs. FHS]{\includegraphics[width=0.32\linewidth, trim=2.7cm 0.3cm 3.9cm 1.2cm, clip]{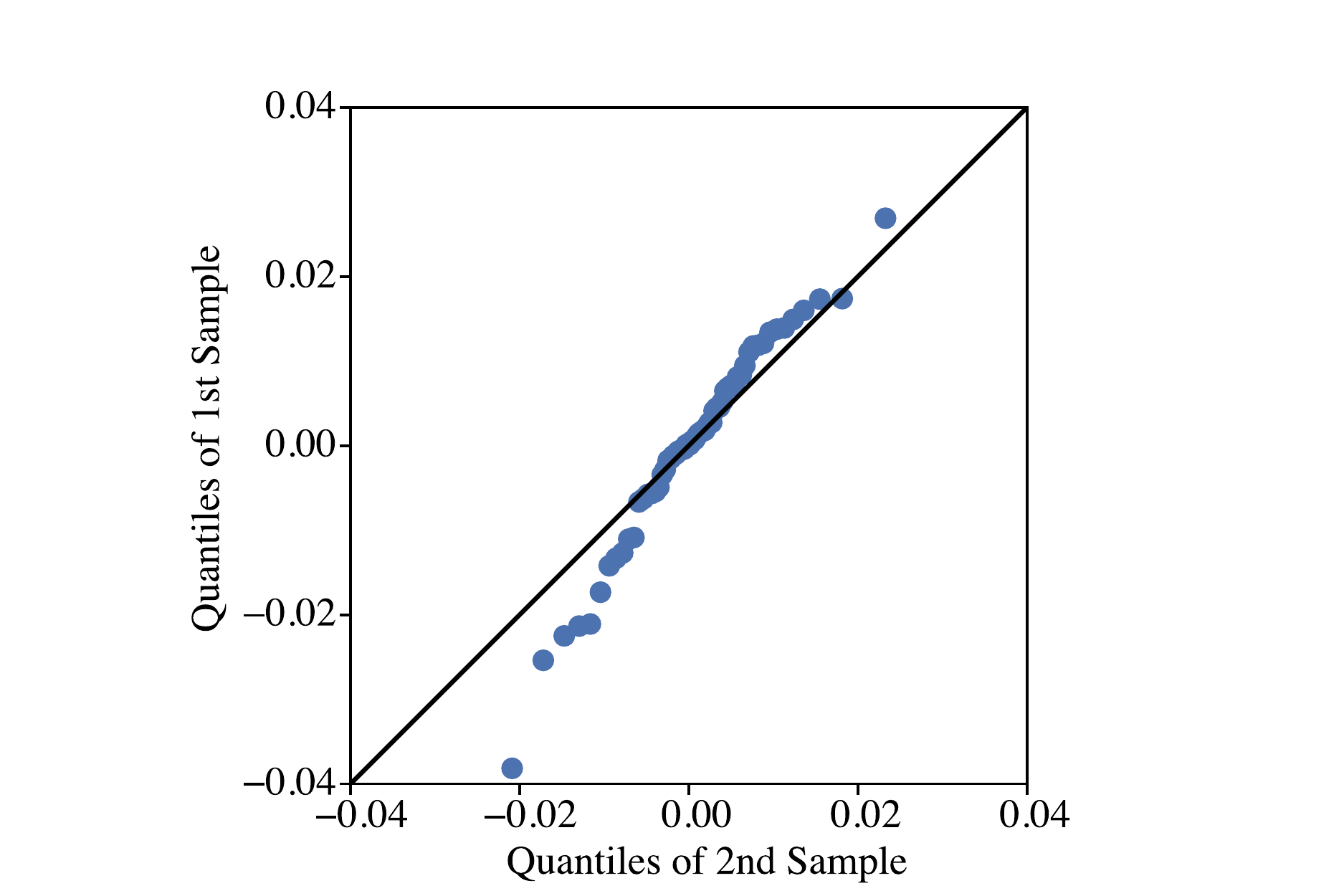}}
\subfigure[Data vs. Normal]{\includegraphics[width=0.32\linewidth, trim=2.7cm 0.3cm 3.9cm 1.2cm, clip]{figures/Real_Data/Stats/realvsnormal.pdf}}
\subfigure[GARCH vs. Normal]{\includegraphics[width=0.32\linewidth, trim=2.7cm 0.3cm 3.9cm 1.2cm, clip]{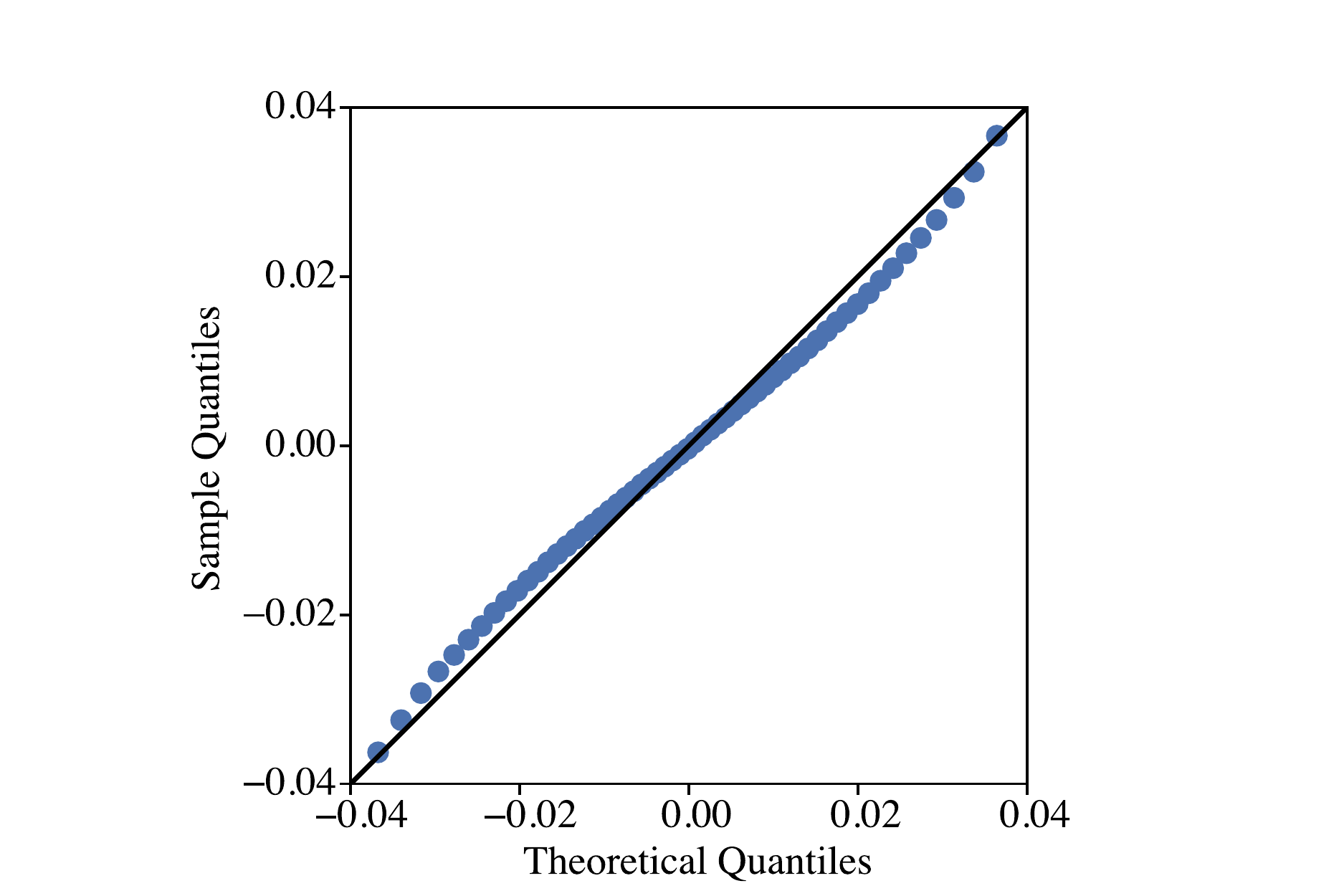}}
\subfigure[Data vs. GARCH]{\includegraphics[width=0.32\linewidth, trim=2.7cm 0.3cm 3.9cm 1.2cm, clip]{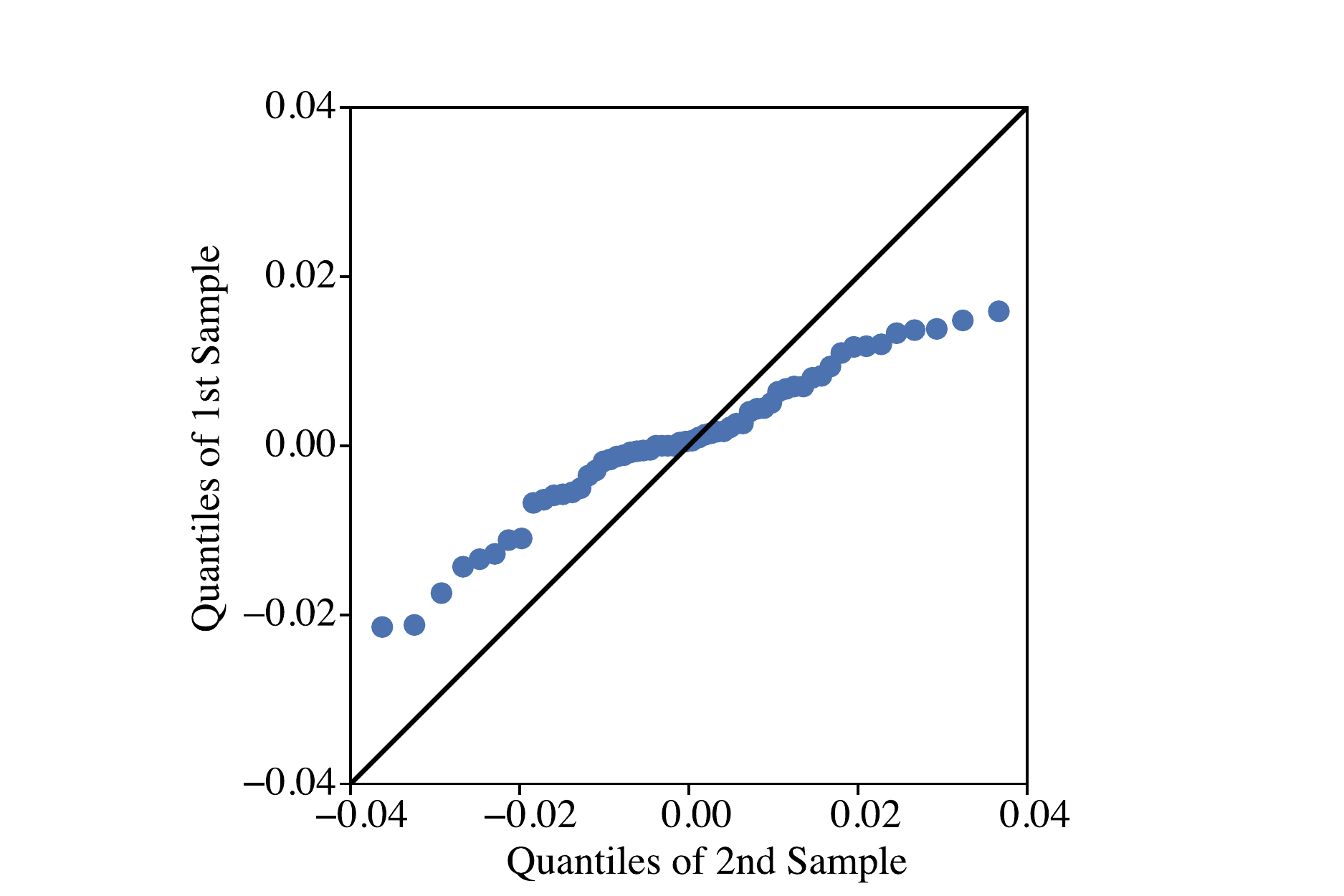}}
\subfigure[Data vs. Normal]{\includegraphics[width=0.32\linewidth, trim=2.7cm 0.3cm 3.9cm 1.2cm, clip]{figures/Real_Data/Stats/realvsnormal.pdf}}
\subfigure[RBM vs. Normal]{\includegraphics[width=0.32\linewidth, trim=2.7cm 0.3cm 3.9cm 1.2cm, clip]{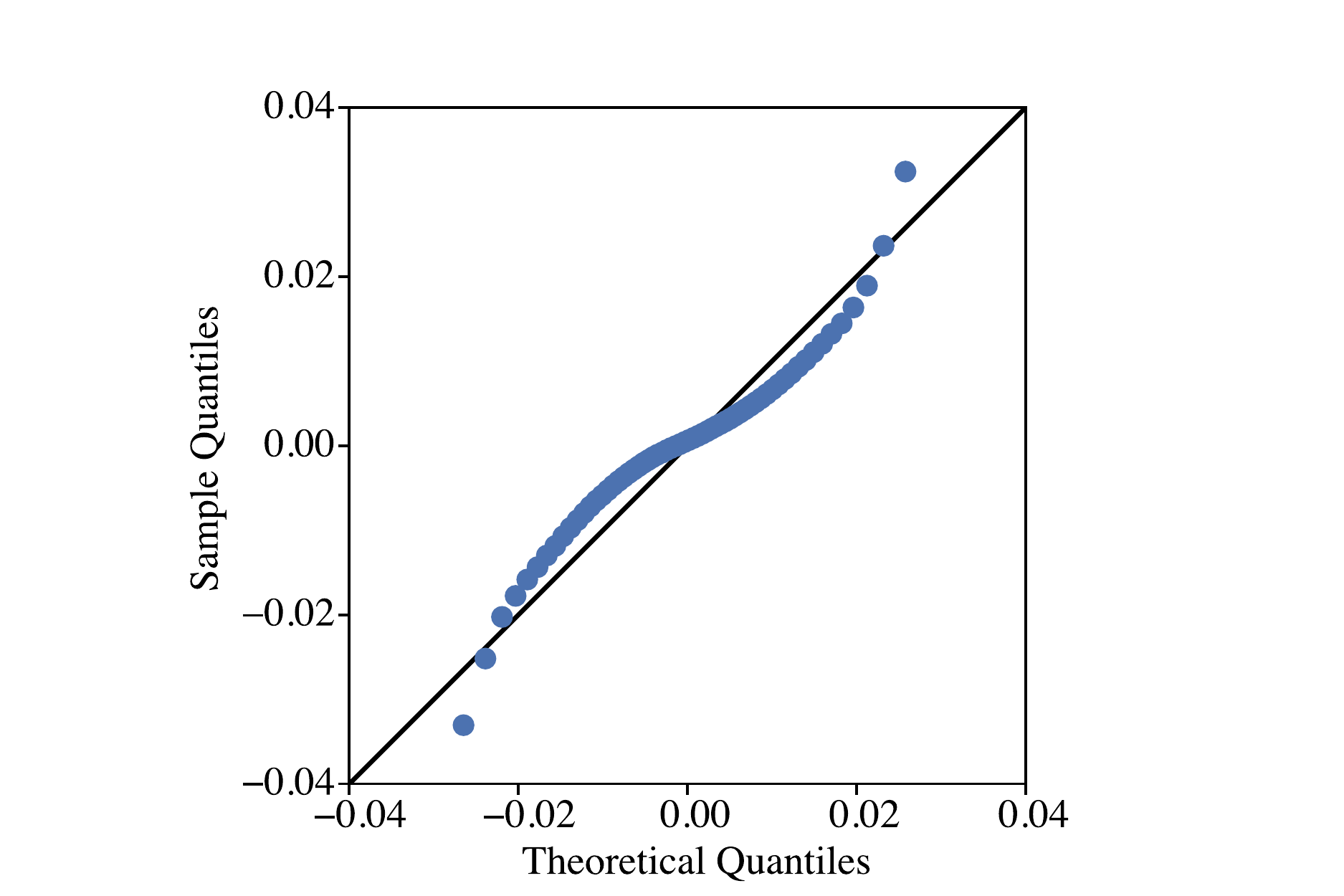}}
\subfigure[Data vs. RBM]{\includegraphics[width=0.32\linewidth, trim=2.7cm 0.3cm 3.9cm 1.2cm, clip]{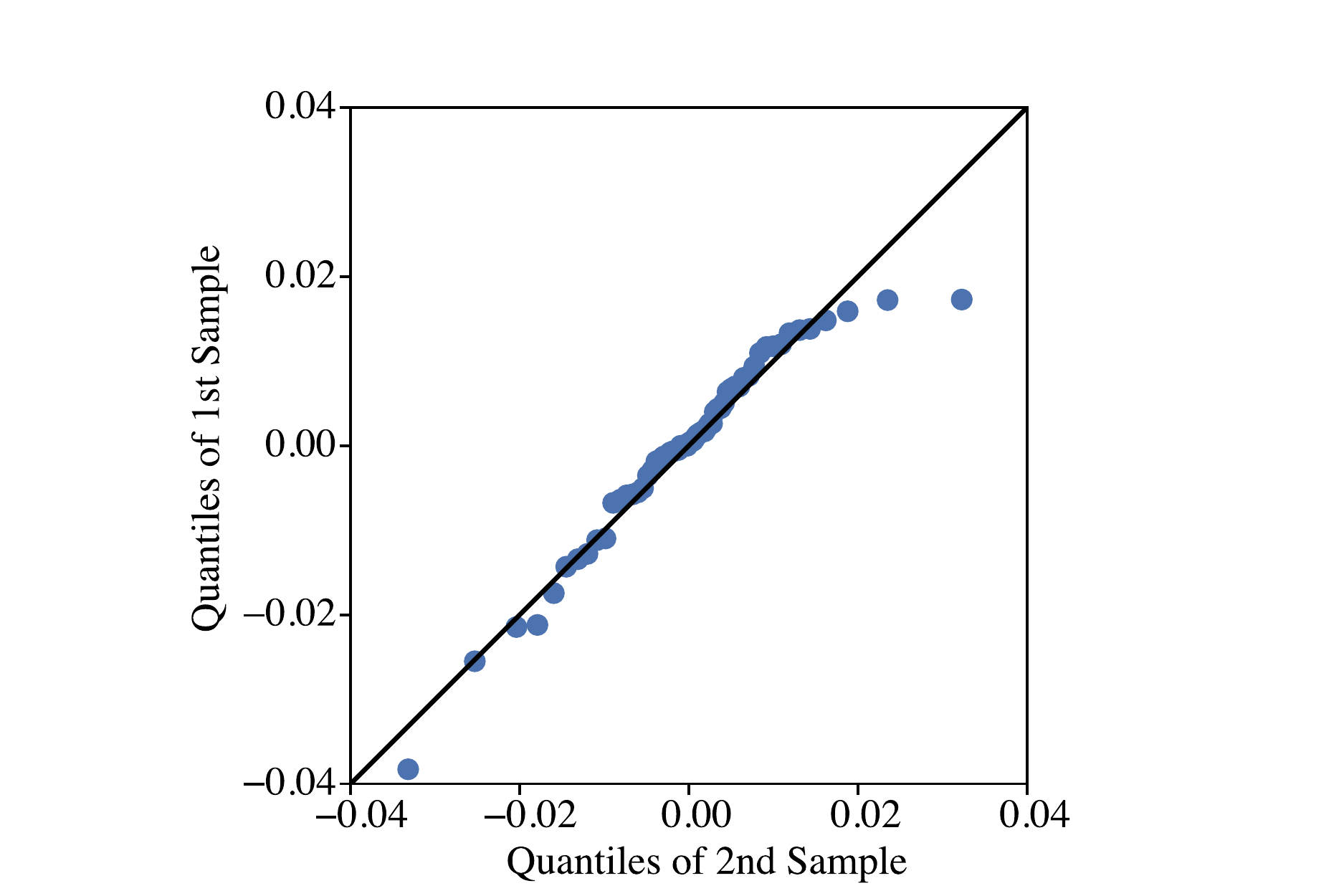}}
\subfigure[Data vs. Normal]{\includegraphics[width=0.32\linewidth, trim=2.7cm 0.3cm 3.9cm 1.2cm, clip]{figures/Real_Data/Stats/realvsnormal.pdf}}%
\subfigure[CVAE vs. Normal]{\includegraphics[width=0.32\linewidth, trim=2.7cm 0.3cm 3.9cm 1.2cm, clip]{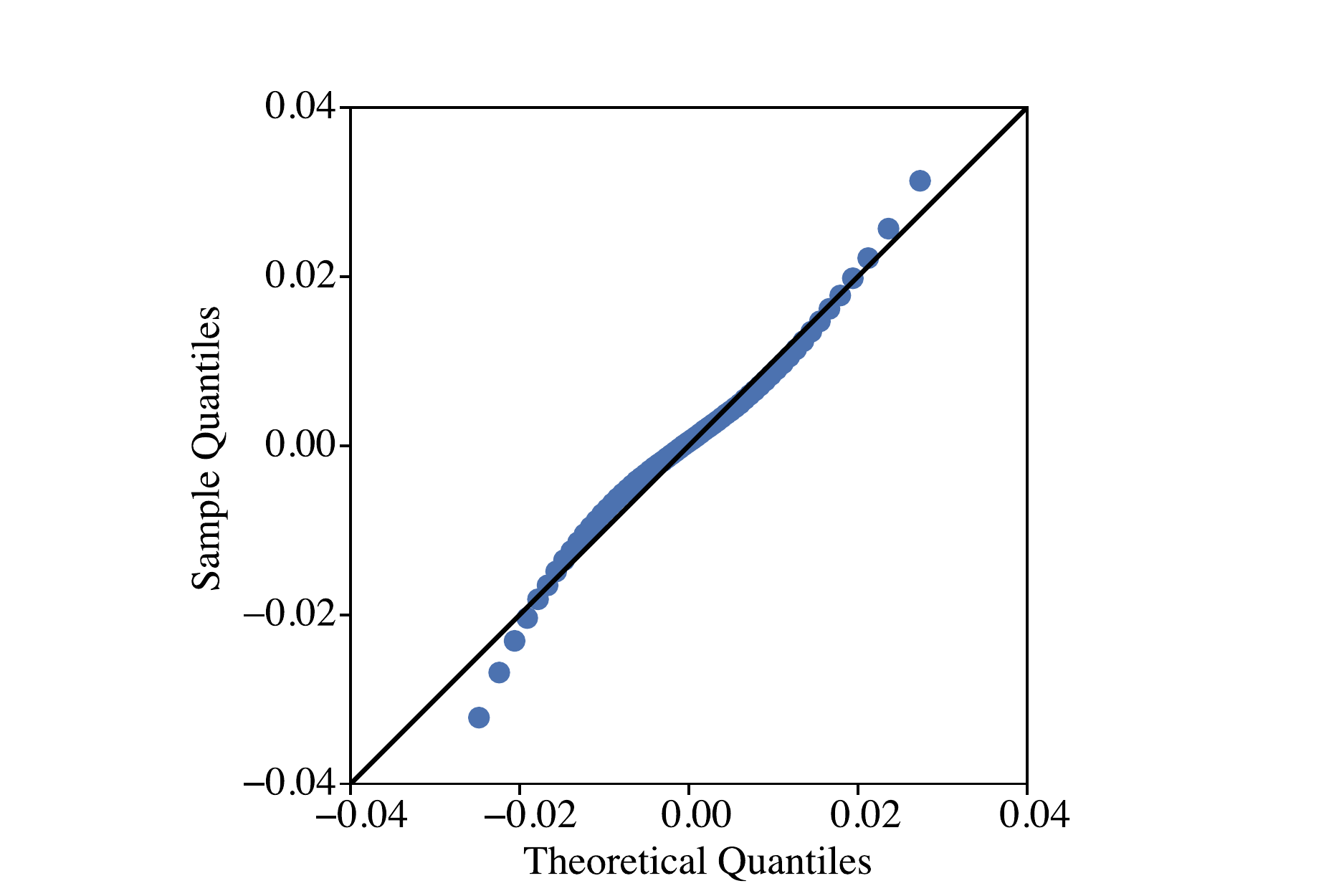}}%
\subfigure[Data vs. CVAE]{\includegraphics[width=0.32\linewidth, trim=2.7cm 0.3cm 3.9cm 1.2cm, clip]{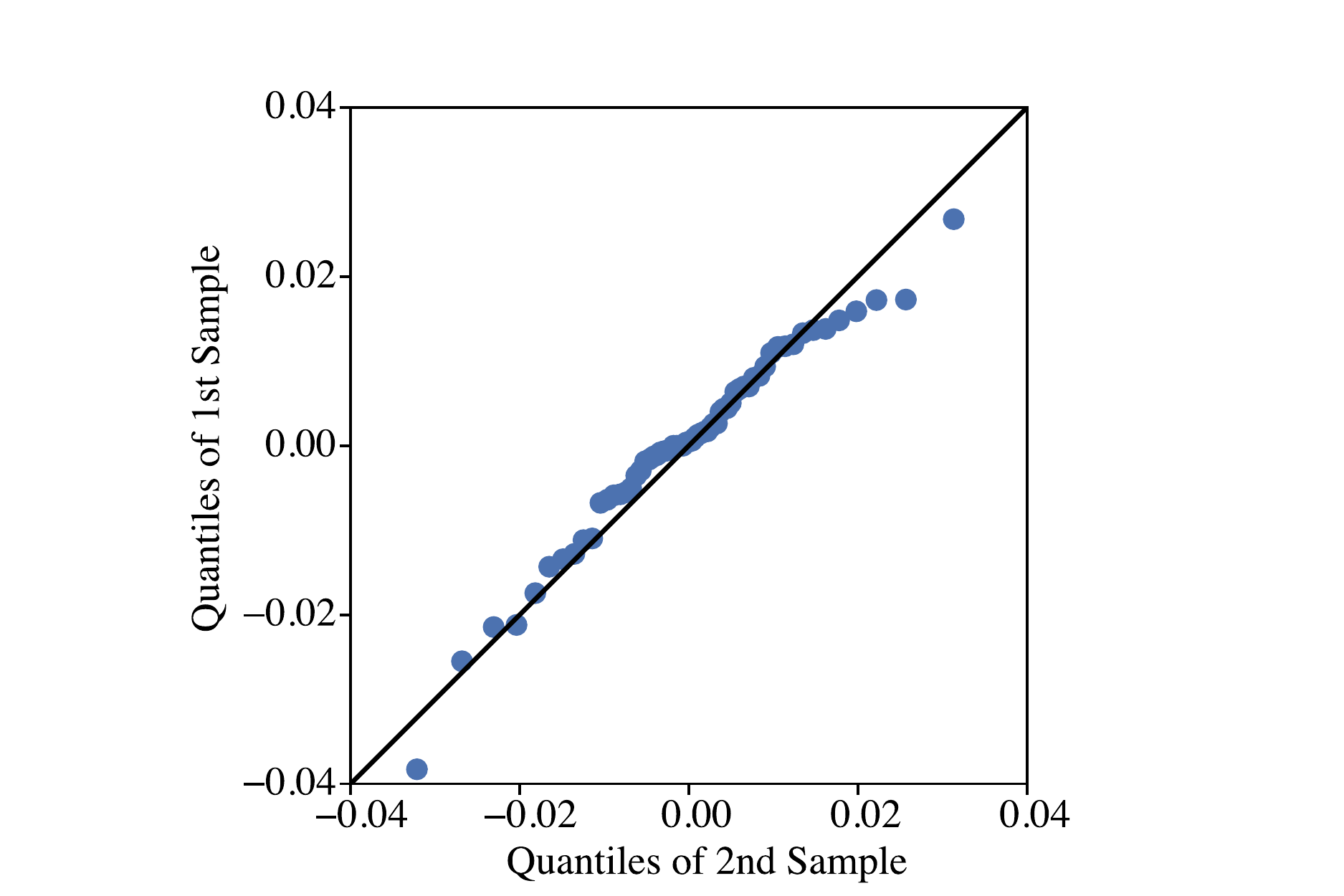}}%
\caption[QQ Plots in real data experiment]{QQ Plots of real Log Returns from 13 weeks test period and generated samples: a)-c): Filtered Historical Simulation; d)-f): GARCH; g)-i): RBM; j)-l): CVAE.}
\label{realQQ}
\end{figure}

Beginning with the summary statistics of the modelled distributions in \ref{statstsreal}, it seems that the both the FHS as well as the GARCH model are unable to capture statistics beyond the first moment. The FHS underestimated the standard deviation by far, while the GARCH model overestimates it. Both RBM and CVAE do better at estimating the first two moments, with the RBM however also showing a tendency to overestimate the standard deviation. Judging by the percentile values of the real data, we see a particularly heavy left tail, which is best modelled by the CVAE, followed by the RBM. The FHS and GARCH struggle to capture this stylized fact. Similar observation is visible in the right tail, where all generations are too light. Out of all tested models, the generative networks with the same amount of training epochs have the best approximations with the CVAE replicating better the second moment and the right tail (though still too light compared to real data). The RBM on the other hand generates a heavier left tail, but the real standard deviation lies well outside the range of two standard deviations from the sample mean and the right tail is worse off than that of the CVAE. 
\begin{table*}[h]
\centering
\begin{adjustbox}{max width=\textwidth}
\begin{tabular}{lrccc}
\toprule
          & \multicolumn{1}{c}{\textbf{Mean}}                  & \multicolumn{1}{c}{\textbf{Standard Deviation}}    & \multicolumn{1}{c}{\textbf{$\text{1}^{\text{st}}$ percentile}} & \multicolumn{1}{c}{\textbf{$\text{99}^{\text{th}}$ percentile}}      \\ \midrule \midrule
\textbf{Real Data} &\multicolumn{1}{c} {-0.0002}             & \multicolumn{1}{c}{0.0121}                & \multicolumn{1}{c}{-0.0395}                & \multicolumn{1}{c}{0.0207}                \\ 
\textbf{FHS}       & 0.0000  ($\pm$ 0.001) & 0.0089  ($\pm$ 0.001) & -0.0237  ($\pm$ 0.006) & 0.0199  ($\pm$ 0.009) \\ 
\textbf{GARCH}  & -0.0002 ($\pm$ 0.003) & 0.0244  ($\pm$ 0.006) & -0.0596 ($\pm$ 0.019)  & 0.0598 ($\pm$ 0.020)  \\ 
\textbf{RBM}       & -0.0001 ($\pm$ 0.002)   &    0.0147 ($\pm$ 0.003) &  -0.0400 ($\pm$ 0.017) &    0.0371 ($\pm$ 0.017)                   \\ 
\textbf{CVAE}      & -0.0002  ($\pm$ 0.001) & 0.0126  ($\pm$ 0.003) & -0.0350  ($\pm$ 0.012)  & 0.0293 ($\pm$ 0.007)\\   \bottomrule                
\end{tabular}
\end{adjustbox}
\caption[Sample statistics on real data experiment, part I]{Summary Statistics over simulation period on real S\&P500 Log Returns vs. generated sample paths over 500 simulations. Format: average ($\pm$ 1 Standard deviation).}
\label{statstsreal}
\end{table*}
In \ref{realQQ} we show different QQ-Plots to compare the real log return versus the generated distributions together with a comparison of both real and generated distribution to a fitted normal distribution as benchmark.
The QQ Plots confirm our findings from \ref{statstsreal}: Both FHS and GARCH struggle to replicate the distribution, with the FHS doing slightly better still. Comparing the Data vs. Normal and GARCH vs. Normal plots, it is clearly visible that the heaviness of the tails can not be replicated by the chosen specification. Similarly the FHS is unable to produce as heavy a tail as in the real data. Comparing the quantiles of both RBM and CVAE against the fitted normal distribution, a heavier left tail similar to that in the real data is visible. Contrary to the real data however, the right tail in both cases is slightly lighter than in the normal distribution.

Another distribution property known to be a stylized fact of the S\&P Index is existing skewness more often negative than positive and a kurtosis higher than 3 \cite{equitystylizedfacts}. Results herefore together with the interquartile range as a measure of deviation are examined in \ref{realstylized}. From the real data we observe indeed a negative skewness, but however a lower kurtosis than 3, possibly resulting from the small size of the data sample. 
\begin{table*}[h]
\centering
\begin{adjustbox}{max width=\textwidth}
\begin{tabular}{lccc} \toprule
                        & \textbf{Interquartile Range}       & \textbf{Skewness}                  & \textbf{Kurtosis}               \\ \midrule \midrule
\textbf{Real Data}               & 0.0120                    & -1.0644                   & 2.1718                 \\ 
\textbf{FHS}                     & 0.0089 ($\pm$ 0.002)      & -0.4301  ($\pm$ 0.7228)   & 2.1655 ($\pm$ 2.4502)  \\ 
\textbf{GARCH}          & 0.0253 ($\pm$ 0.004)      & -0.0026  ($\pm$ 1.0698)   & 3.1676 ($\pm$ 0.6641) \\ 
\textbf{RBM}                     & 0.0117 ($\pm$ 0.003)      & -0.2056 ($\pm$ 1.5342)    &  6.6834 ($\pm$ 4.9421)   \\ 
\textbf{CVAE}                    & 0.0123 ($\pm$ 0.003)      & -0.3782 ($\pm$ 0.5822)    & 2.0284 ($\pm$ 1.9172)     \\          
\bottomrule
\end{tabular}
\end{adjustbox}
\caption[Sample statistics on real data experiment, part II]{Statistics comparisons over simulation period on real S\&P500 Log Returns and generated sample paths over 500 simulations, Format: average ($\pm$ 1 Standard deviation).}
\label{realstylized}
\end{table*}
We first observe that all models seem to struggle with the skewness, with only the RBM managing to capture the real value in a range of one standard deviation from the sample mean, possible simply from the fact that its standard deviation is so high. It produced a kurtosis far to large with a standard deviation almost as big as the value itself. We see again that the  GARCH model does significantly worse than all other models in all statistics, FHS and CVAE provide similar results for both skewness and kurtosis, with the CVAE however having lower standard deviation and an IQR closer to the real one. 

\begin{figure}[h]
\centering  
\subfigure[FHS]{\includegraphics[width=0.475\linewidth]{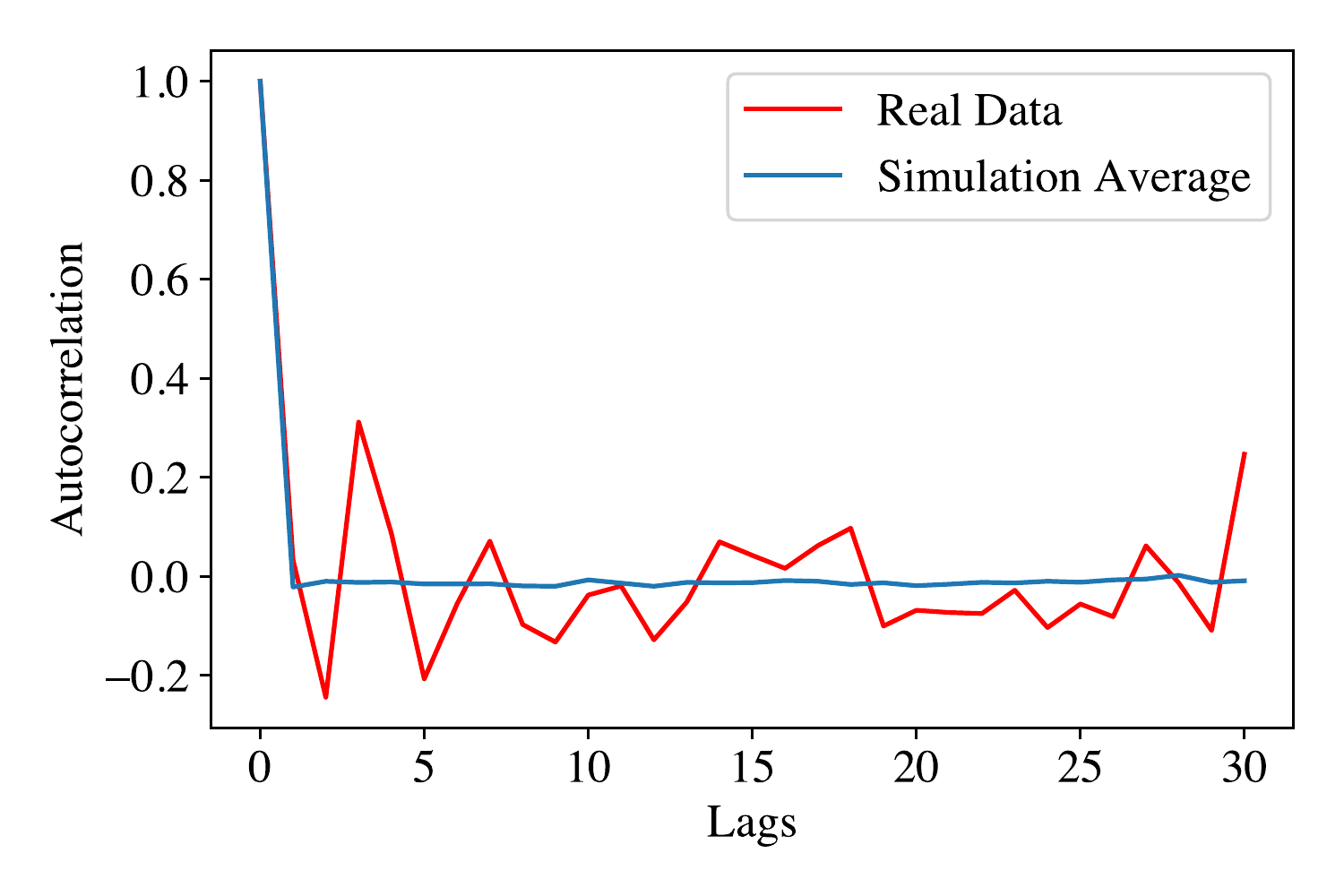}}
\subfigure[GARCH]{\includegraphics[width=0.475\linewidth]{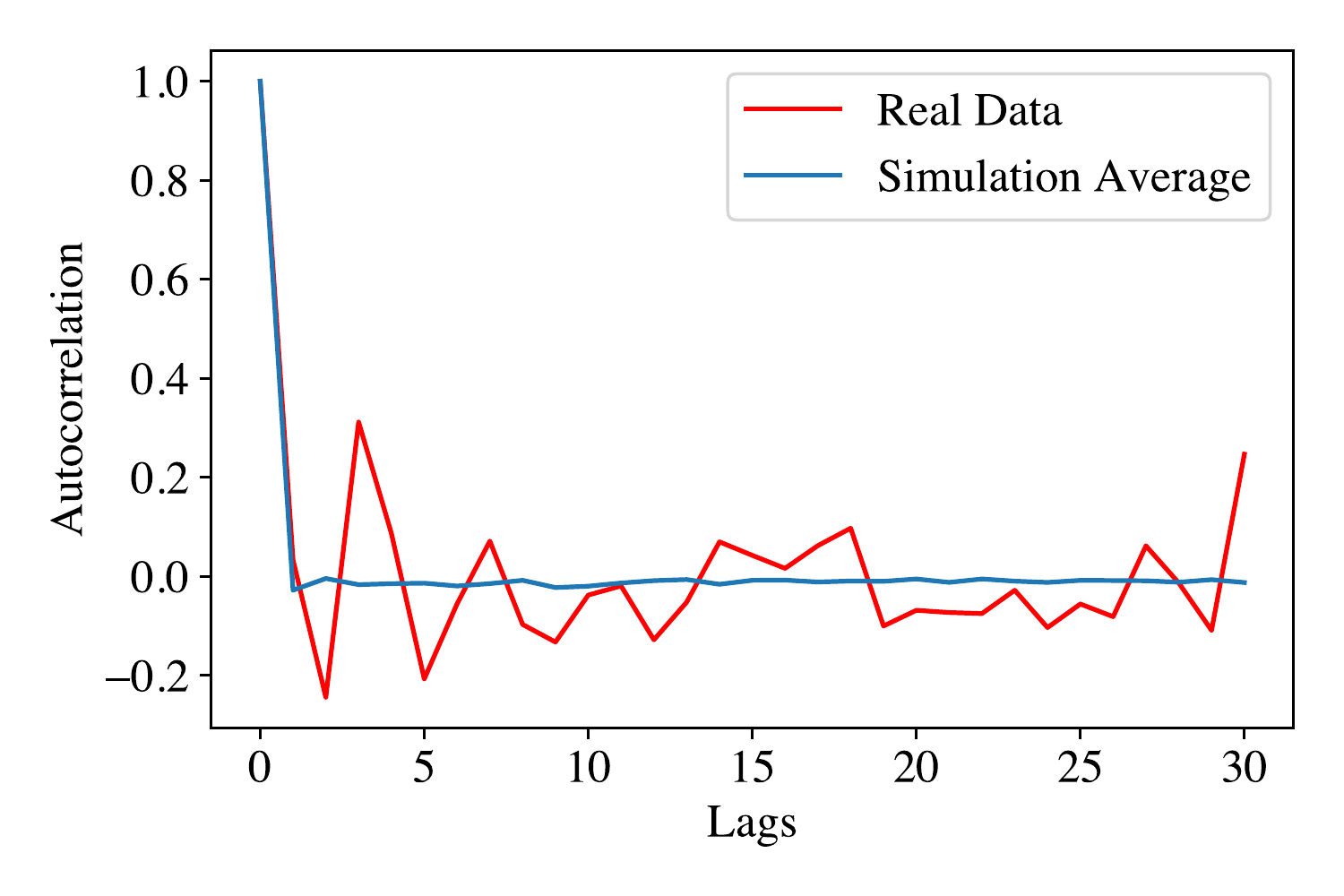}}
\subfigure[RBM]{\includegraphics[width=0.475\linewidth]{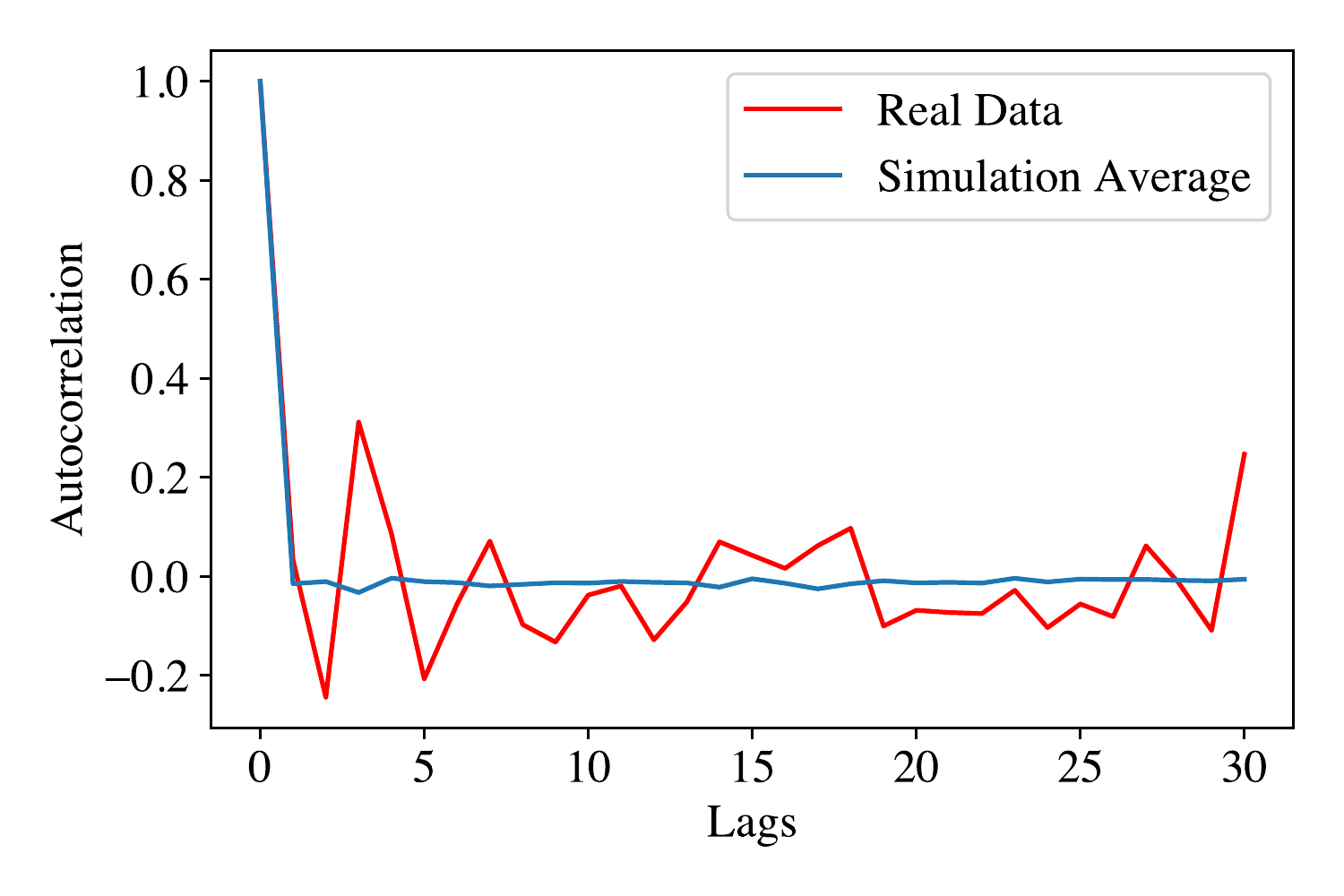}}
\subfigure[CVAE]{\includegraphics[width=0.475\linewidth]{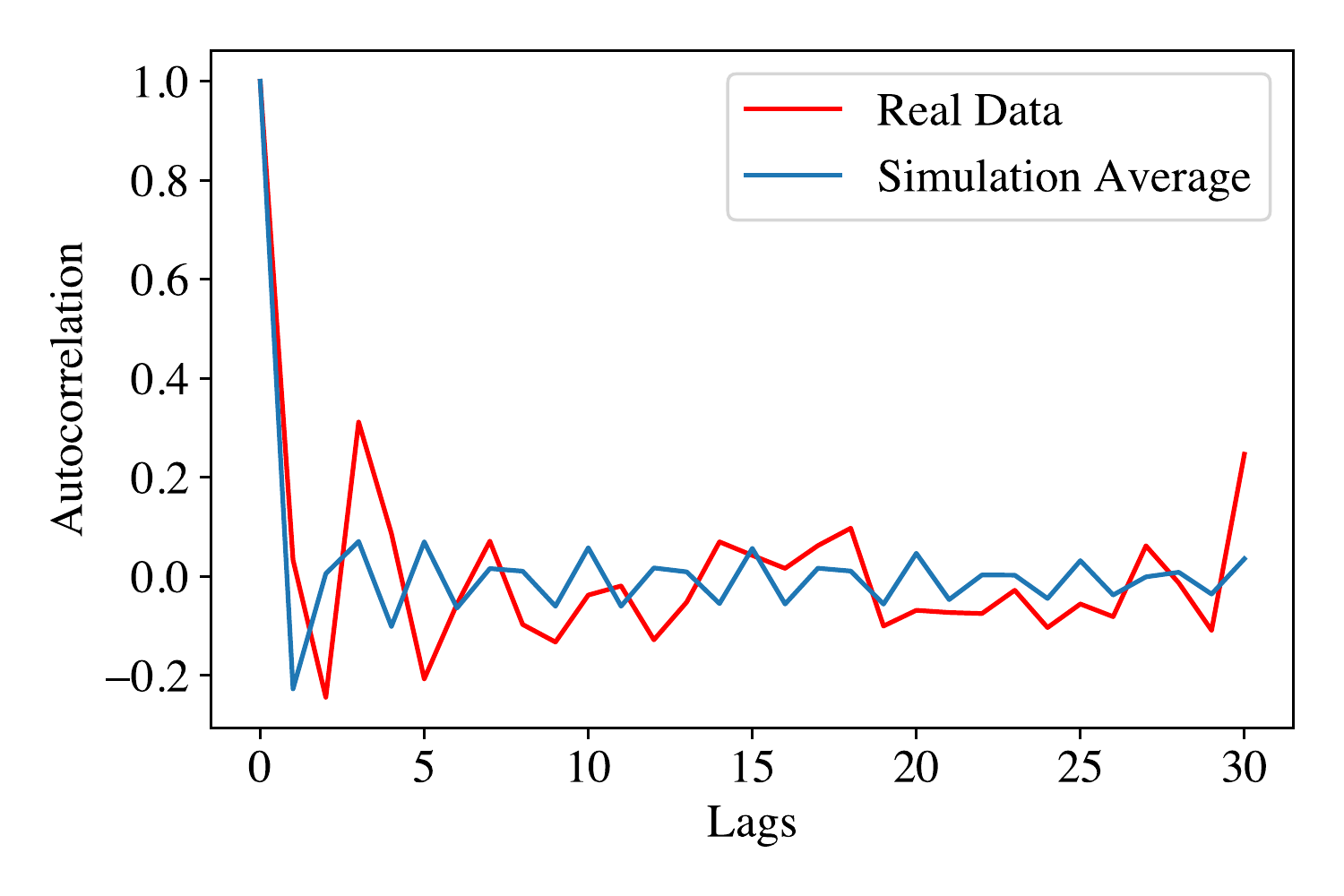}}
\caption[Autocorrelation plots in real data experiment]{Average autocorrelation of generated returns vs. autocorrelation of real S\&P500 Log Returns.}
\label{acfreal}
\end{figure}

\begin{figure}[h]
\centering  
\subfigure[FHS]{\includegraphics[width=0.475\linewidth]{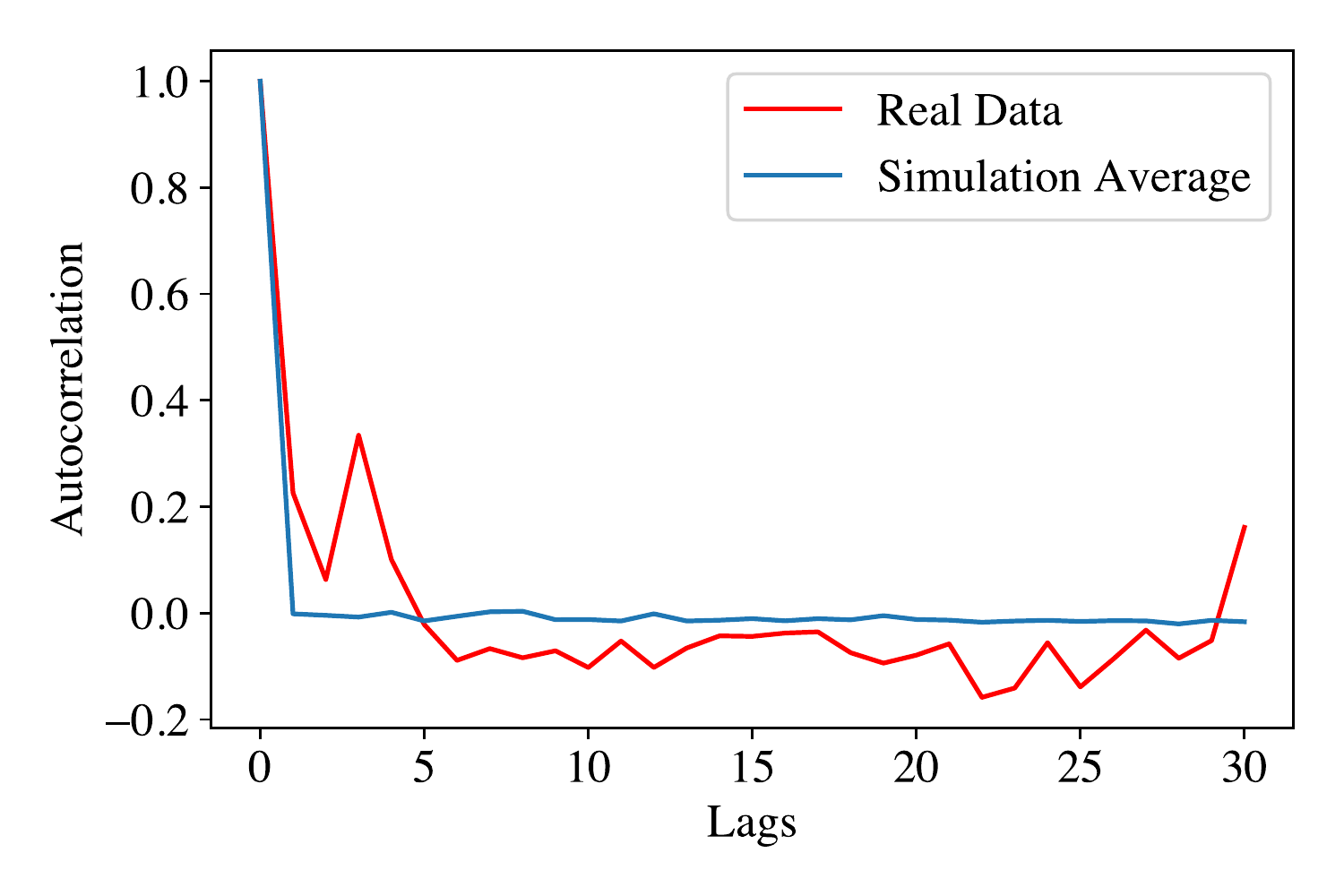}}
\subfigure[GARCH]{\includegraphics[width=0.475\linewidth]{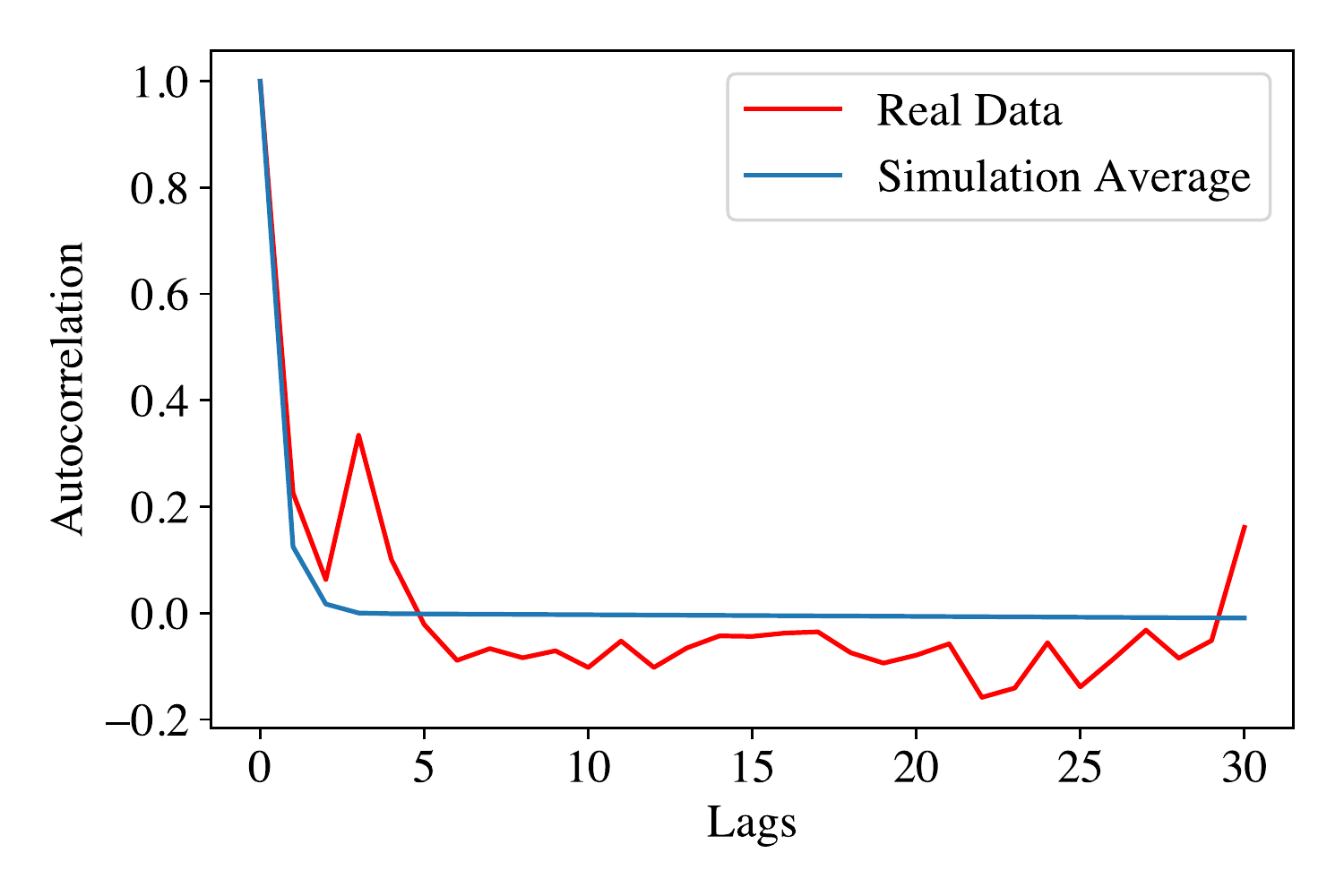}}
\subfigure[RBM]{\includegraphics[width=0.475\linewidth]{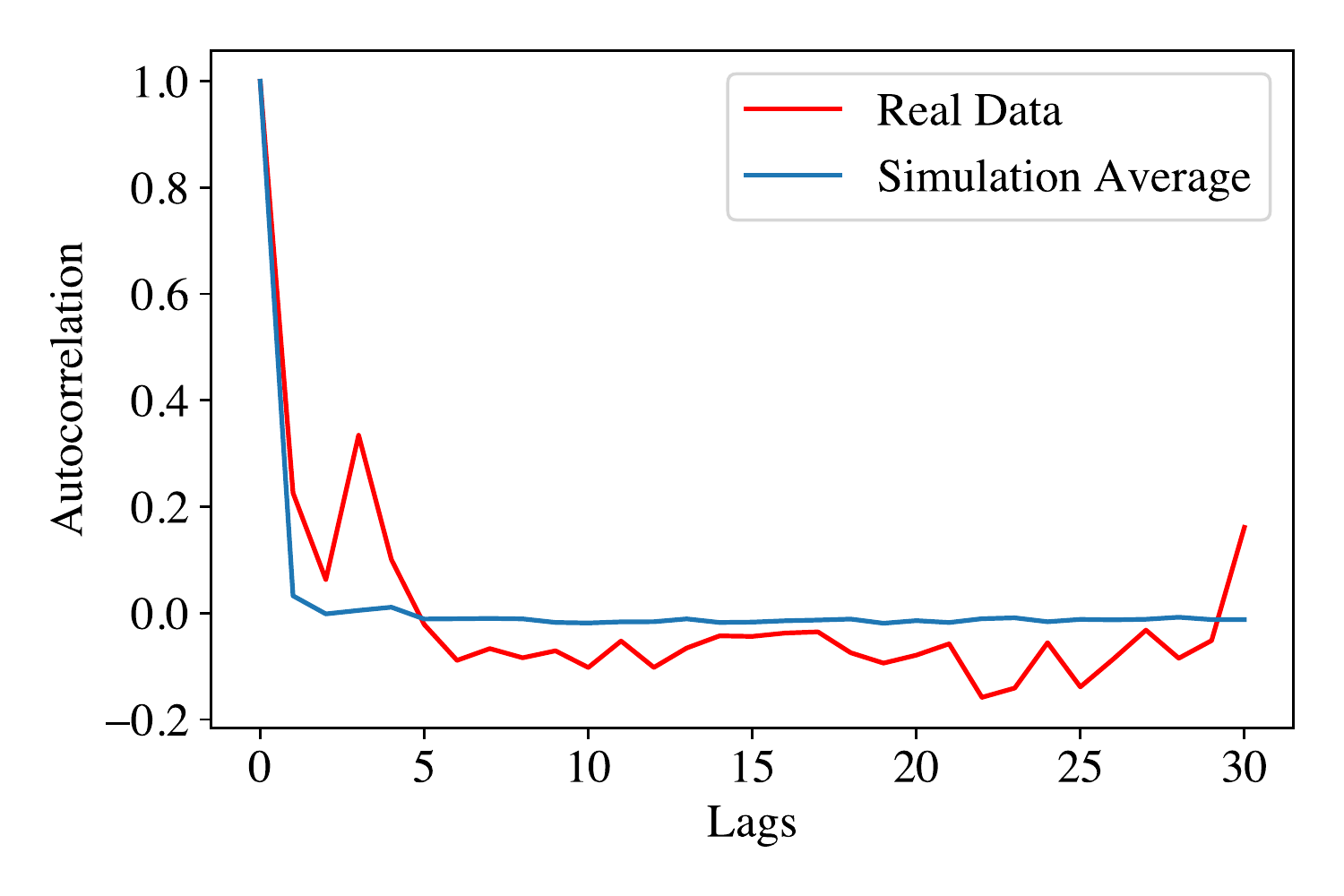}}
\subfigure[CVAE]{\includegraphics[width=0.475\linewidth]{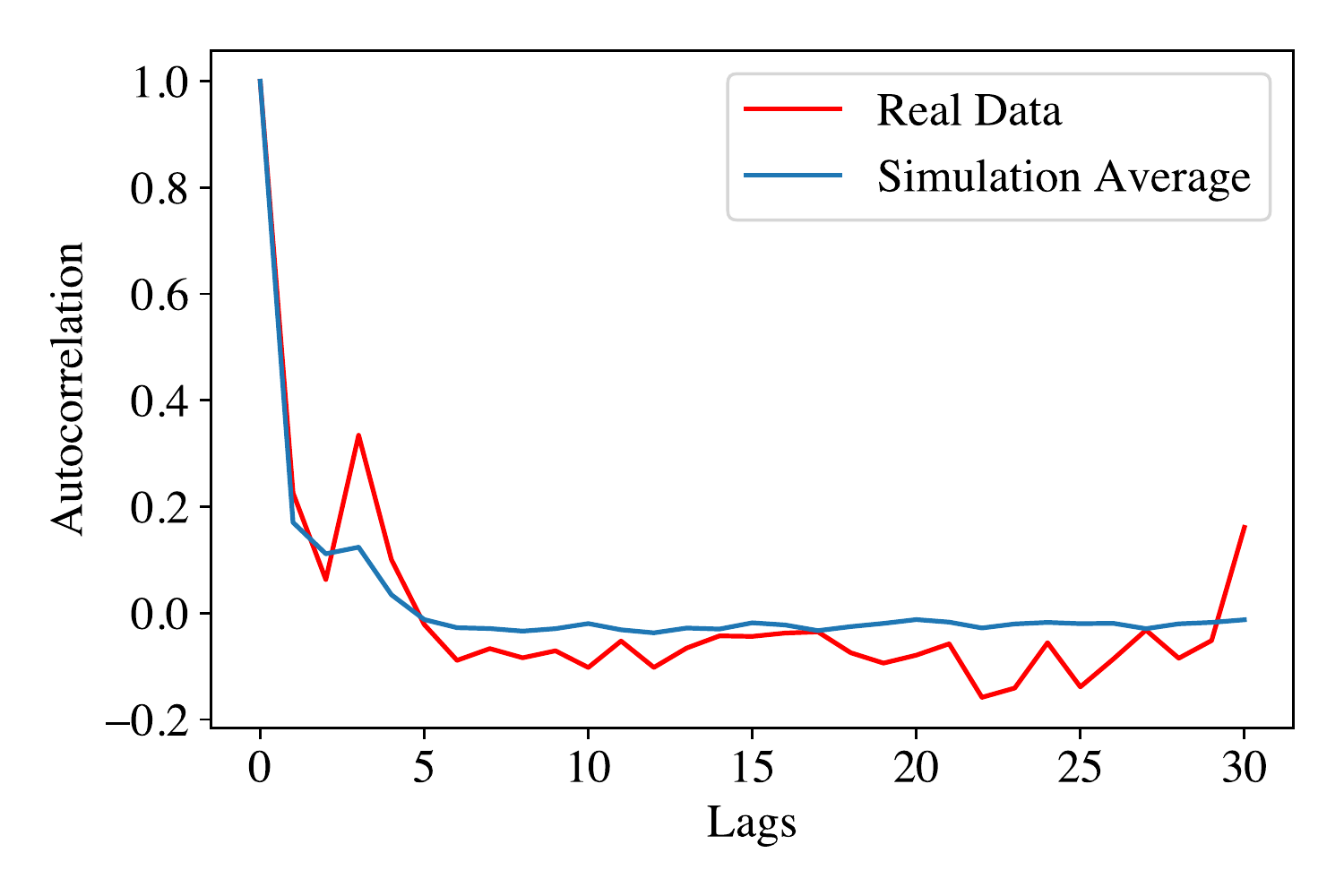}}
\caption[Autocorrelation plots of squared samples in real data experiment]{Average autocorrelation of squared generated returns vs. autocorrelation of squared S\&P500 Log Returns.}
\label{sqacfreal}
\end{figure}

Additional stylized facts to verify are no/low autocorrelations in the samples and the existing autocorrelations in the squared samples, which show volatility clustering. Keeping in mind that the modelled distribution in RBM and CVAE is that of weekly data, the autocorrelation of these two models beyond lag five does not reflect any modelling qualities. Contrary to the GARCH model, which produces an entire path of sample length, but given our choice of using only lags of 1, we can only judge the first lag performance. In \ref{acfreal} and \ref{sqacfreal} we plot the autocorrelation in both real and squared real data versus the average autocorrelation function over all simulations of the generated samples. From \ref{acfreal} we see indeed some autocorrelation in the real data in the first four lags as well as the thirtieth one, which we blame on the short sampling period of only 13 weeks\footnote{Indeed, further analysis with shorter/longer testing periods in \ref{projection} show lowwer autocorrelation for the longer testing periods of 26 and 52 weeks, coinciding with findings in the literature.}, which the FHS, GARCH and RBM replicate with zero average autocorrelation. Only the CVAE samples show some variation of autocorrelation different from almost constant zero. 
From \ref{sqacfreal}, it seems that no model is on average able to perfectly replicate the autocorrelation structure visible in the real squared data. The squared FHS samples have on average basically constant 0 autocorrelation, similarly with the RBM. The GARCH model, with only one lag in the specification, does replicates some low autocorrelation in the first lag following the pattern observed in real data. Apart from the GARCH, only the CVAE  shows on average an ability to produce some autocorrelation noticeably higher than constant 0 within the first five lags, showing a similar structure as in the real data in the first two lags.
\subsubsection{Summary statistical analysis}
In this section we perform statistical tests to analyze the model fits in a 65-day sample period. We see from the test results that the FHS is unable to replicate key statistical measures of the data such as the first two moments. It also overestimates the heaviness of the left tail, but manages to replicate the right tail as well as the kurtosis. Apparently the proposed scaling of daily Log Returns with the VIX index is unable to satisfy the strong i.i.d. assumption needed for the FHS, as it also shows bad fits for any time dependent properties of the data. The GARCH process with joint likelihood on the other hand is by design able to replicate the time dependent data properties, which we do observe in the average autocorrelation of the squared samples until lag one given our model specification. However, distributional stylized facts cannot be replicated, possibly a result of an unfit innovation distribution of $t_{4}$: simulated standard deviation and tails do not fit well and as a symmetric innovation distribution it is unable to model the data’s negative skewness. These observations are in line with the drawback of high model risk depending on the a priori model assumptions in parametric models as described in the introduction section.

In comparison, the machine learning models both do better in the distribution approximation,  with the RBM modelling the left tail better and the right tail worse than the CVAE but the CVAE having a better tradeoff when measuring the interquartile range. We note however, that the CVAE does better on the other distributional statistics, the modelled first two moments lie well within a one standard deviation range of the real values. Similarly, the generated kurtosis of the CVAE is more realistic than that of the RBM, although the skewness is further off than that of the RBM. It is however noticeable, that the CVAE sample statistics have a lower standard deviation across the generated samples. This possibly results from the fact that the CVAE converged better or the design of its sampling process, which is more stable than that of the RBM. Again, we see that the RBM seems to be unable to reproduce time dependent properties with our training and tuning: the samples and their squares show no autocorrelation on average. In contrast, the CVAE seems to be better at reproducing the autocorrelation structure in the squared samples.  

Same analysis for different projection periods of 4 weeks, 26 weeks and 52 weeks in \ref{projection} show similar findings: all models show low autocorrelation, which coincides for longer periods with the observed real autocorrelation. Volatility clustering is only replicated by the GARCH (in first lag) and the CVAE (with similar first lag performance as the GARCH, but higher than 0 in the lags until 5). From the point of view of the distributional approximation the FHS suffers from a too light left tail in all projection periods and the GARCH distribtuion specifications are unfit for the data. Interestingly the RBM overestimates the left tail for longer periods and has a too light right tail for all projection periods. On the contrary, the CVAE tails become more fitting for longer projection horizons, showing the best tail fits out of all models for the 26- and 52-week projections.

However, no model showed perfect fit for the to be replicated distribution, though the machine learning approach performed better than the parametric and nonparametric models and the CVAE performing better than the RBM. The problem of the RBM for the time dependent properties may lie in the infeasible heuristic approach of inferring a conditional distribution, as the distributional metrics were approximated with the second best overall quality. For both machine learning models deviations in the modelled distribution statistics could result from the performed training and parameter search being insufficient. It is possible that the CVAE could perform better in both distributional approximation and replication of time dependent properties given more training/ better fitting parameters.

\subsection{Portfolio Performance}\label{realpf}
We now test the forecasting ability of our models by using out-of-sample forecast generations to build an indicator for a stop loss strategy on a buy-and-hold strategy on S\&P500 futures. This way we can judge the model forecasting qualities by looking at the portfolio performance, as their value is directly dependent on the forecast quality: the more realistic the forecasts (or rather the used statistics from the forecast included in the indicator) the better the portfolio performance based on this forecast should be. Additionally, this procedure also serves as a partial robustness analysis with regard to robustness against different (shorter or longer depending on the starting date) training data.

Above explanation implies that the choice of indicator plays a substantial role in this evaluation part, as it should include enough important information from the modelled distribution itself as well as reflect the correlation and influence of the forward looking conditioning factor on the simulated risk factor, which is the main source of the model's forecasting ability. Given our choice to model the S\&P500 Index given the VIX Index, our stop loss threshold should check on the value as well as the volatility of the projected S\&P500 Log-Returns as explained in the justification of factor choice in the experiment setup.

Combining all these thoughts, we introduce the general framework of our simplified dynamic stop loss strategy $\mathcal{S}(L,H, a(t,L^{*}))$ with $0/1$ asset allocation rule ${s_{t}}$ at time $t$ between the S\&P500 Index Futures and a risk free asset\footnote{We assume the risk free asset to have return 0.} based on the stop-loss strategy provided in \cite{KAMINSKI2014234}:
\begin{equation}
  s_{t}=
  \begin{cases}
    \! 
    \begin{alignedat}{2}
      & 1, && \quad \mathbb{E}^{t, t+H}(r|\text{VIX}_{t-1}) \geq \mathbb{E}^{t-L, t}(r)- a(t,L^{*})*\sigma^{t-L, t}(r)
      \\
      &0, && \quad \text{else} 
    \end{alignedat}
  \end{cases}
  \label{stategy}
\end{equation}
with $L$ denoting the look-back period for which we determine a mean of the historical daily Log-Returns as a general benchmark; $H$ the projection length of our model forecasts and our main extension beyond the strategy of \cite{KAMINSKI2014234}:  $a(t,L^{*})$ a function controlling the tightness the stop loss benchmark boundary with values depending on the current forecast at time $t$ as well as the special look-back period $L^{*}$.

The main underlying idea is to vary the stop loss bound tightness depending on the projected volatility: For a given forecasted volatility, we decide whether it is high or low and take a tighter or looser stop loss boundary respectively. Thus the function $a(t,L)$ has an inverse relation to the forecasted volatility. This idea is based on the very nature of the modelled distribution, as the conditioning factor influences the volatility of the S\&P500 Log-Returns, and on the goal to benefit from potential model misfits: If the model is inaccurate in modelling the underlying relation between S\&P500 Log-Returns and previous VIX Index, then the forecasted volatility will be off. Given a simplistic differentiation of volatility into high and low, a bad model would produce high volatility when in reality the volatility is low and vice versa. The choice of the tightness-regulating function $a(t,L^{*})$ then results in a tight stop loss rule in the first case despite low real volatility, thus missing out on potential profits; and a loose stop loss rule in the second case, thus risking high losses; giving a clear distinction between accurate and inaccurate models simply by judging the strategy performance. Hence our values for $a(t,L^{*})$ should differ rather highly between the two distinct cases of volatility. 

Based on above intuition, we design the function $a(t,L)$ in the following way:
\begin{equation}
a(t,L^{*})=
  \begin{cases}
    \! 
    \begin{alignedat}{2}
      & 0.5, && \quad \sigma^{t, t+H}(r|\text{VIX}_{t-1}) \geq \sigma^{t-L^{*},t}(r)
      \\
      & 3 , && \quad \text{else} 
    \end{alignedat}
  \end{cases}
  \label{stategy}
\end{equation}
with $L^{*}$ the look-back period from which we determine the distinction between high and low volatility.  The values of $a(t, L^{*})$ vary greatly and are conservative to avoid losses due to large negative movements to focus more on returns made during low volatile periods. Since we know the S\&P500 Log-Returns to have a heavier left tail than the standard normal Gaussian (which has $1^{\text{st}}$ percentile of around -2), we choose the the tightness factor in case of a low projected volatility to be 3. 

In practice we choose $L=L^{*}=260$, i.e. have one year look back periods; and $H=5$, i.e. produce one week forecasts. We ignore transaction costs and make in addition the following assumptions:
\begin{enumerate}
    \item We can only decide to go enter the strategy before in the first week of every simulation period;
    \item If we decide to go enter, we buy at the closing price of the Monday in the first simulation week;
    \item Exiting the strategy can only be done at the Friday closing price of the latest holding week and always involves 100\% of the investment;
    \item After exiting the strategy we do not reenter.
\end{enumerate}
The above assumptions may seem restrictive and unrealistic for a true trading strategy, but we keep in mind that our final goal is the judgement of our distribution model for which we only need the strategy to be based on reasonable economic decisions, so that a poor strategy performance is indeed retraceable to a poor model fit. We note however that the second assumption in particular disadvantages the FHS model, as it produces daily forecasts and we cannot update the volatility filtering for 5 day forecasts - thus only having constant volatility samples throughout each simulation week. This is an important limitation to keep in mind when judging the portfolio performances.
\begin{figure}[h]
\centering  
\subfigure[FHS]{\includegraphics[width=0.475\linewidth]{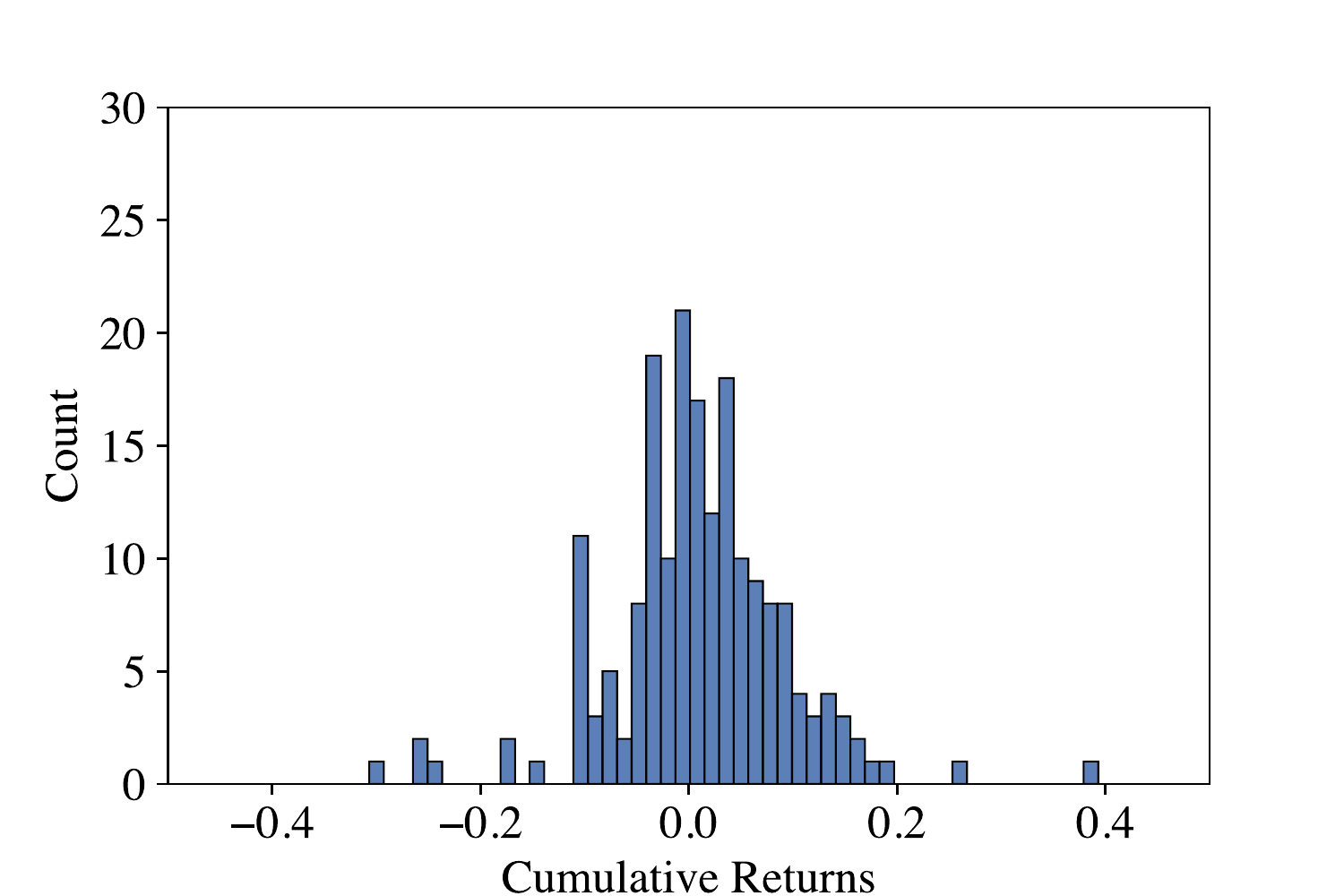}}
\subfigure[GARCH]{\includegraphics[width=0.475\linewidth]{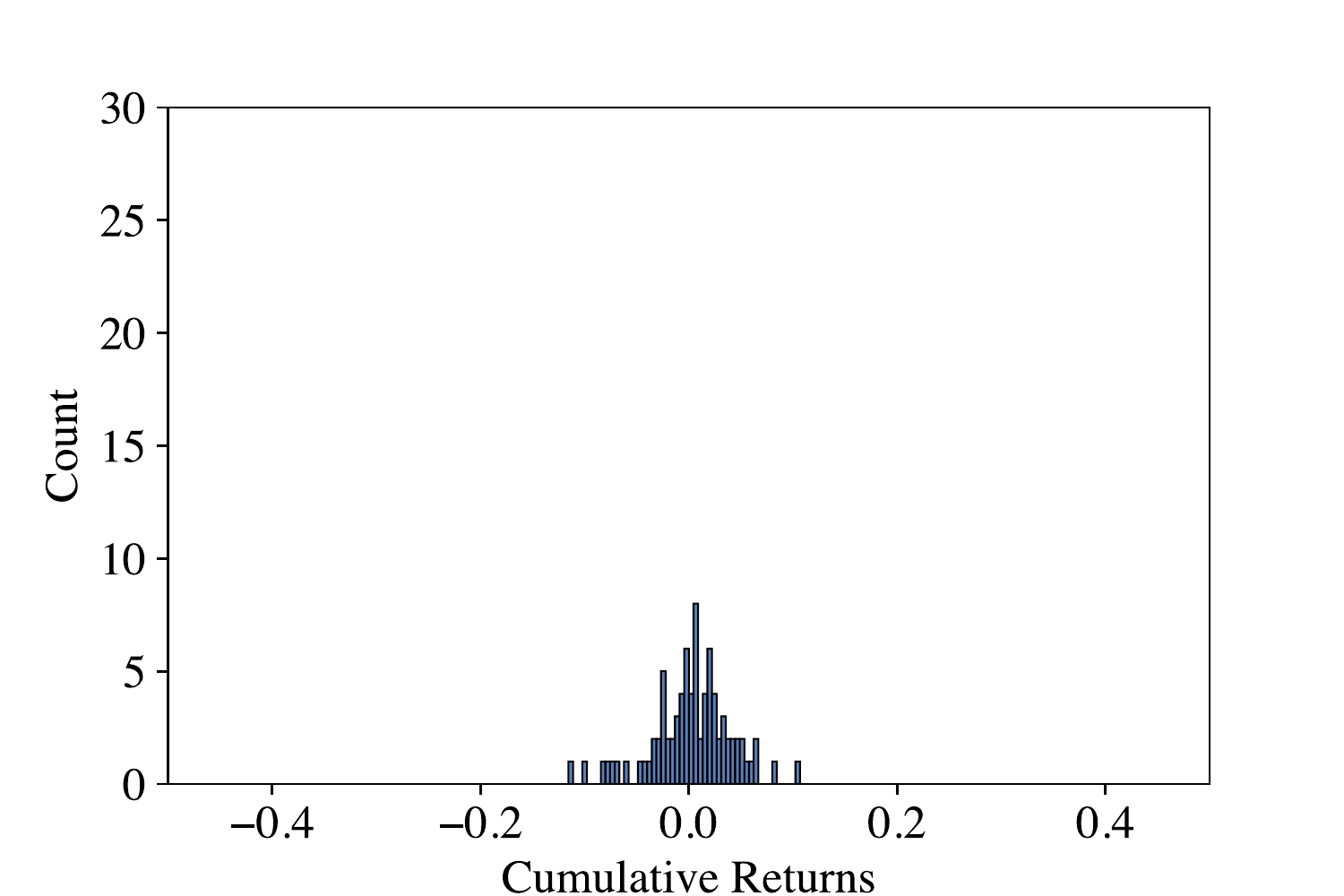}}
\subfigure[RBM]{\includegraphics[width=0.475\linewidth]{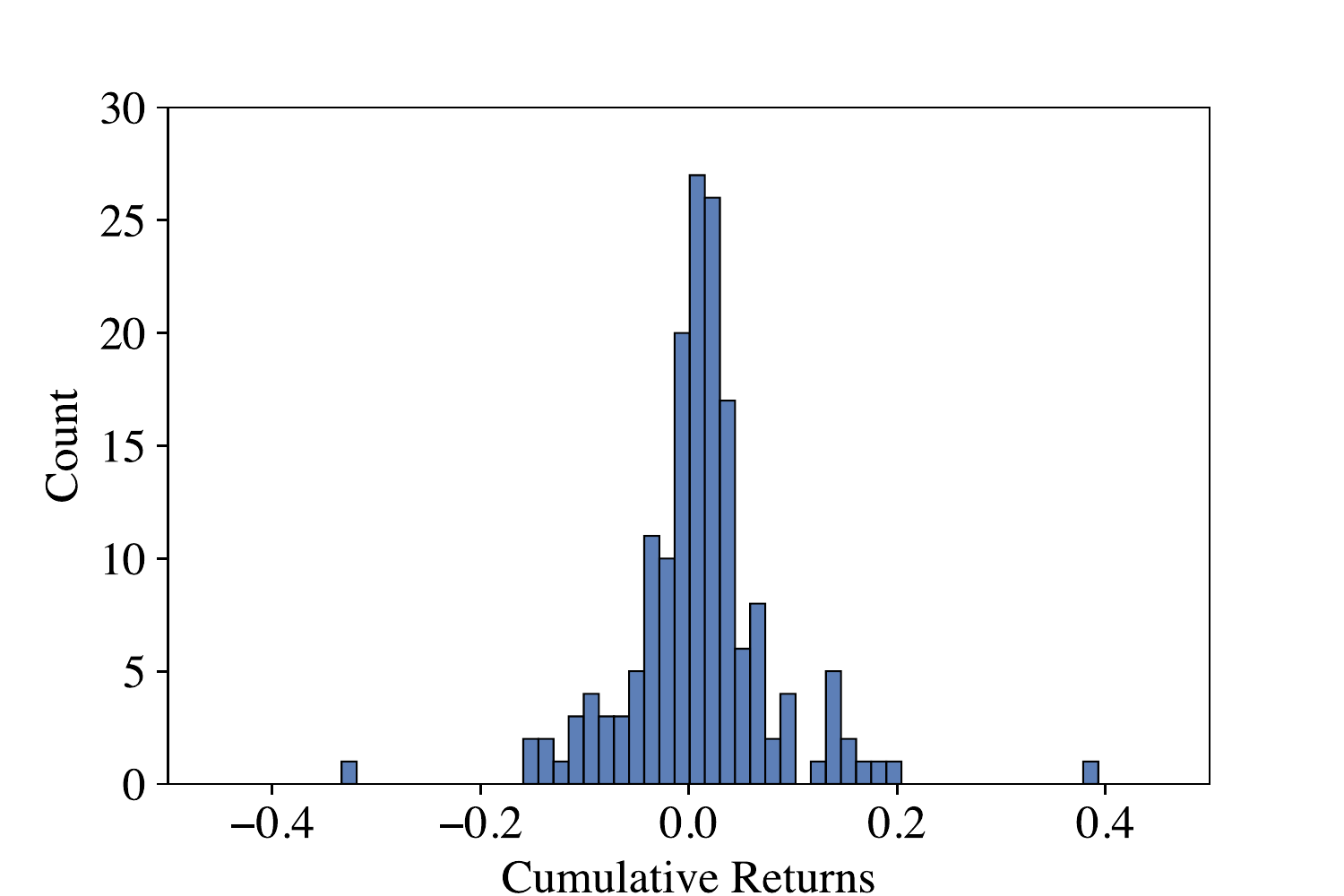}}
\subfigure[CVAE]{\includegraphics[width=0.475\linewidth]{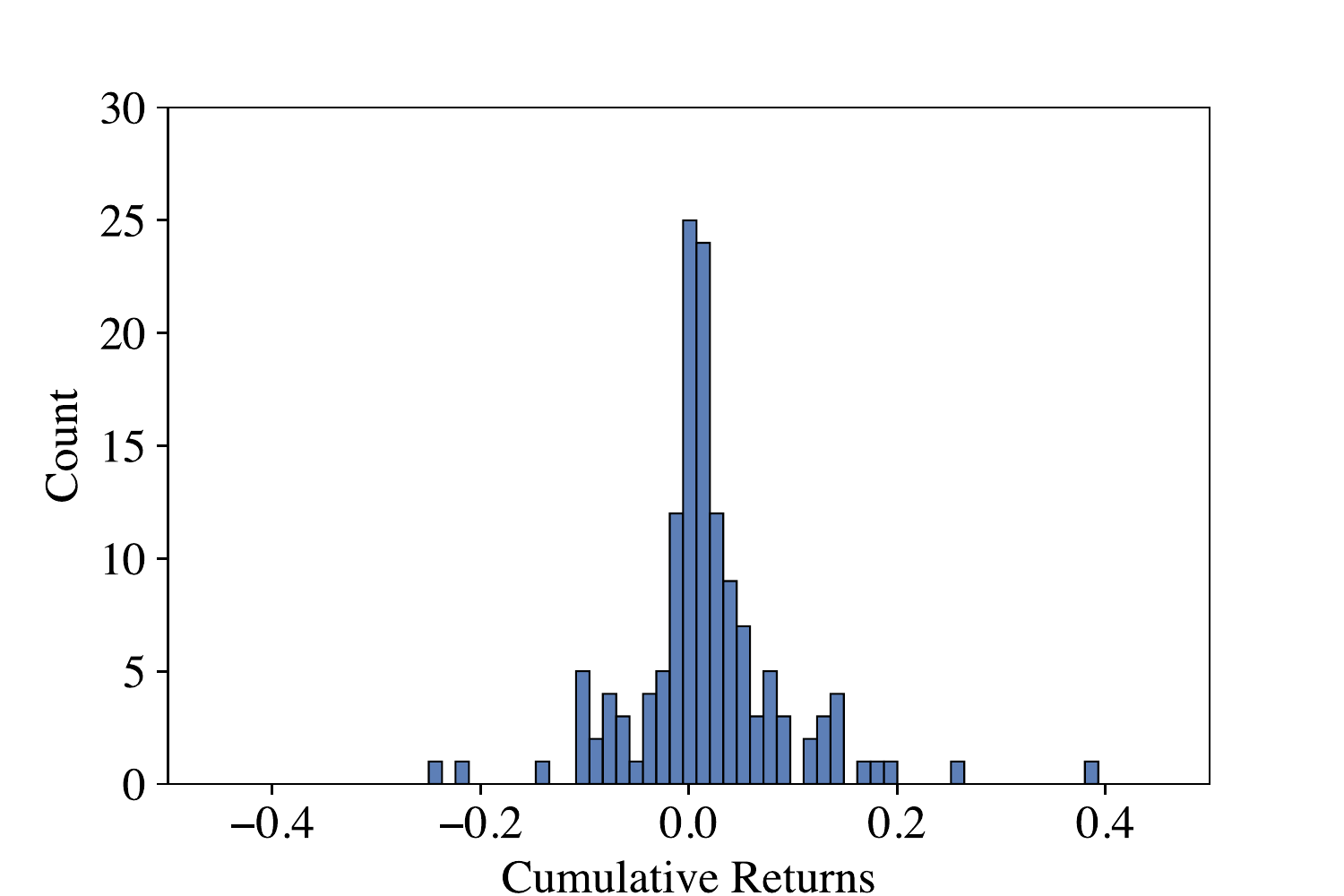}}
\caption[Portfolio Profit and Loss histograms]{Histogram of portfolio Profit-and-Loss using indicators across all modelling approaches.}
\label{pfpnl}
\end{figure} 
We run the above described experiment with all models on 200 randomly chosen starting dates, from which 100 come from the time period 2007-2009 to include the equity crisis and the leftover 100 starting dates uniformly chosen from 2003-2007 and 2009 until one simulation period before the end of available data. We take a backtest length of 13 weeks for every starting point, as too long a period would result in too many overlaps in the projection periods and too short a period loses information about the strategy performance. After simulation, we calculate the distribution of the 13 week portfolio Profit-and-Loss (P\&L), reporting their probability distribution in \ref{pfpnl} as well as their their first two distribution moments in \ref{pnltable}.

From the P\&L distributions we notice immediately that the portfolio using indicators from the GARCH model has the smallest returns range and the least amount of performed strategy entries. This behaviour can be traced back to an overestimation of the volatility in the model forecast which leads more to projected log returns below the entrance threshold than above and is in line with the previous statistical results in \ref{statstsreal}. An interesting connection of our conservative strategy to a disadvantage also in regulatory conservatism is also clearly visible: Too conservative models/rules potentially heavily limit the upside potential. Similar behaviour is observed in the portfolio using the RBM, whose returns all strictly lie in the range of [-0.2,0.2] with only two exceptions. We deem the number of returns outside this range as too little and transition of the P\&L distribution in the tail to be too abrupt to consider these exceptions as part of the consistent RBM performance. Thus the histogram of the portfolio using the RBM also indicates an overestimation of the volatility, similar as observed in \ref{statstsreal}. On the other hand, wide return ranges are visible for portfolios using the FHS and CVAE, however with the CVAE portfolio returns being more limited on the losses than the FHS, indicating a well fitted stop loss criterion, contrary to the FHS.
\begin{table*}[h]
\centering
\begin{adjustbox}{max width=\textwidth}
\begin{tabular}{lrrrrr} \toprule
                   & \textbf{Buy-and-Hold} & \textbf{FHS}    & \textbf{GARCH} & \textbf{RBM} & \textbf{CVAE}   \\ \midrule \midrule
\textbf{Mean}               & -0.98981 & 0.00955 & 0.00294  &  0.01011   & 0.01696 \\
\textbf{Standard Deviation} & 0.09590 & 0.08461 & 0.03850   &  0.07226 & 0.07509 \\ 
\textbf{Sharpe Ratio}       & -10.3213 & 0.11284 & 0.07631 &  0.13993 &  0.22586 \\
\textbf{Max. Drawdown}       & -40.65\% & -34.96\% & -22.23\% & -40.65\% & -27.63\\
\bottomrule
\end{tabular}
\end{adjustbox}
\caption[Risk measures of portfolio performance]{Risk measures of portfolio performance across modelling approaches.}
\label{pnltable}
\end{table*}

In \ref{pnltable} we provide the first two moments of the P\&L distribution shown in the histograms of \ref{pfpnl} and additionally as risk measures the Sharpe Ratio and the Maximum Drawdown. We benchmark all these values with the corresponding value of a Buy-and-Hold strategy. From \ref{pnltable} it is notable that the Buy-and-Hold strategy over our chosen dates and time period has a negative mean return, possibly resulting from our emphasis on starting points during the financial crisis. Judging by the Sharpe Ratio, all approaches outperform the Buy-and-Hold strategy, taking low but positive values. However, one disadvantage of the Sharpe Ratio is the use of volatility, which doesn't differentiate between upside and downside movements of the P\&L distribution. For this reason we include the maximum drawdown of each portfolio approach across all starting dates to gain a different view on the portfolio performance. A well designed stop loss strategy should have a low max. drawdown, as it exits the strategy in case of bad performance, thus limiting the portfolio losses. Thus unsurprisingly, the max. drawdown of the Buy-and-Hold strategy is the highest. However, the portfolio using the RBM outputs has the same max. drawdown, indicating that the RBM generations were unable to accurately forecast the poor index performance in this extreme case. Interestingly, the portfolio using the GARCH model has the lowest max. drawdown. However, this might not necessarily reflect a better portfolio performance, but could simply result from the heavy volatility overestimation by the GARCH model, which together with the strict conservatism of our strategy triggers more and ealier exits, thus lowering the max. drawdown. Portfolios using the CVAE and FHS have max. drawdowns in between, possibly indicating reasonably triggered stop losses more due to accurate forecasting than to estimation trends that benefit from the strategy design.

From this experiment we deduce that the CVAE seems to capture the relationship of the VIX on the first two moments of the daily Log Returns of the S\&P500 Index particularly well throughout different starting points out of sample. It seems that the GARCH and RBM both overestimate the standard deviation, with the GARCH even more so. Additionally we observe that GARCH performance in terms of convergence and parameter estimation is highly dependent on the initial values, showing the nonrobustness of this approach against random seeds. The FHS on the other hand shows robust performance throughout the experiment, yielding results in line with observations from the previous section despite the design disadvantage.

\chapter{Conclusions} \label{conc}
We compare four models from the non-parametric, parametric as well as Machine Learning family for their performance as a PiT ESG generating realizations of the S\&P500 Index Log-Returns conditional on the last available VIX value. The generator quality is judged by a: their approximation of distributional metrics and b: of time dependent properties; c: their robustness across different modelling periods as well as d: their of sample performance as strategy indicator. Based on these evaluation criteria, our experiments indicate that the CVAE exhibits the best combination of results out of all our tested approaches. 

The CVAE has the best approximation of distributional metrics such as the first four moments and tail behaviour and shows signs of being able to replicate time dependent properties such as volatility clustering. Also the robustness of performance quality is confirmed in out of sample tests and data generations over different time periods. In comparison, the poor performance of the FHS likely results from a violation of the i.i.d. data assumption, meaning that the filtering process with the VIX is not enough as a data transformation. It appears that our GARCH specifications are unfit for the data, resulting in poor distributional approximation. Additionally, model results show a tendency to overestimate volatility, associated with lower projected returns, and model performance is highly sensitive to the initial starting seeds during optimization. The RBM as the other tested generative network exhibits better performance in statistical tests, but a decline in approximation quality over longer time periods and difficulties in replicating time dependent properties. 

From the point of view of computational efficiency, the generative networks are heavier than the other tested models. However the CVAE still does better than the RBM, which requires possibly hundreds of forward and backward passes during both training and sampling, that cannot be done in parallel due to the dependence on the previous sample. 

In summary, it appears that there is a lot of potential in the generative networks, but the trade off between training complexity and improvements in modelling capabilities needs to be carefully considered. Model performance in case of the RBM could possibly be improved by the usage of a conditional RBM as introduced in \cite{crbmori}. Similarly, both RBM and CVAE might produce better results with more fitting parameters. Improvements in this area are however left for future works.

\appendix
\chapter{Appendix}
\section{Conversion between real-valued samples and binary features} \label{appendix}
In this section we present the pseudo code for the algorithm provided in \cite{RBM} for the conversion between real valued samples and 16-digit binary features. \\
\begin{algorithm}[H]
\SetAlgoLined
\KwData{Real-valued data set $X_{\text{real}}$ with N samples}
\KwResult{16-digit binary representation of real values sample}
$\epsilon \geq 0$ \\
$X_{\text{min}} \xleftarrow{} \text{min}(X)-\epsilon$\\
$X_{\text{max}} \xleftarrow{} \text{max}(X)+\epsilon$ \\
 \For{l=\text{1,...,N}}{
  $X_{int}^{(l)} \xleftarrow{} \text{int}(65535 \text{ x } (X_{\text{real}}^{(l)}-X_{\text{min}})\text{/}(X_{\text{max}}-X_{\text{min}}))
  X_{bin}^{(l)} \xleftarrow{} \text{binarize}(X_{int}^{(l)})$
 }
\caption{Transformation real-valued samples to 16 digit binary}
\end{algorithm}

\begin{algorithm}[H]
\SetAlgoLined
\KwData{16 digit binary sample $X=(X_{1},...,X_{16})$}
\KwResult{real valued sample from 16-digit binary sample}
 $X_{int} \xleftarrow{} 0$ \\
 \For{i=\text{1,...,16}}{
  $X_{int} \xleftarrow{} X_{int} + 2^{i-1} \text{ x } X_{16-i}$
 }
 $X_{\text{real}} \xleftarrow{} X_{\text{min}}+X_{int}\text{ x }(X_{\text{max}}-X_{\text{min}})/65535$
 \caption{Transformation 16-digit binary to real-valued samples }
\end{algorithm}

\section{Parameter search}\label{paramgrid}
Table \ref{paramgridtable} displays the hyperparameter grid tested and chosen for the RBM and CVAE for all results achieved in \ref{realdatamodelling}. The listed tested values only indicate the initial search grid, which were subsequently fine tuned and the parameters resulting in the best results were chosen. The hyperparameter searches were performed on the ETH Euler Scientific Compute Cluster.
\begin{table}[]
\centering
\begin{adjustbox}{max width=\textwidth}
\begin{tabular}{|l|l|l|l|}
\hline
Model & Hyperparameters                                                                                                                                                                                                                 & Tested values                                                                                                                                                                                                                      & Final choice                                                                                                    \\ \hline
RBM   & \begin{tabular}[c]{@{}l@{}}learning rate\\ batch size\\ network activation function\\ optimizer\\ number of epochs\\ Gibbs sampling steps\\ contrastive divergence steps $k$\\ network architecture\end{tabular}                & \begin{tabular}[c]{@{}l@{}}1e-4, 1e-3, 1e-2\\ 50, 100\\ Sigmoid\\ Adam\\ 50k, 100k\\ 500, 800,1000\\ 1, 5, 10\\ 16-32, 16-48, 16-64\end{tabular}                                                                       & \begin{tabular}[c]{@{}l@{}}1e-4\\ 50\\ Sigmoid\\ Adam\\ 50k\\ 800\\ 5\\ 16-32\end{tabular}                      \\ \hline
CVAE  & \begin{tabular}[c]{@{}l@{}}learning rate\\ batch size\\ network activation function\\ optimizer\\ number of epochs\\ KL weight factor\\ encoder architecture \\ decoder architecture\end{tabular} & \begin{tabular}[c]{@{}l@{}}1e-4,1e-3, 1e-2\\ 1\\ ReLu/leaky ReLu+Sigmoid\\ Adam\\ 50k, 100k\\ 1e-3,1e-4\\ 6-4, 6-30-4, 6-50-4, 6-20-50-4 \\ 4-6, 4-30-6, 4-50-6, 4-20-50-6\end{tabular} & \begin{tabular}[c]{@{}l@{}}5e-4\\ 1\\ leaky ReLu+Sigmoid\\ Adam\\ 50k\\ 1e-3\\ 6-30-4\\ 4-30-6\end{tabular} \\ \hline
\end{tabular}
\end{adjustbox}
\caption[Hyperparameter search grid]{Initial hyperparameter search grid for results reported on real data in \ref{realdatamodelling}}
\label{paramgridtable}
\end{table}

\section{Anaysis over different projection periods} \label{projection}
After the statistical analysis on a period of 13 weeks, we report in this chapter graphical results of the same analysis starting from 2018-01-1 over periods of 4 (1 month), 26 (6 months) and 52 (1 year) weeks to judge the influence of the projection period length on the model fit. 
\begin{figure*}
\centering  
\subfigure[Data vs. Normal]{%
\includegraphics[width=0.32\linewidth, trim=1.5cm 0.1cm 3.2cm 1.2cm, clip]{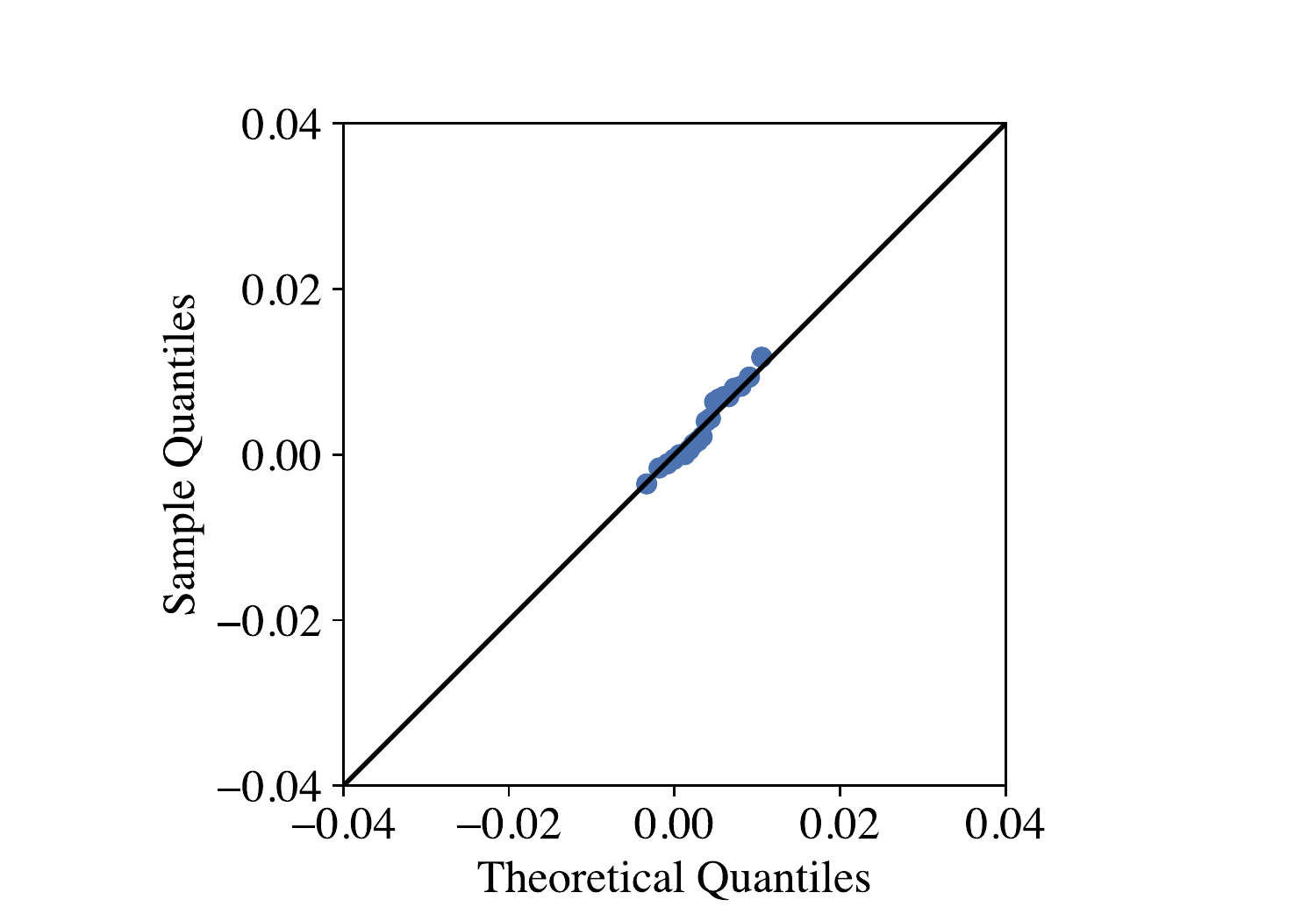}%
}%
\hspace*{\fill}
\subfigure[FHS vs. Normal]{%
\includegraphics[width=0.32\linewidth, trim=2cm 0.1cm 3.9cm 1.55cm, clip]{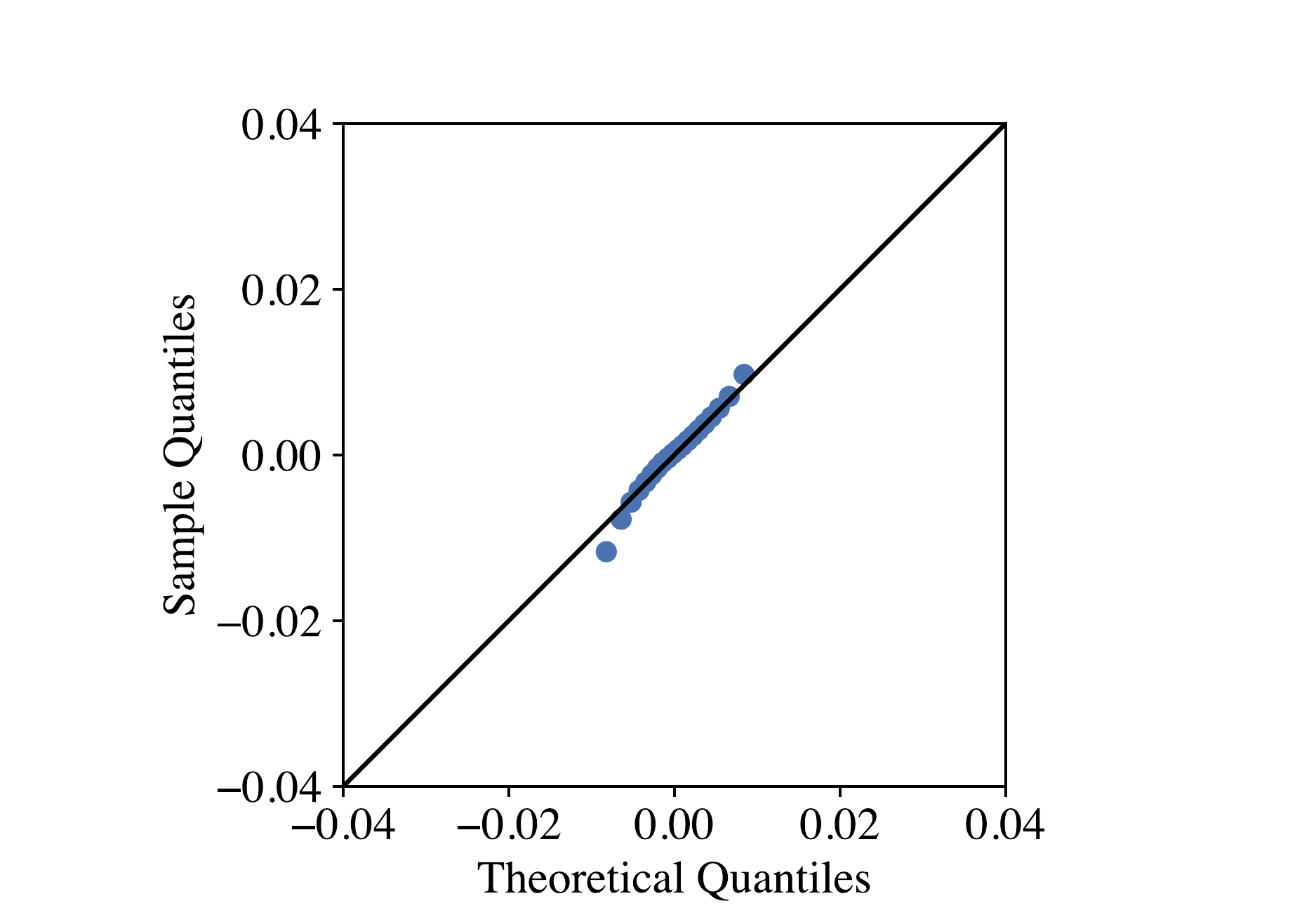}%
}%
\hspace*{\fill}
\subfigure[Data vs. FHS]{%
\includegraphics[width=0.32\linewidth,trim=1.5cm 0.1cm 3.2cm 1.2cm, clip]{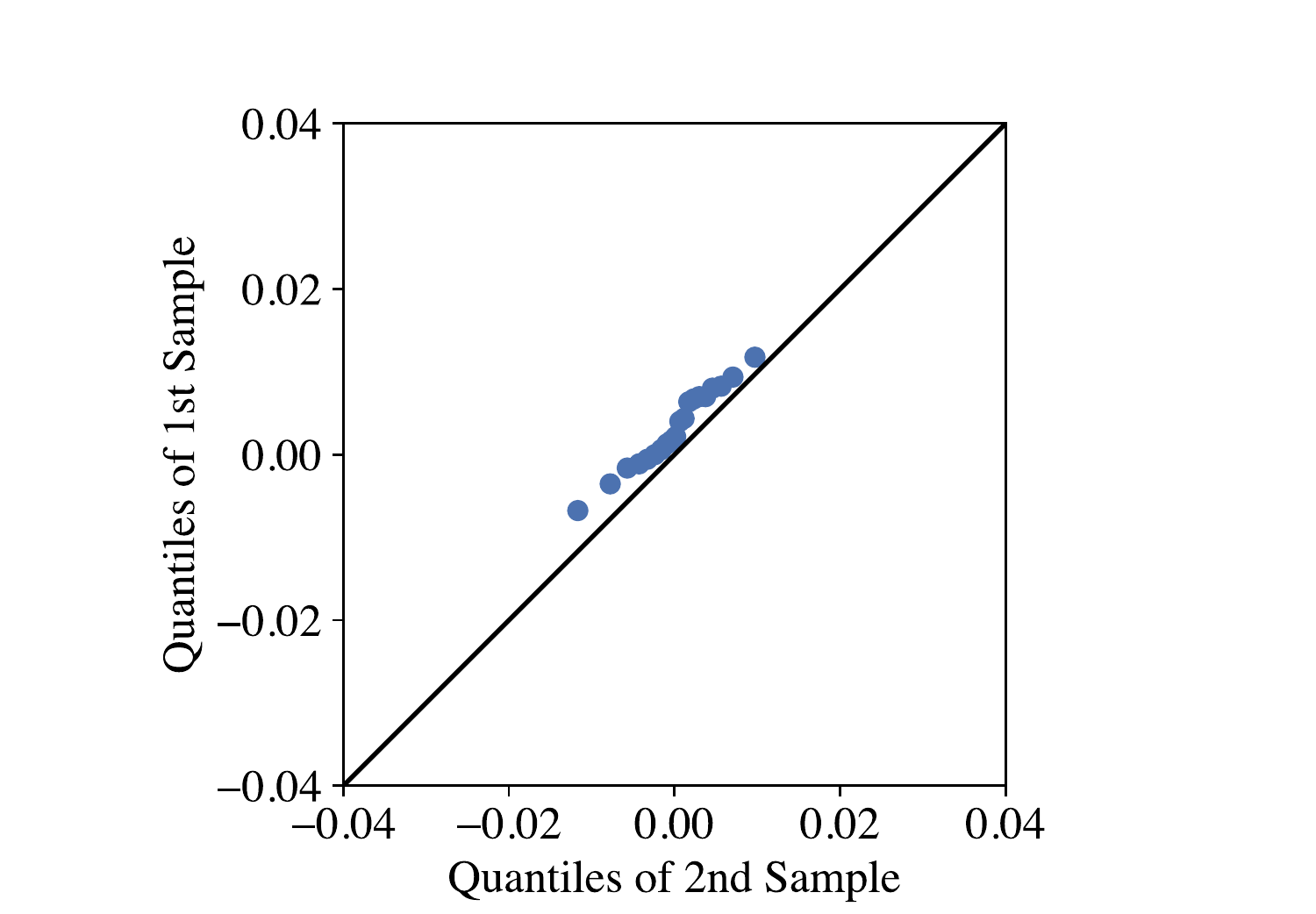}%
}
\subfigure[Data vs. Normal]{%
\includegraphics[width=0.32\linewidth,trim=1.5cm 0.1cm 3.2cm 1.2cm, clip]{figures/Real_Data/robust/qq/4real.pdf}%
}%
\hspace*{\fill}
\subfigure[GARCH vs. Normal]{%
\includegraphics[width=0.32\linewidth,trim=1.5cm 0.1cm 3.2cm 1.2cm, clip]{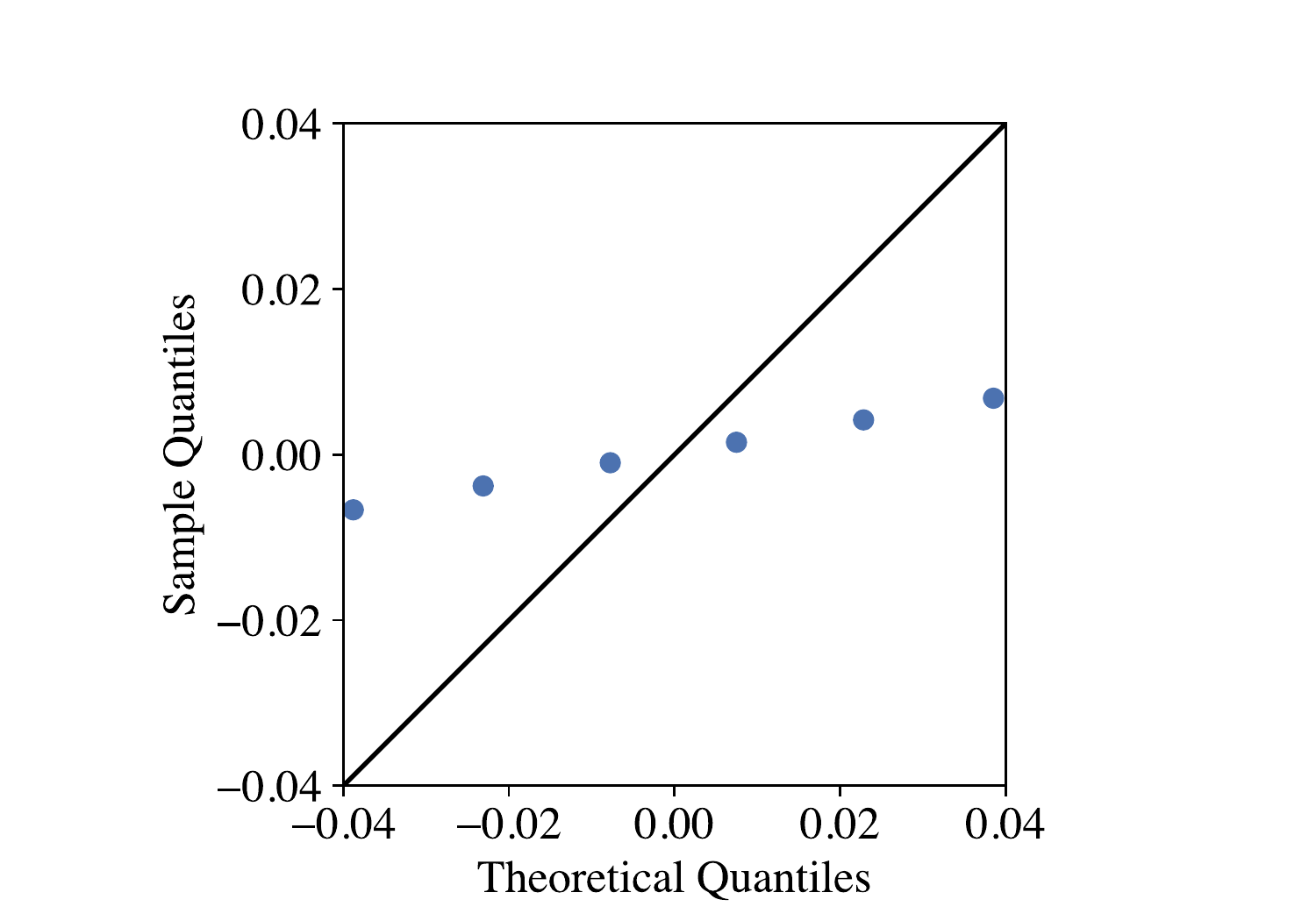}%
}%
\hspace*{\fill}
\subfigure[Data vs. GARCH]{%
\includegraphics[width=0.32\linewidth, trim=1.5cm 0.1cm 3.2cm 1.2cm, clip]{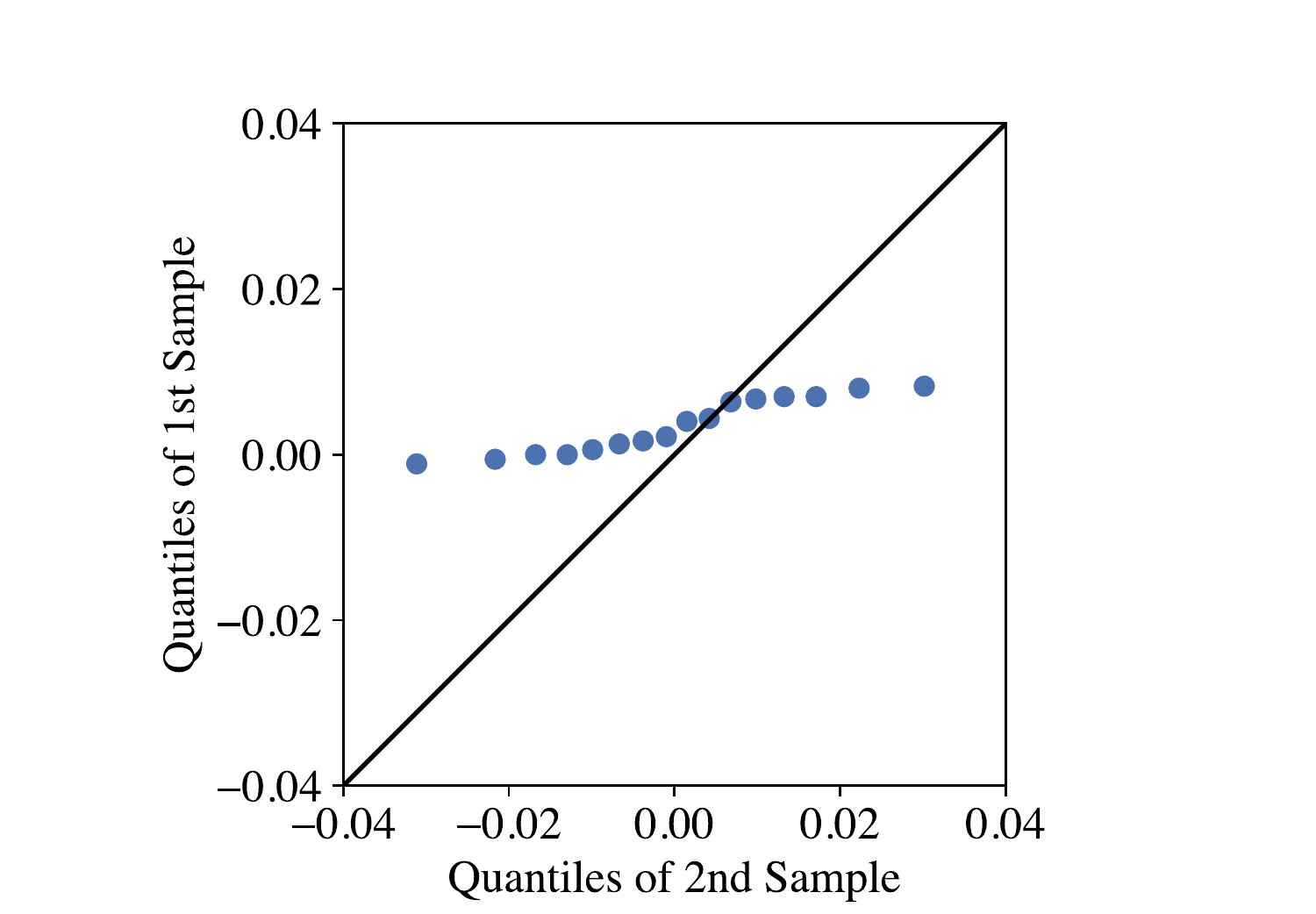}%
}
\subfigure[Data vs. Normal]{%
\includegraphics[width=0.32\linewidth,trim=1.5cm 0.1cm 3.2cm 1.2cm, clip]{figures/Real_Data/robust/qq/4real.pdf}%
}%
\hspace*{\fill}
\subfigure[RBM vs. Normal]{%
\includegraphics[width=0.32\linewidth, trim=1.5cm 0.1cm 3.2cm 1.2cm, clip]{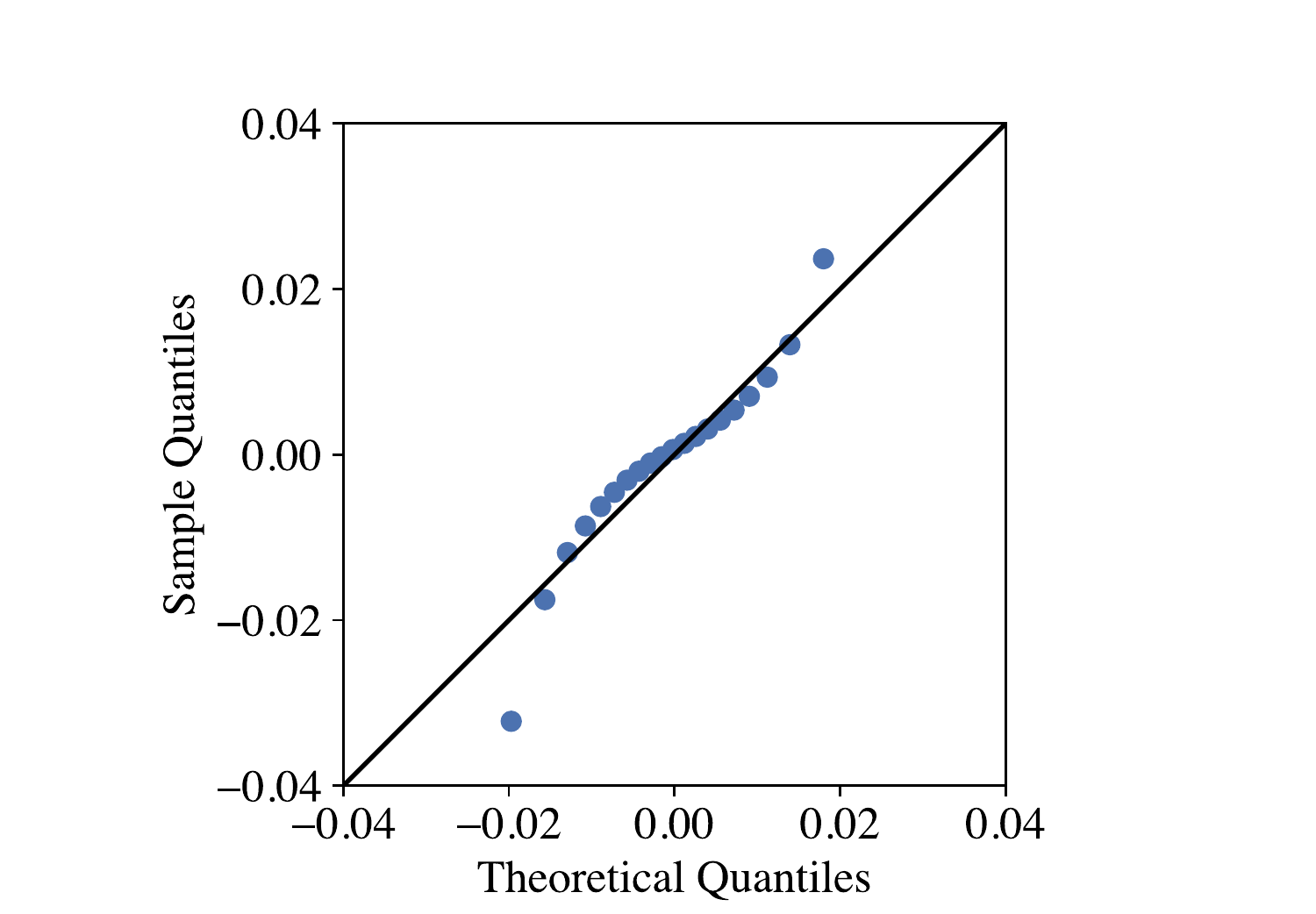}%
}%
\hspace*{\fill}
\subfigure[Data vs. RBM]{%
\includegraphics[width=0.32\linewidth, trim=1.5cm 0.1cm 3.2cm 1.2cm, clip]{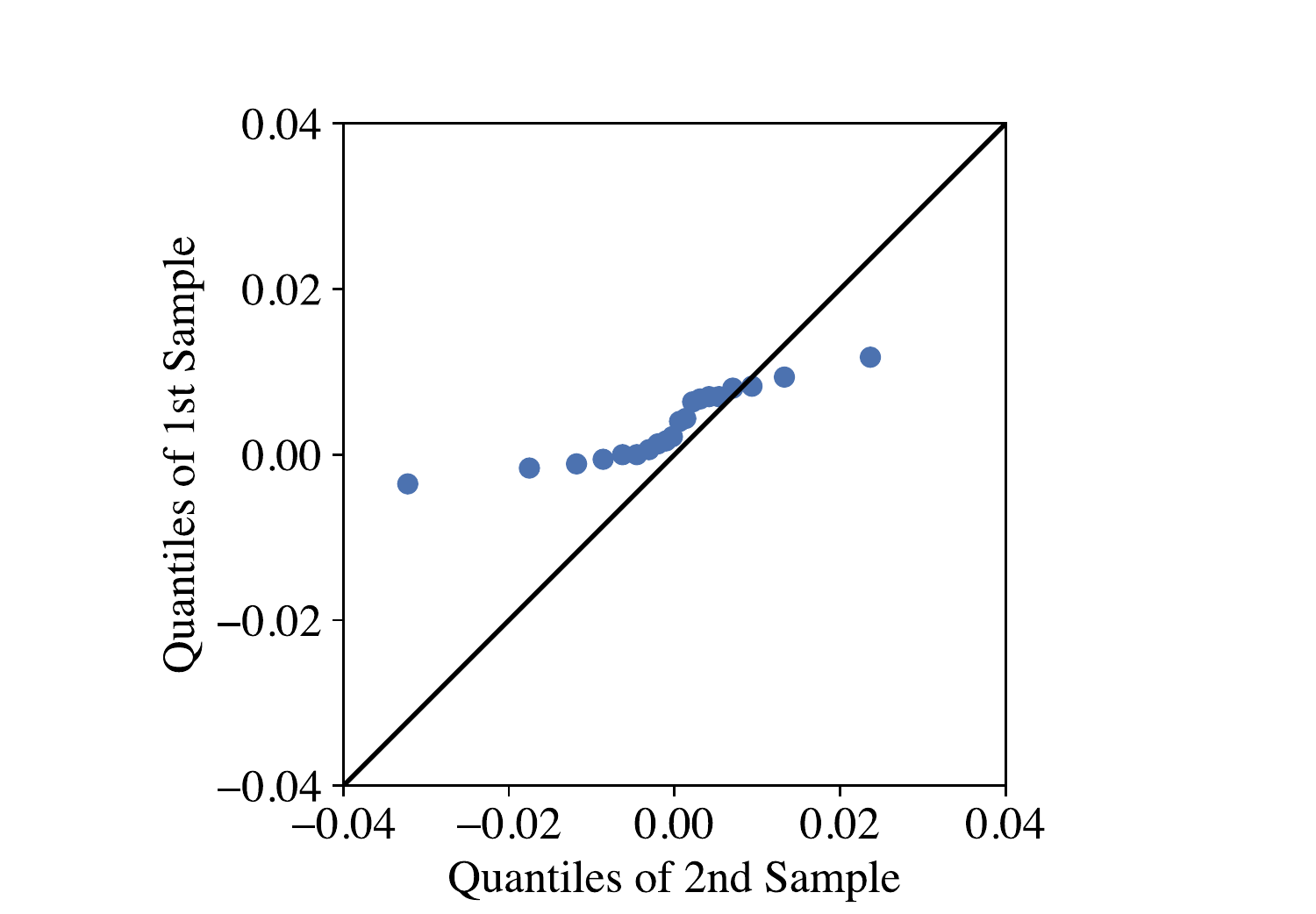}%
}
\subfigure[Data vs. Normal]{%
\includegraphics[width=0.32\linewidth,trim=1.5cm 0.1cm 3.2cm 1.2cm, clip]{figures/Real_Data/robust/qq/4real.pdf}%
}%
\hspace*{\fill}
\subfigure[CVAE vs. Normal]{%
\includegraphics[width=0.32\linewidth, trim=1.5cm 0.1cm 3.2cm 1.2cm, clip]{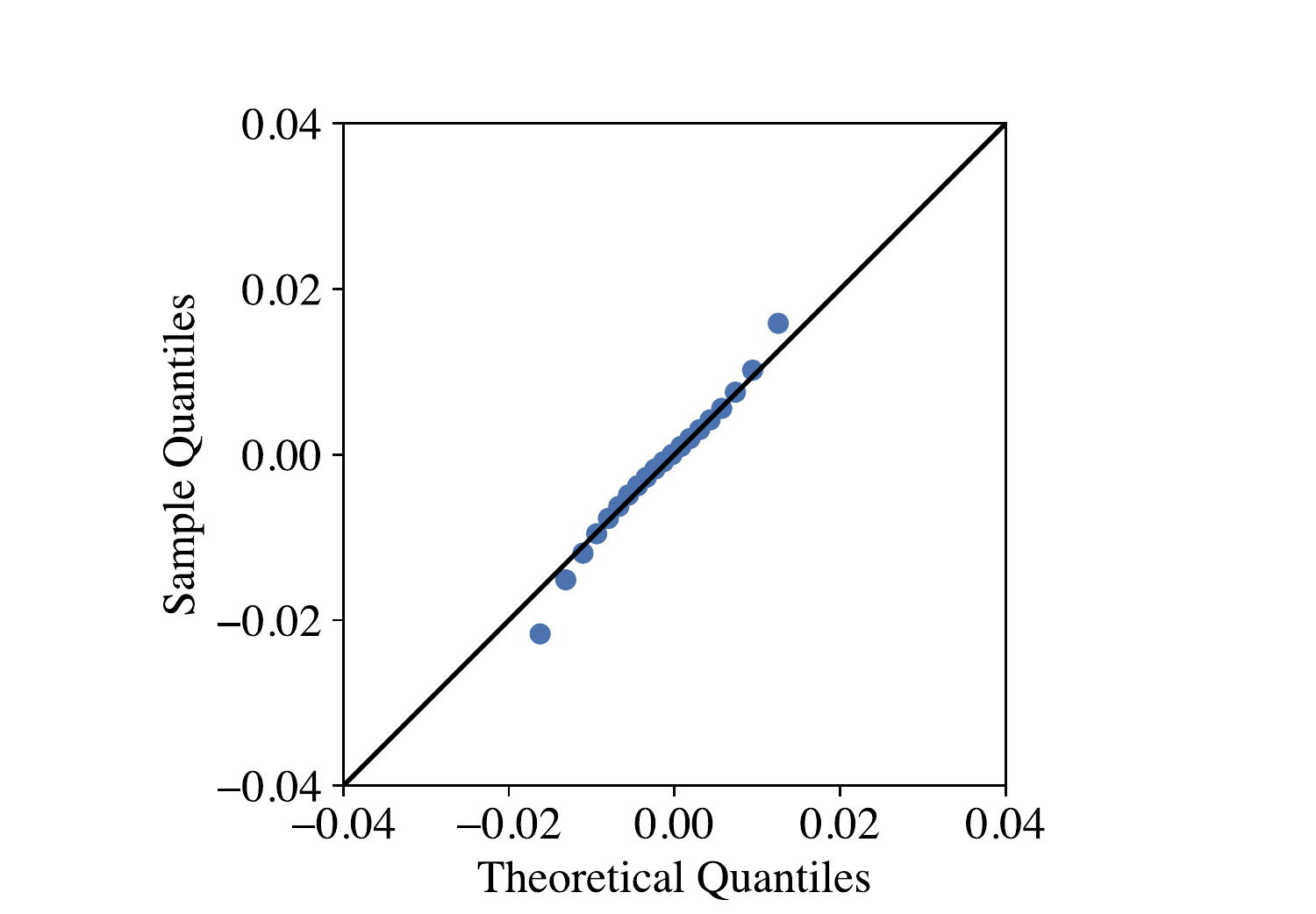}%
}%
\hspace*{\fill}
\subfigure[Data vs. CVAE]{%
\includegraphics[width=0.32\linewidth, trim=1.5cm 0.1cm 3.2cm 1.2cm, clip]{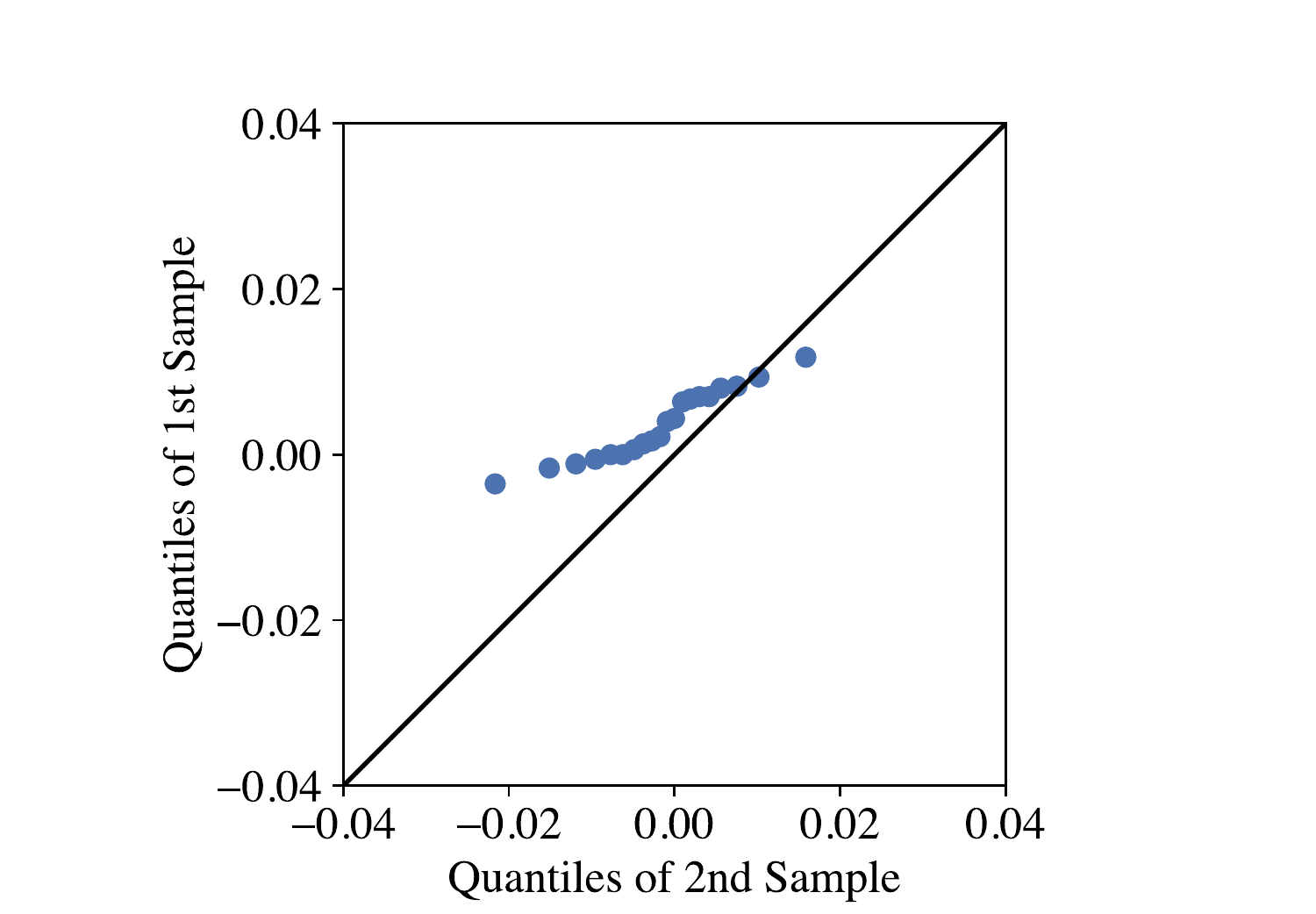}%
}
\caption[QQ Plots for 4 weeks test period]{QQ Plots of real Log Returns and generated samples over 4 weeks period: a)-c): Filtered Historical Simulation; d)-f): GARCH; g)-i): RBM; j)-l): CVAE.}
\label{realQQfour}
\end{figure*}

\begin{figure}[tbh]
\centering
\begin{adjustbox}{minipage=\textwidth, scale=0.87}
\subfigure[FHS]{%
\includegraphics[width=0.5\linewidth]{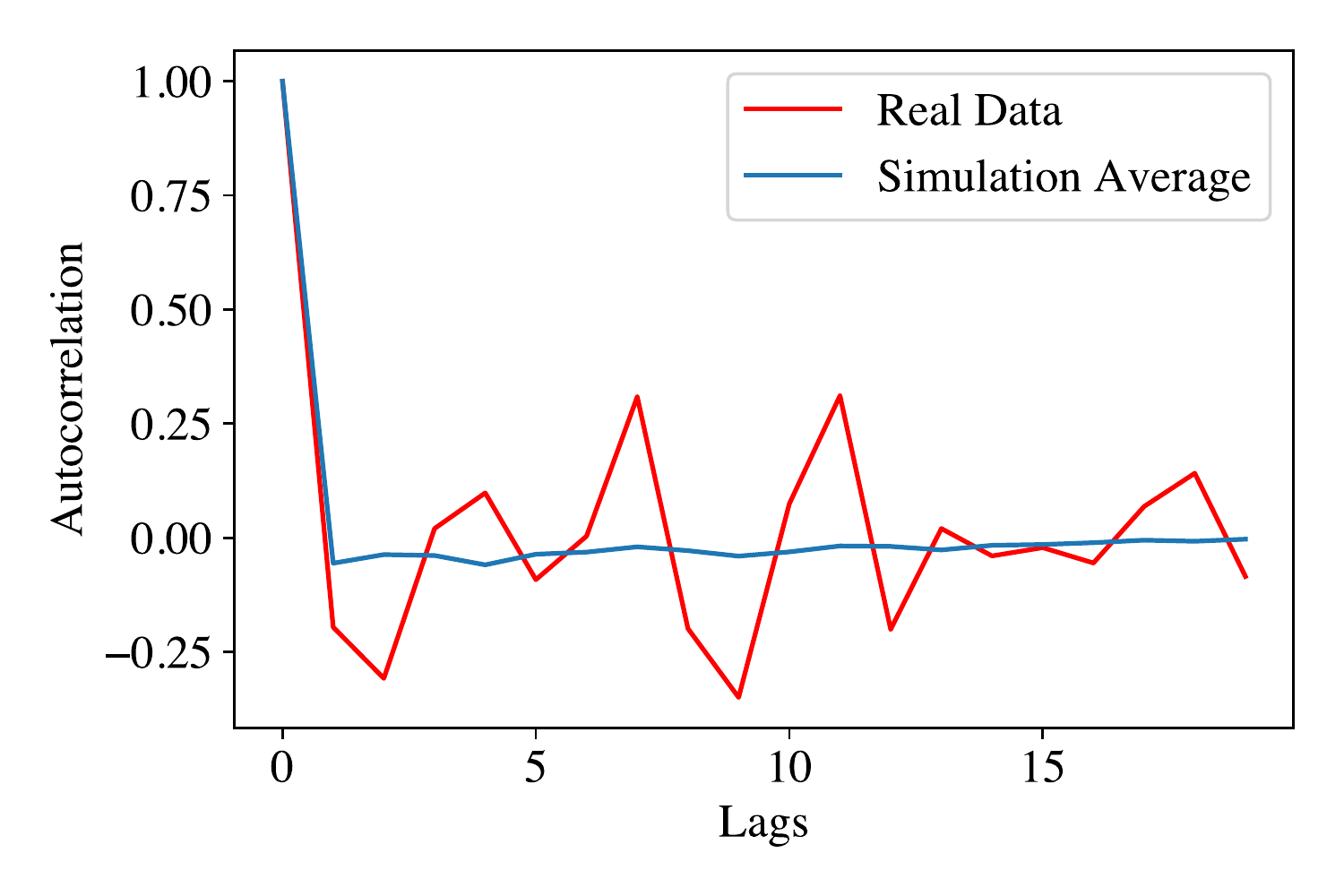}%
}%
\hspace*{\fill}
\subfigure[GARCH]{%
\includegraphics[width=0.5\linewidth]{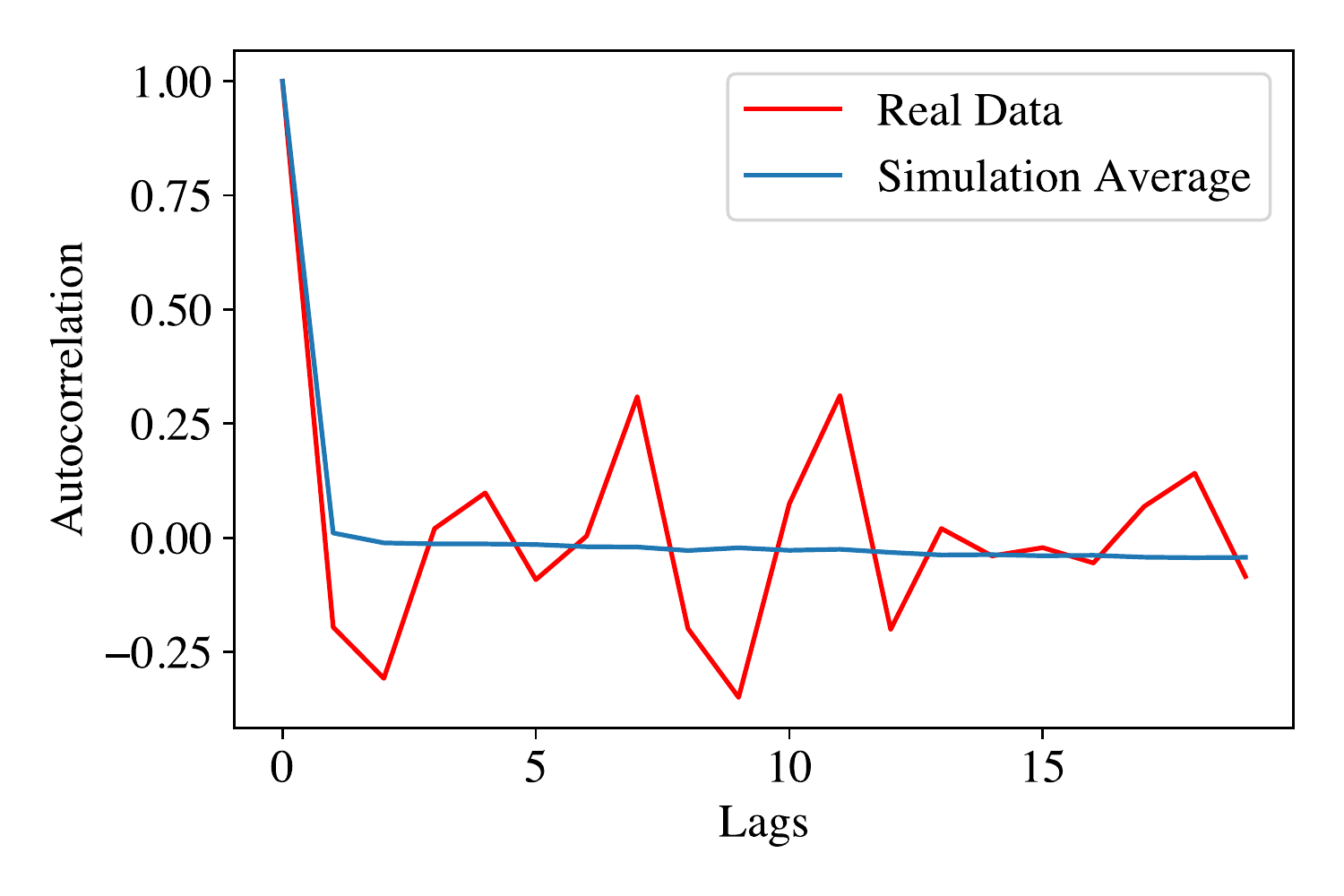}%
}
\subfigure[RBM]{%
\includegraphics[width=0.5\linewidth]{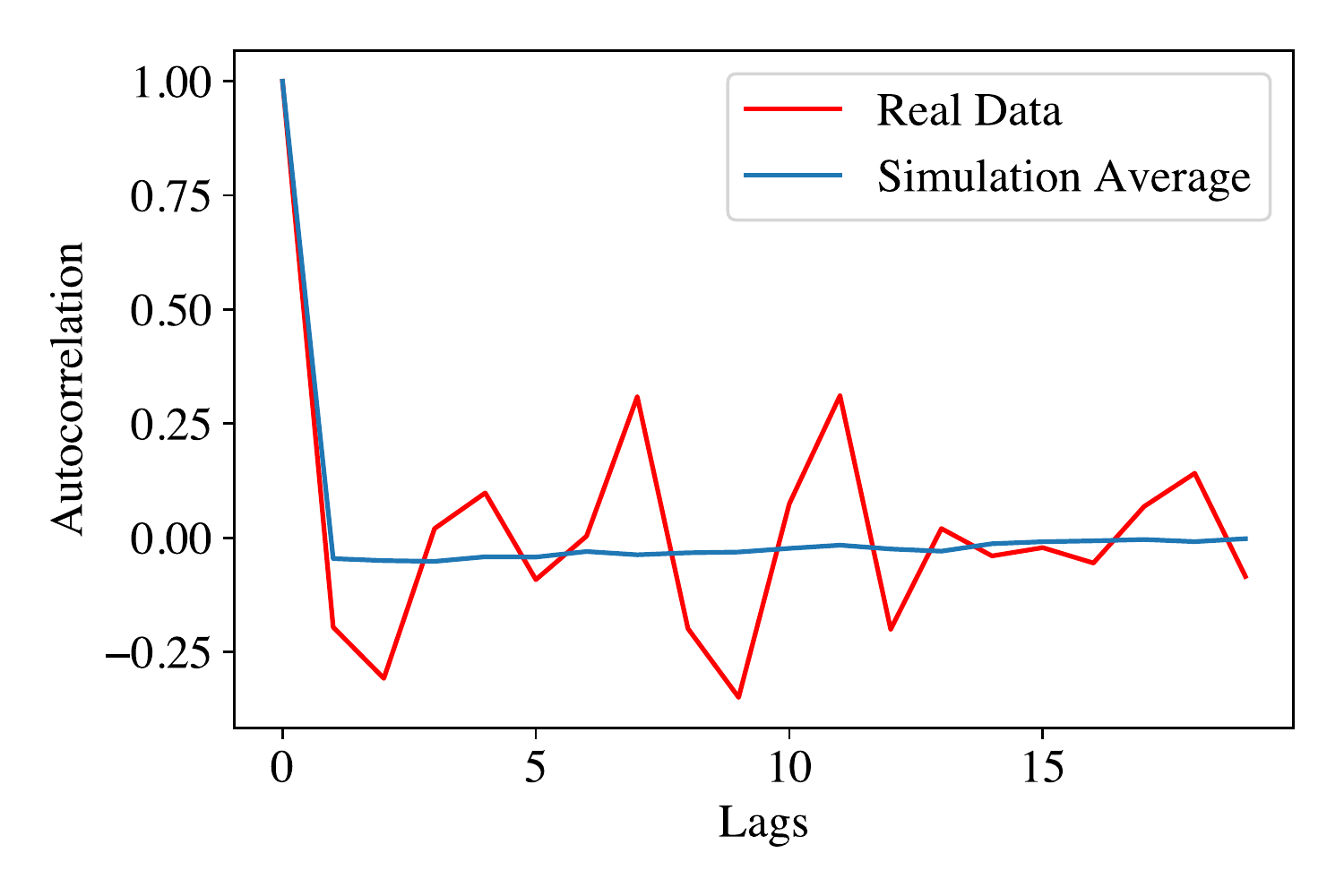}%
}%
\hspace*{\fill}
\subfigure[CVAE]{%
\includegraphics[width=0.5\linewidth]{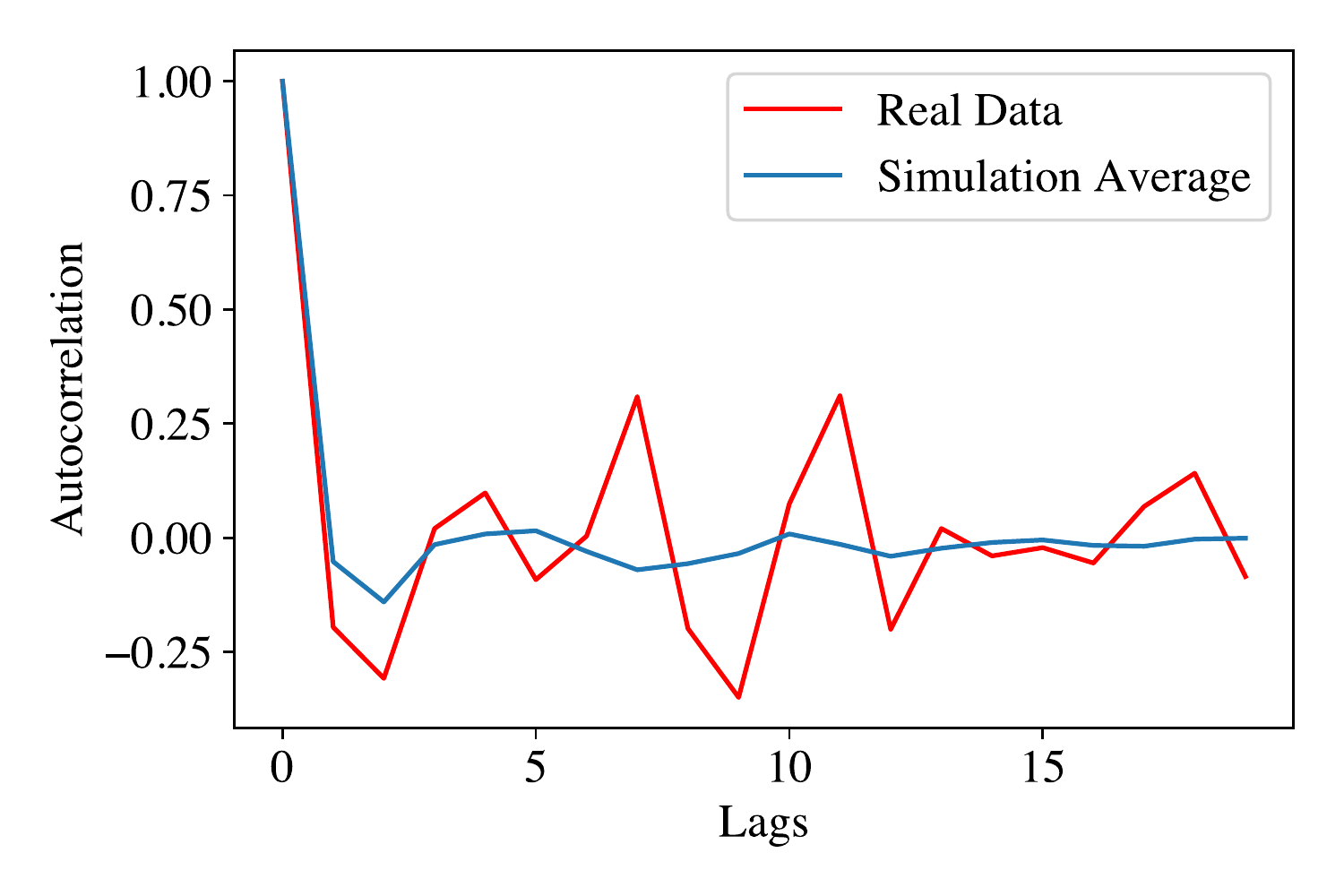}%
}
\caption[Autocorrelation plots for 4 weeks test period]{Average autocorrelation over 4 weeks period of generated returns vs. autocorrelation of real S\&P500 Log Returns.}
\subfigure[FHS]{%
\includegraphics[width=0.5\linewidth]{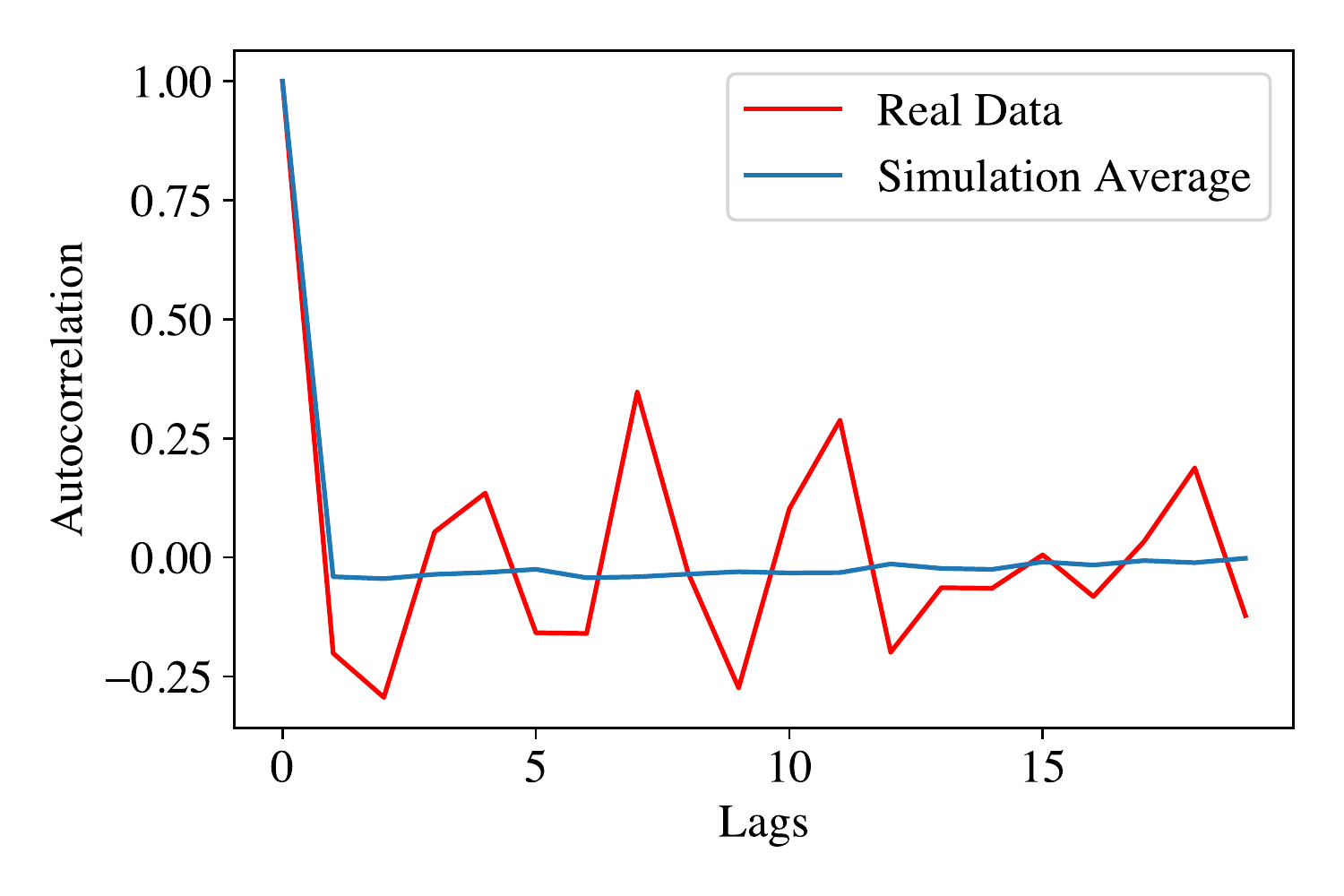}%
}%
\hspace*{\fill}
\subfigure[GARCH]{%
\includegraphics[width=0.5\linewidth]{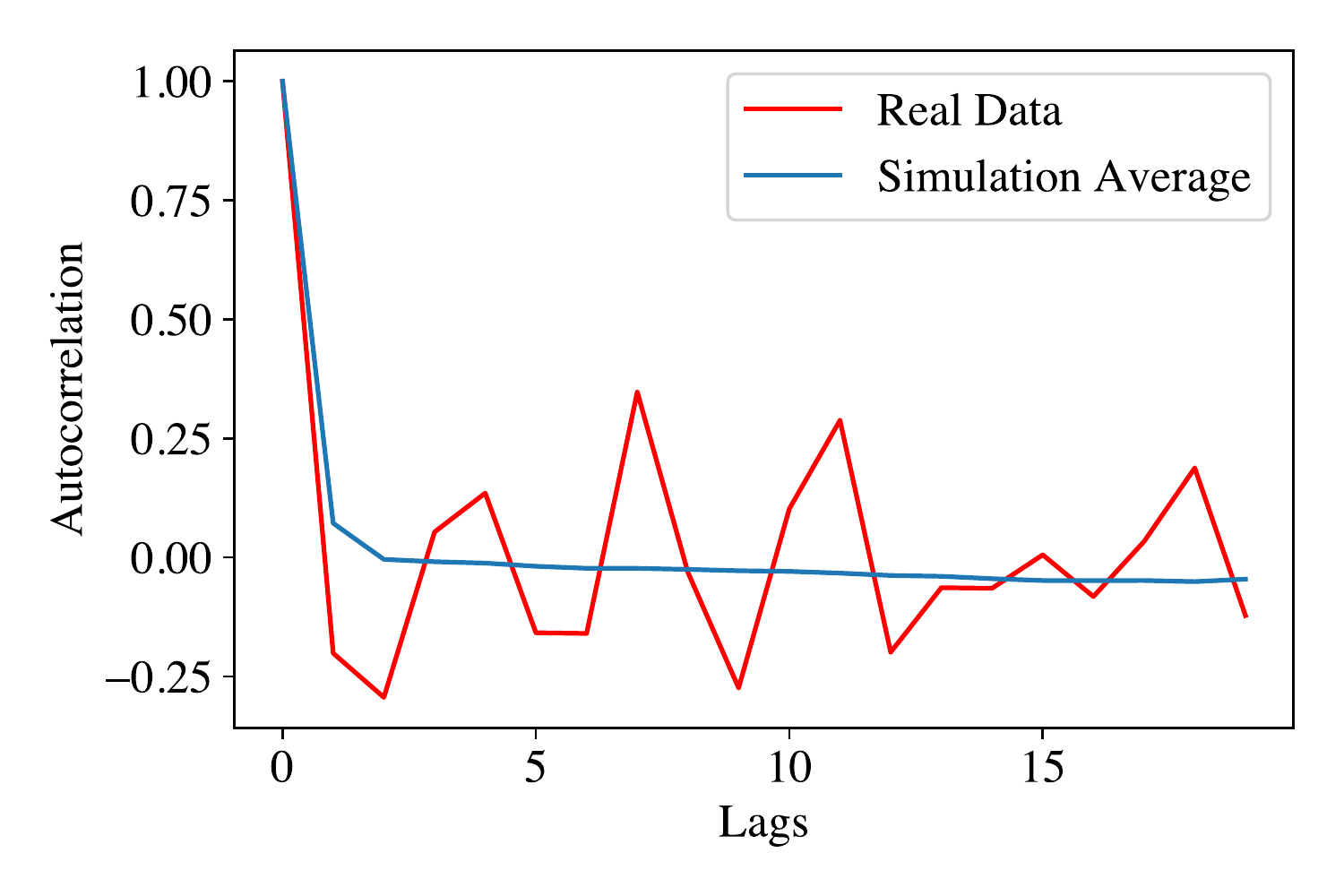}%
}
\subfigure[RBM]{%
\includegraphics[width=0.5\linewidth]{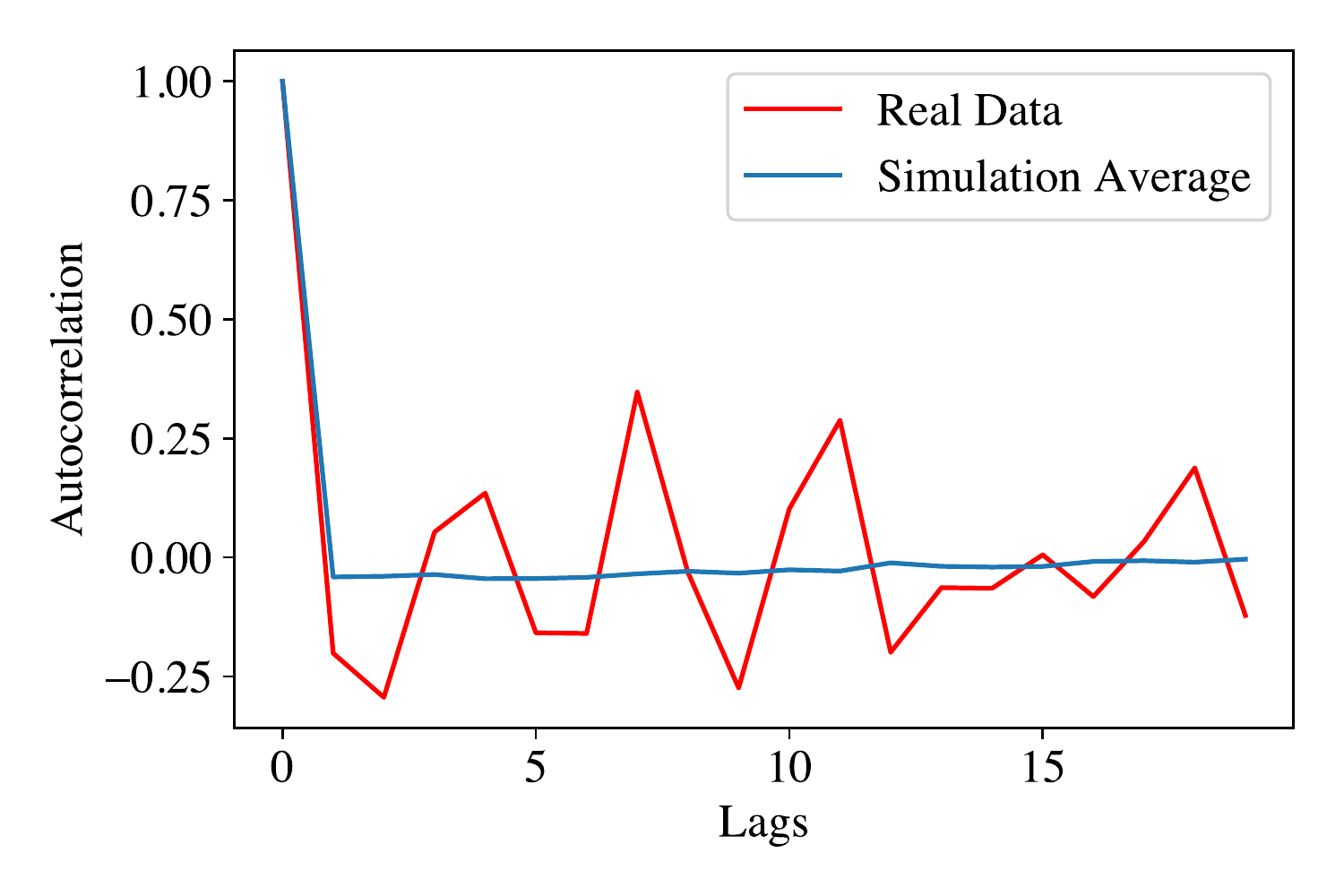}%
}%
\hspace*{\fill}
\subfigure[CVAE]{%
\includegraphics[width=0.5\linewidth]{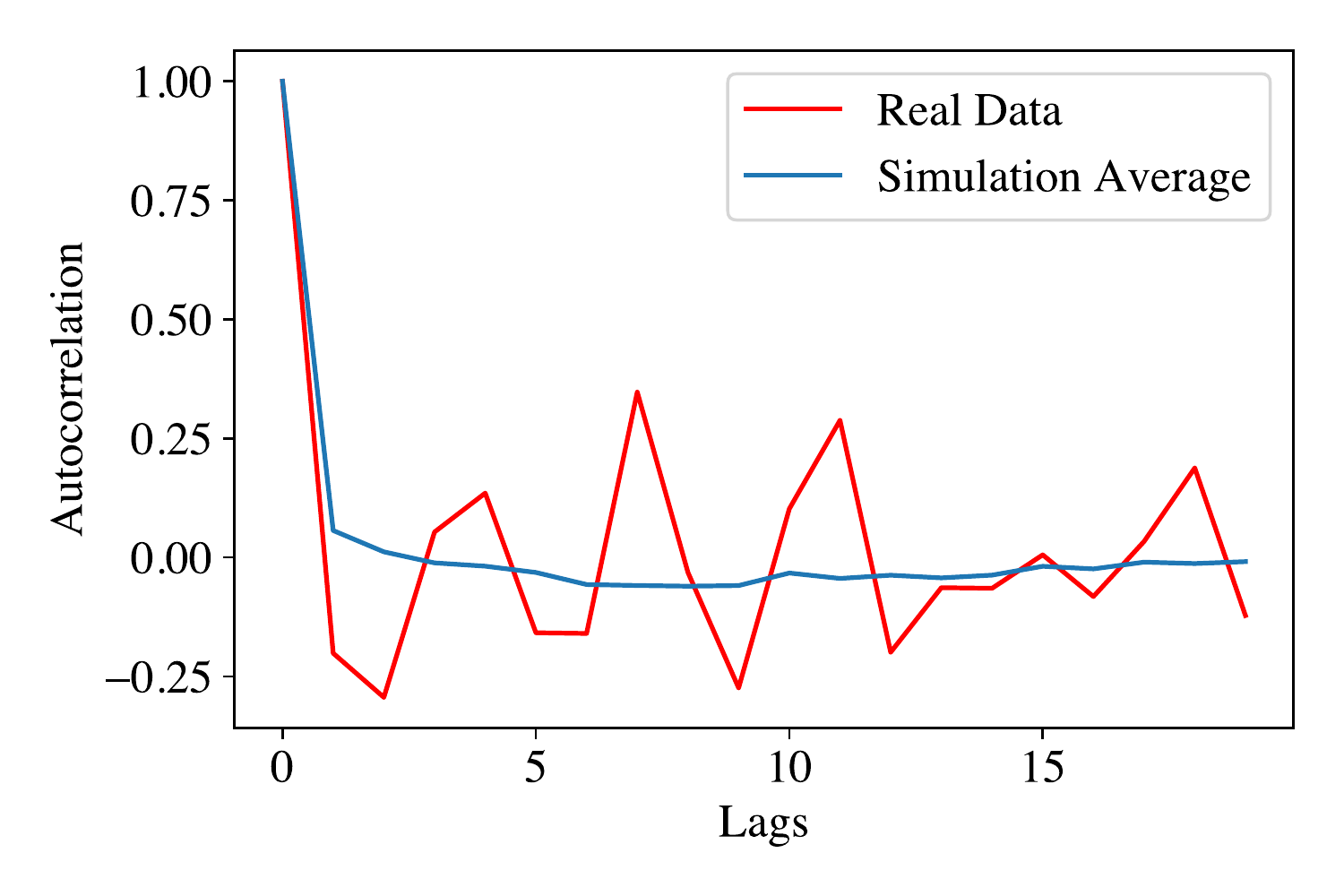}%
}
\caption[Autocorrelation plots of squared samples for 4 weeks test period]{Average autocorrelation over 4 weeks period of squared generated returns vs. autocorrelation of squared S\&P500 Log Returns.}
\label{sqacfrealfour}
\end{adjustbox}
\end{figure} 

\begin{figure*}
\centering  
\SetFigLayout{4}{3}
  \subfigure[Data vs. Normal]{%
\includegraphics[width=0.32\linewidth, trim=2cm 0.1cm 3.9cm 1.55cm, clip]{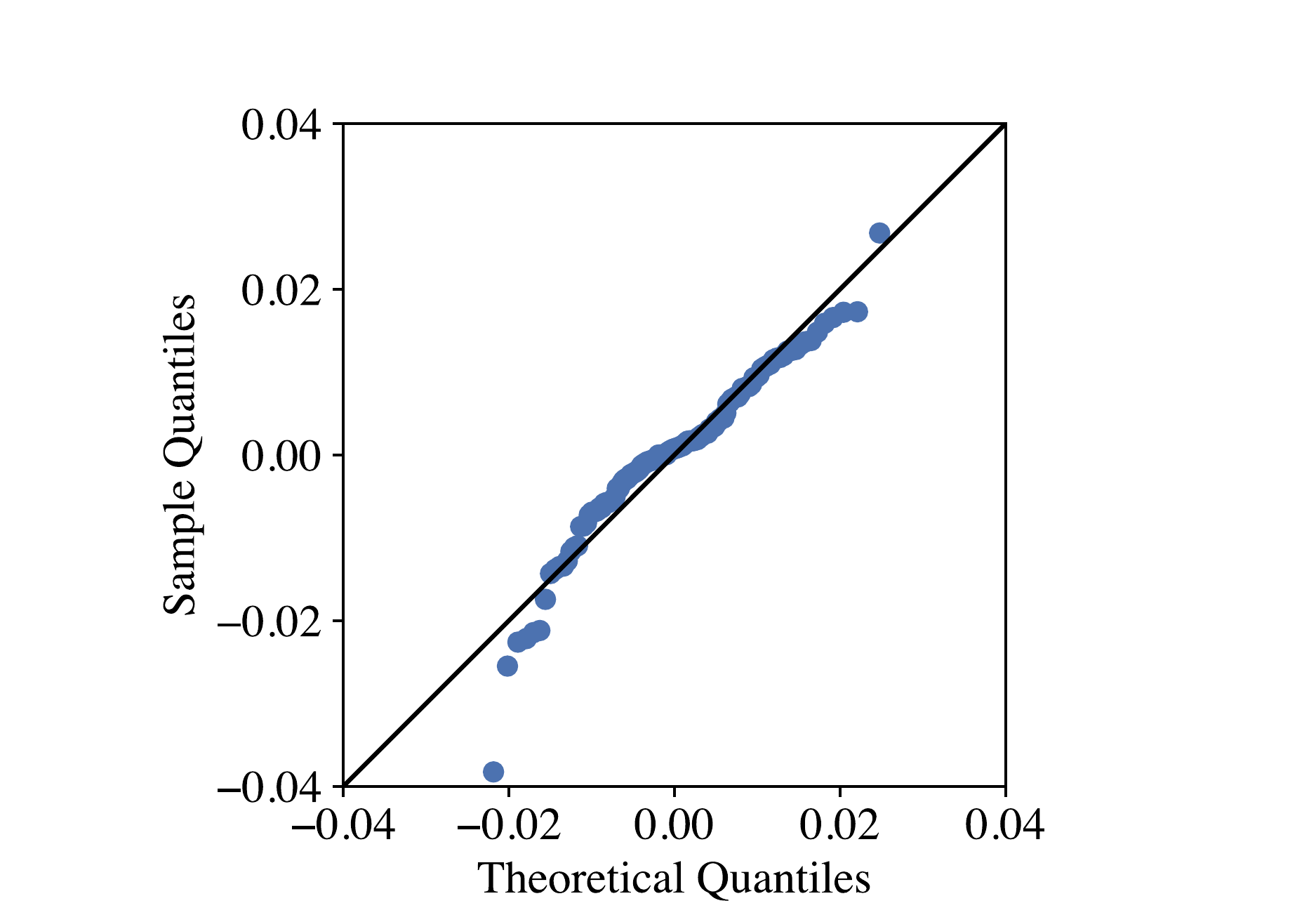}%
}%
\hspace*{\fill}
\subfigure[FHS vs. Normal]{%
\includegraphics[width=0.32\linewidth,trim=2cm 0.1cm 3.9cm 1.55cm, clip]{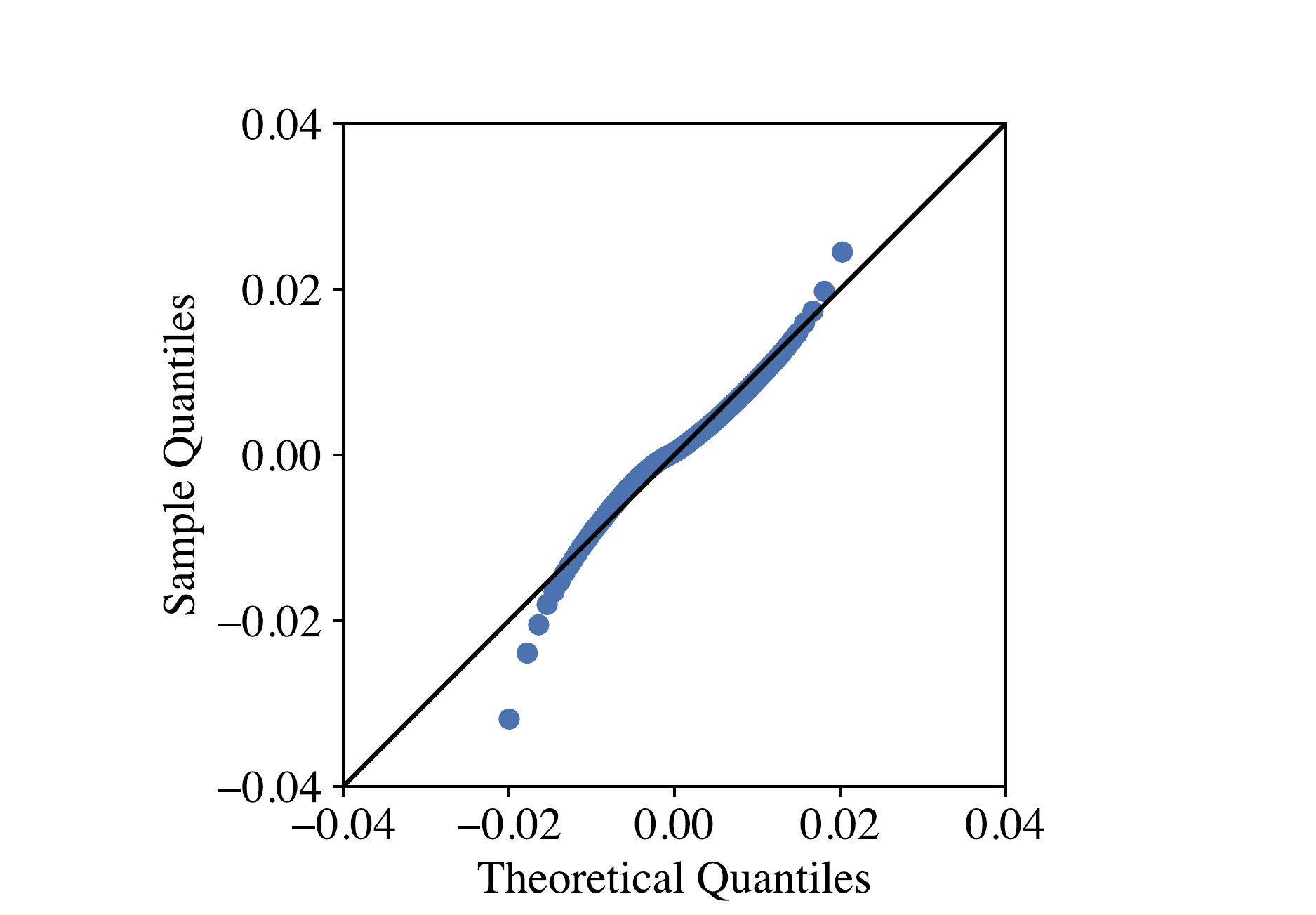}%
}%
\hspace*{\fill}
\subfigure[Data vs. FHS]{%
\includegraphics[width=0.32\linewidth,trim=2cm 0.1cm 3.9cm 1.55cm, clip]{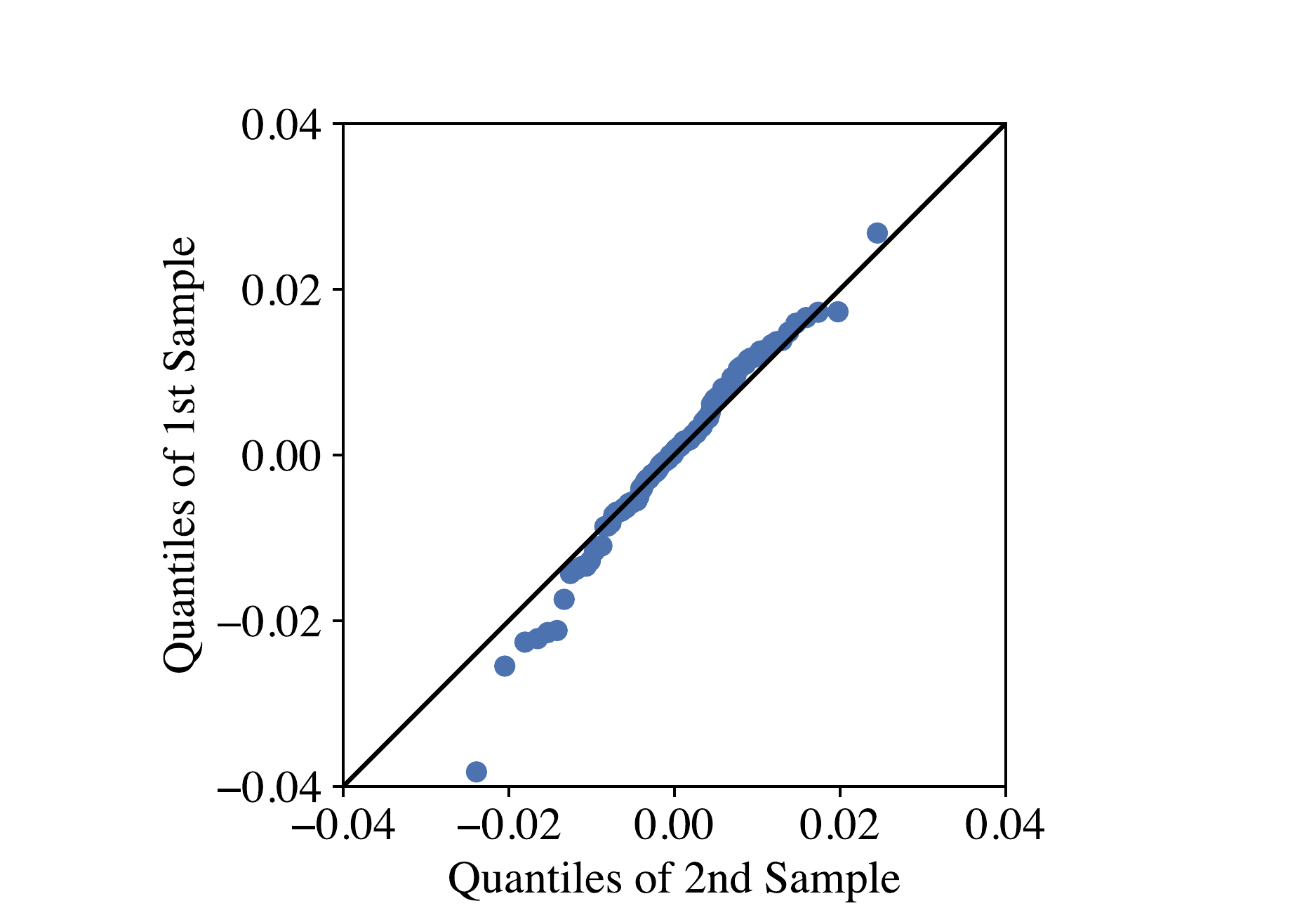}%
}
\subfigure[Data vs. Normal]{%
\includegraphics[width=0.32\linewidth, trim=2cm 0.1cm 3.9cm 1.55cm, clip]{figures/Real_Data/robust/qq/26real.pdf}%
}%
\hspace*{\fill}
\subfigure[GARCH vs. Normal]{%
\includegraphics[width=0.32\linewidth, trim=2cm 0.1cm 3.9cm 1.55cm, clip]{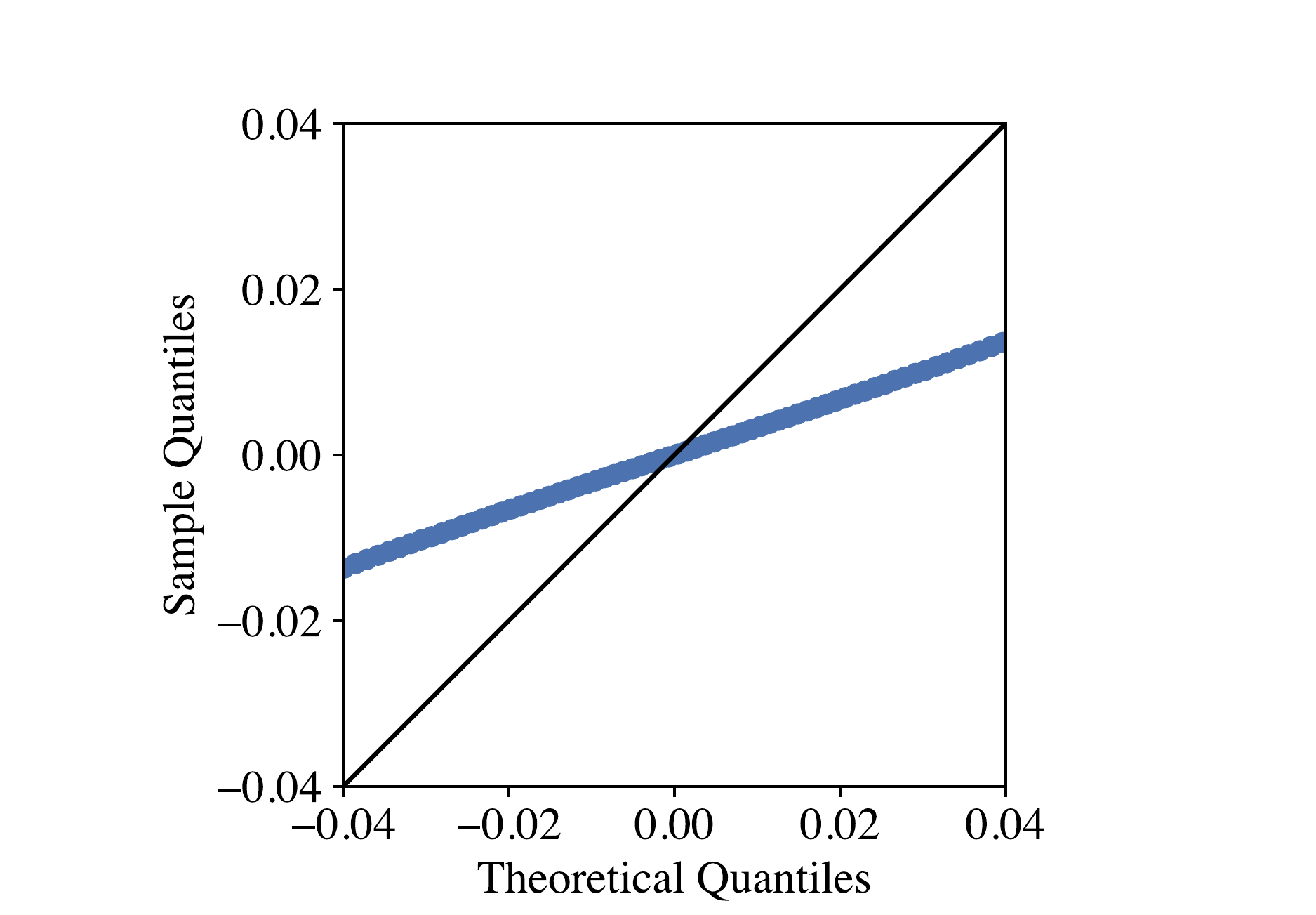}%
}%
\hspace*{\fill}
\subfigure[Data vs. GARCH]{%
\includegraphics[width=0.32\linewidth, trim=2cm 0.1cm 3.9cm 1.55cm, clip]{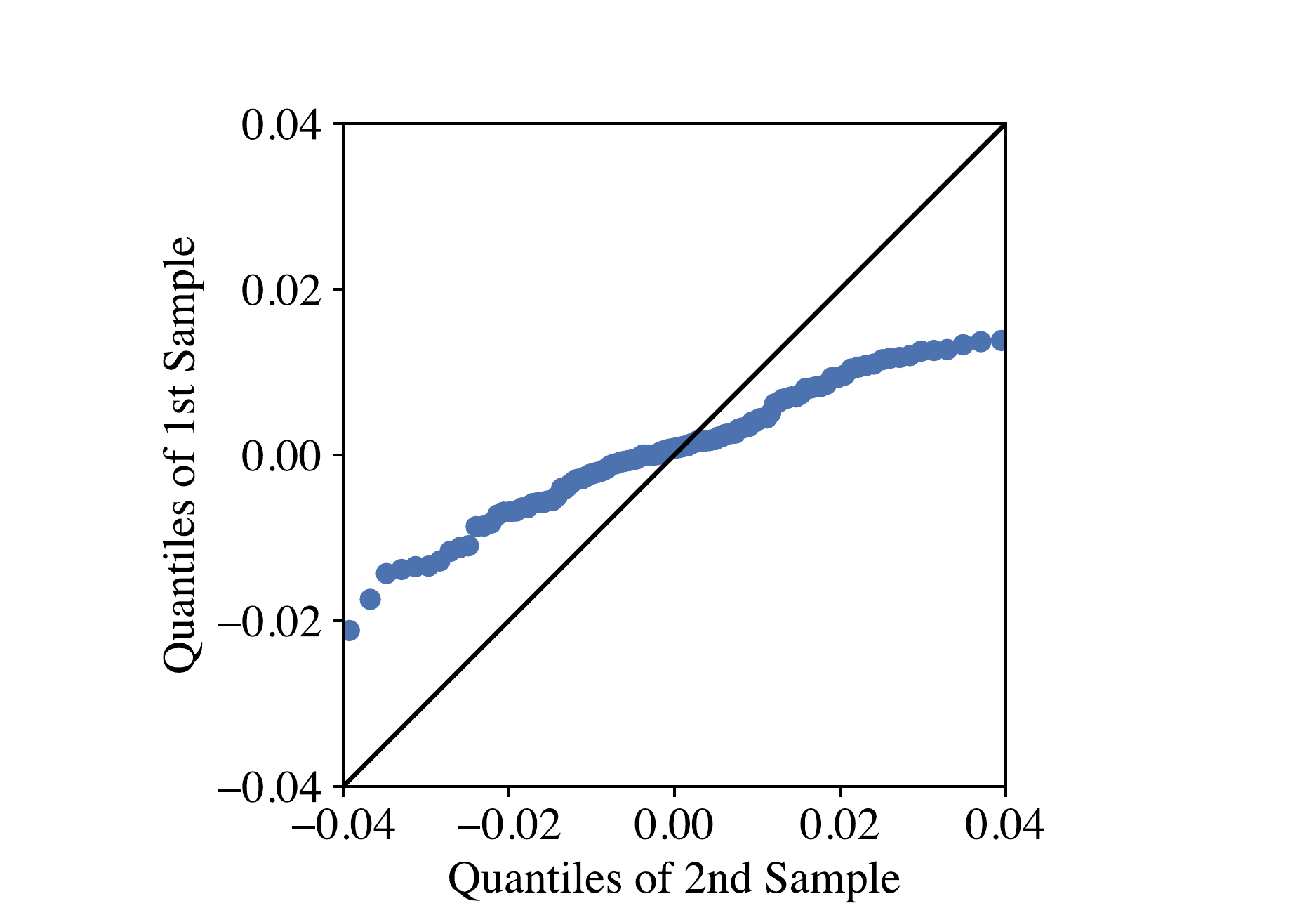}%
}
\subfigure[Data vs. Normal]{%
\includegraphics[width=0.32\linewidth, trim=2cm 0.1cm 3.9cm 1.55cm, clip]{figures/Real_Data/robust/qq/26real.pdf}%
}%
\hspace*{\fill}
\subfigure[RBM vs. Normal]{%
\includegraphics[width=0.32\linewidth, trim=1.8cm 0.1cm 3cm 0.5cm, clip]{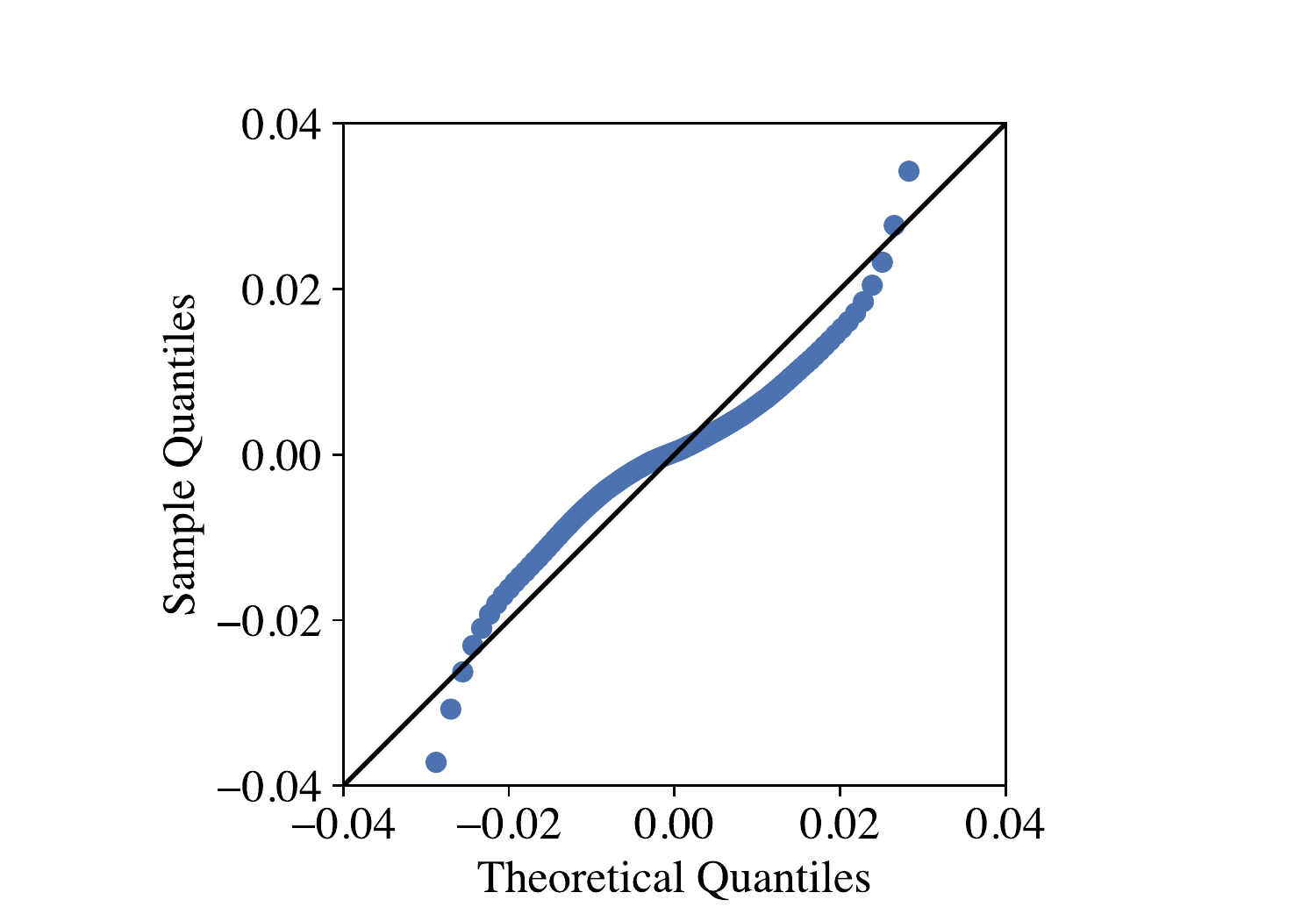}%
}%
\hspace*{\fill}
\subfigure[Data vs. RBM]{%
\includegraphics[width=0.32\linewidth, trim=1.8cm 0.1cm 3cm 0.5cm, clip]{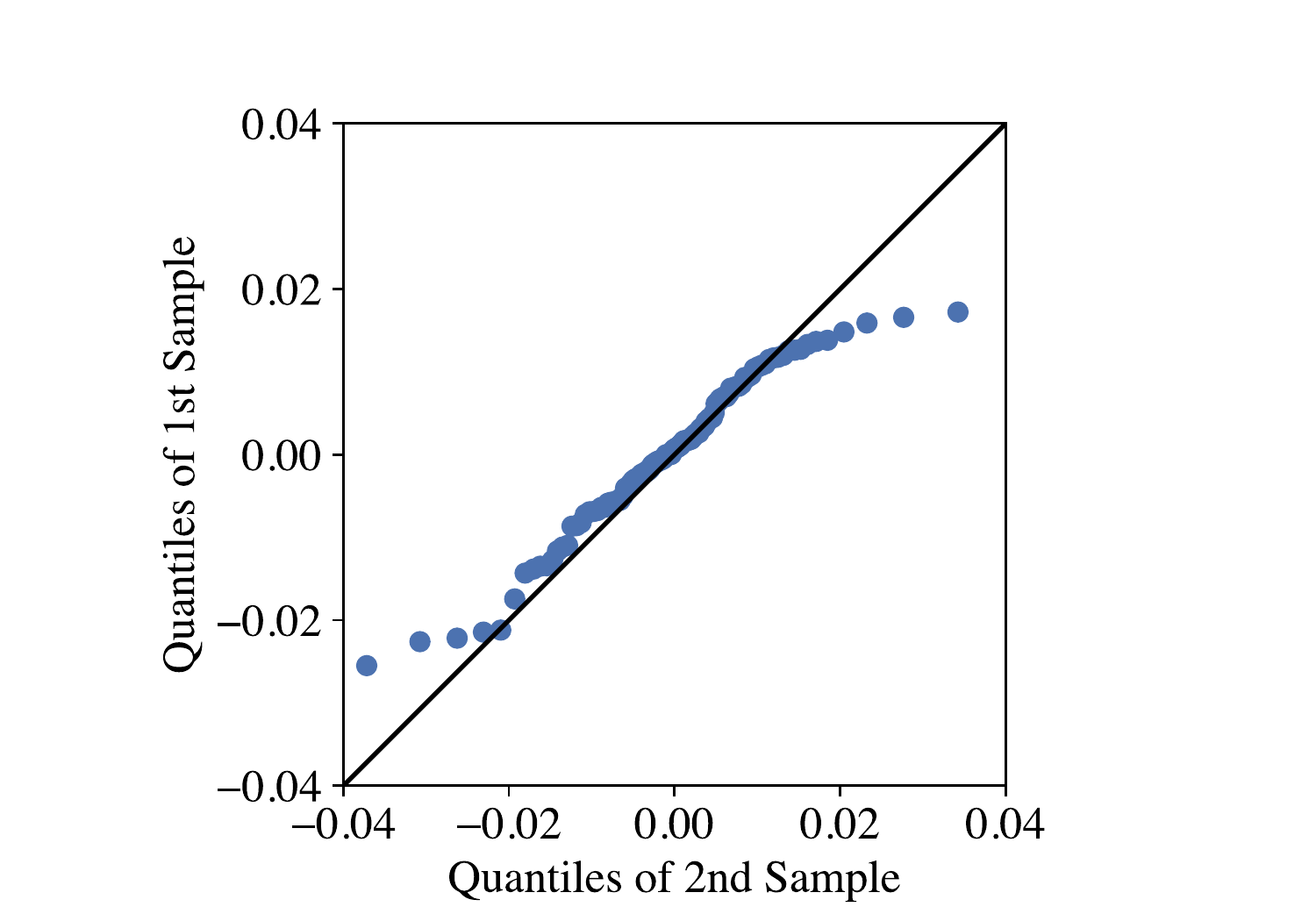}%
}
\subfigure[Data vs. Normal]{%
\includegraphics[width=0.32\linewidth, trim=2cm 0.1cm 3.9cm 1.55cm, clip]{figures/Real_Data/robust/qq/26real.pdf}%
}%
\hspace*{\fill}
\subfigure[CVAE vs. Normal]{%
\includegraphics[width=0.32\linewidth, trim=2cm 0.1cm 3.9cm 1.55cm, clip]{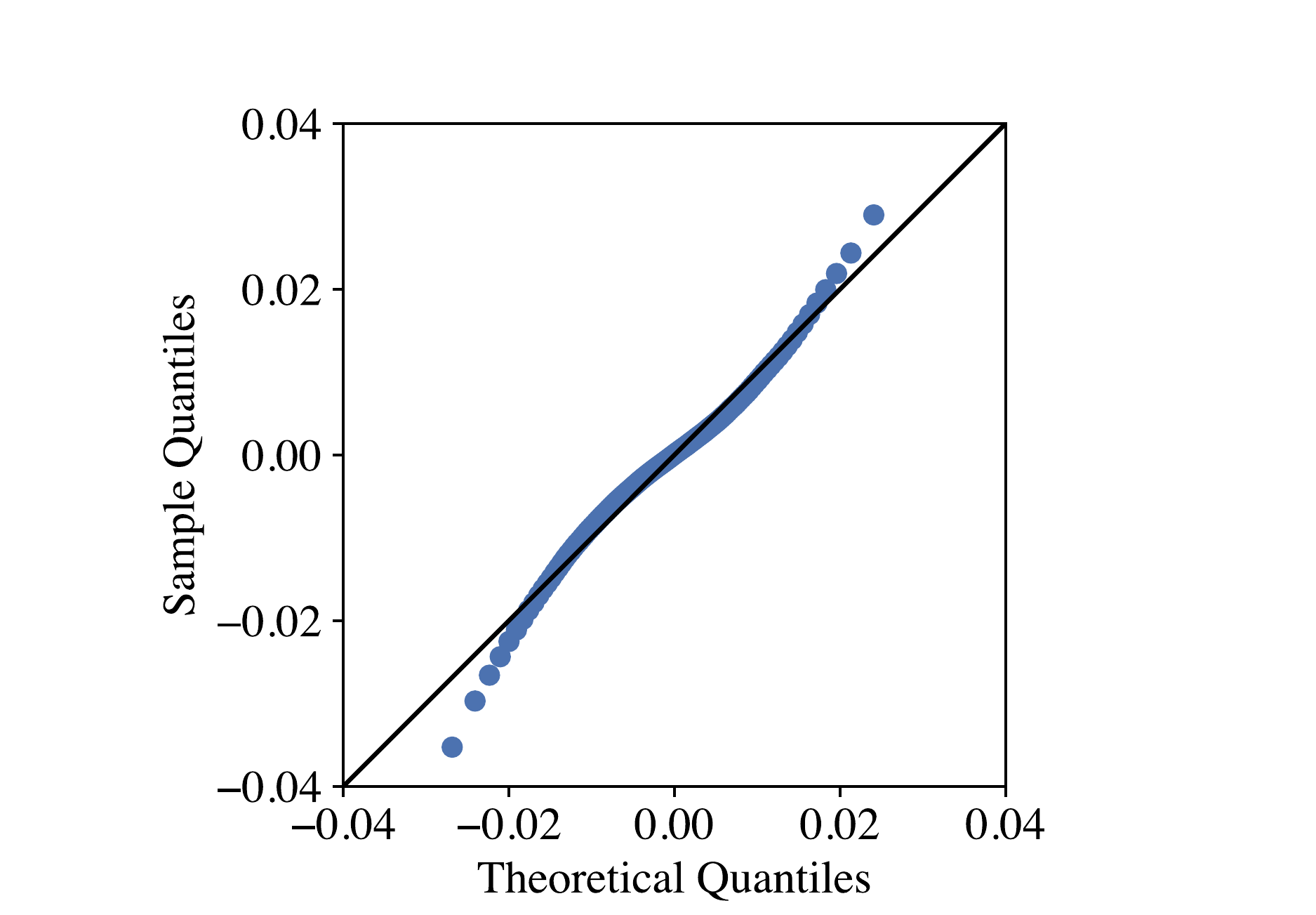}%
}%
\hspace*{\fill}
\subfigure[Data vs. CVAE]{%
\includegraphics[width=0.32\linewidth, trim=2cm 0.1cm 3.9cm 1.55cm, clip]{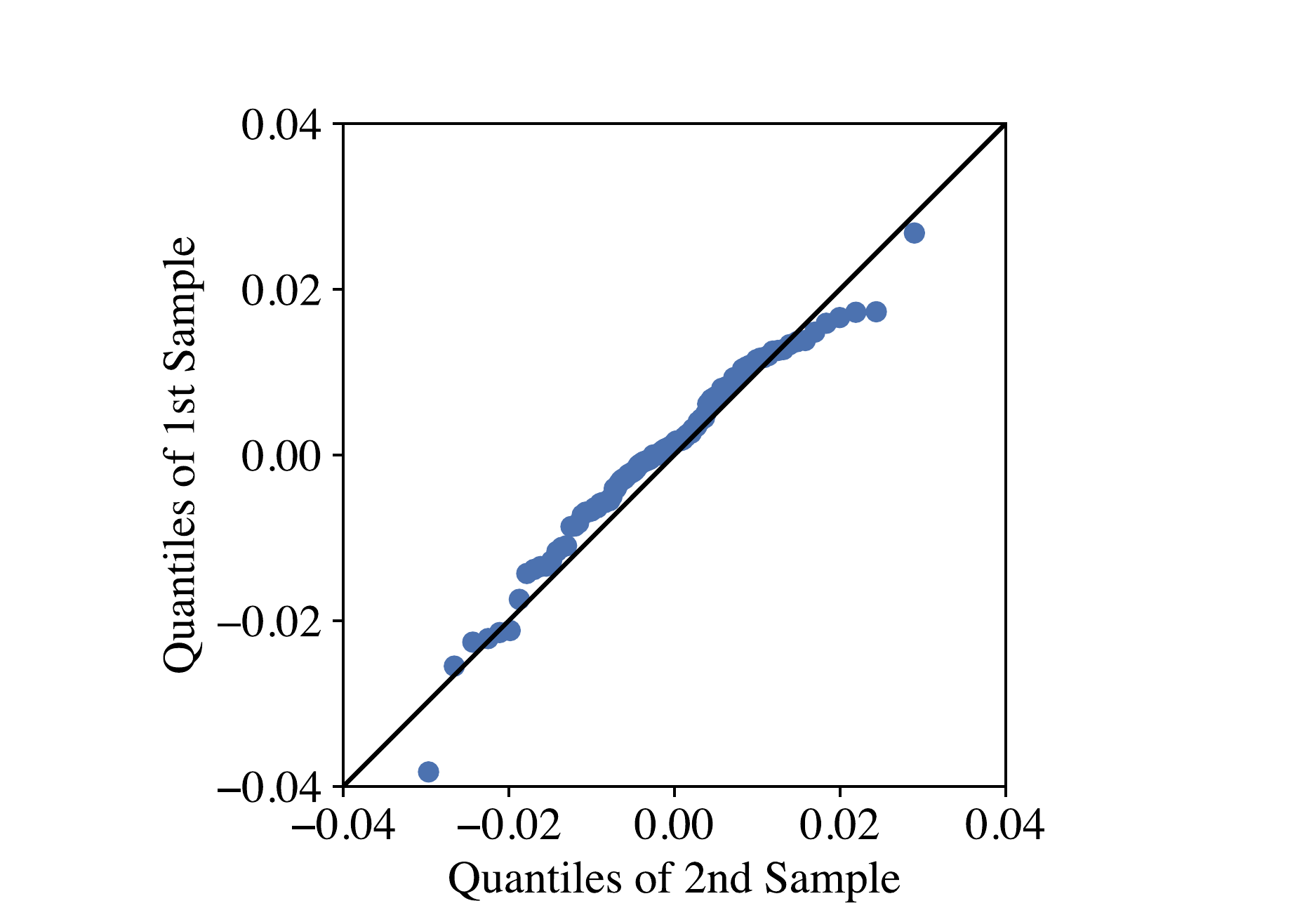}%
}
\caption[QQ Plots for 26 weeks test period]{QQ Plots of real Log Returns and generated samples over 26 weeks period: a)-c): Filtered Historical Simulation; d)-f): GARCH; g)-i): RBM; j)-l): CVAE.}
\label{sqacfreal26}
\end{figure*}

\begin{figure}[tbh]
\centering
\begin{adjustbox}{minipage=\textwidth, scale=0.86}
\subfigure[FHS]{%
\includegraphics[width=0.5\linewidth]{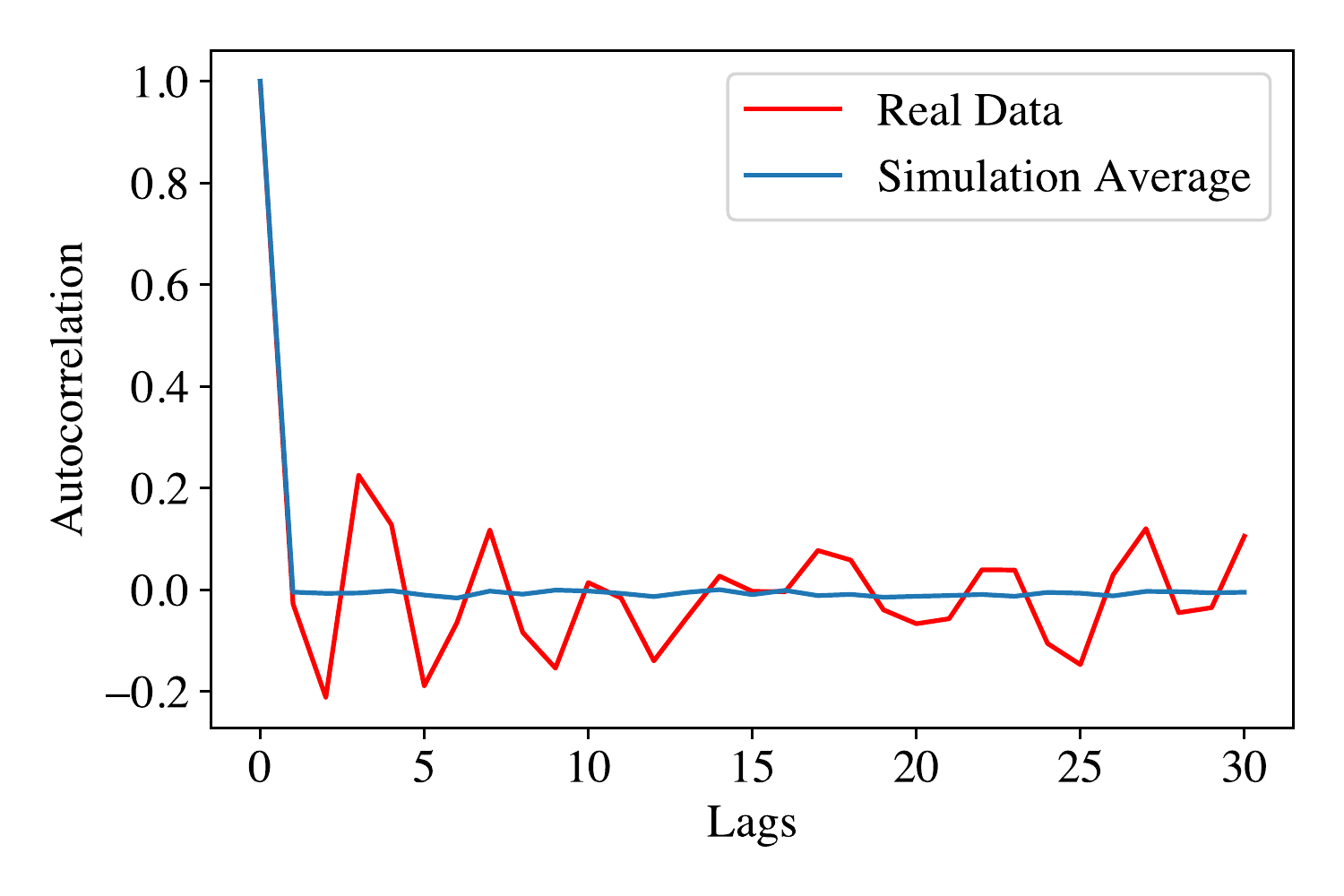}%
}%
\hspace*{\fill}
\subfigure[GARCH]{%
\includegraphics[width=0.5\linewidth]{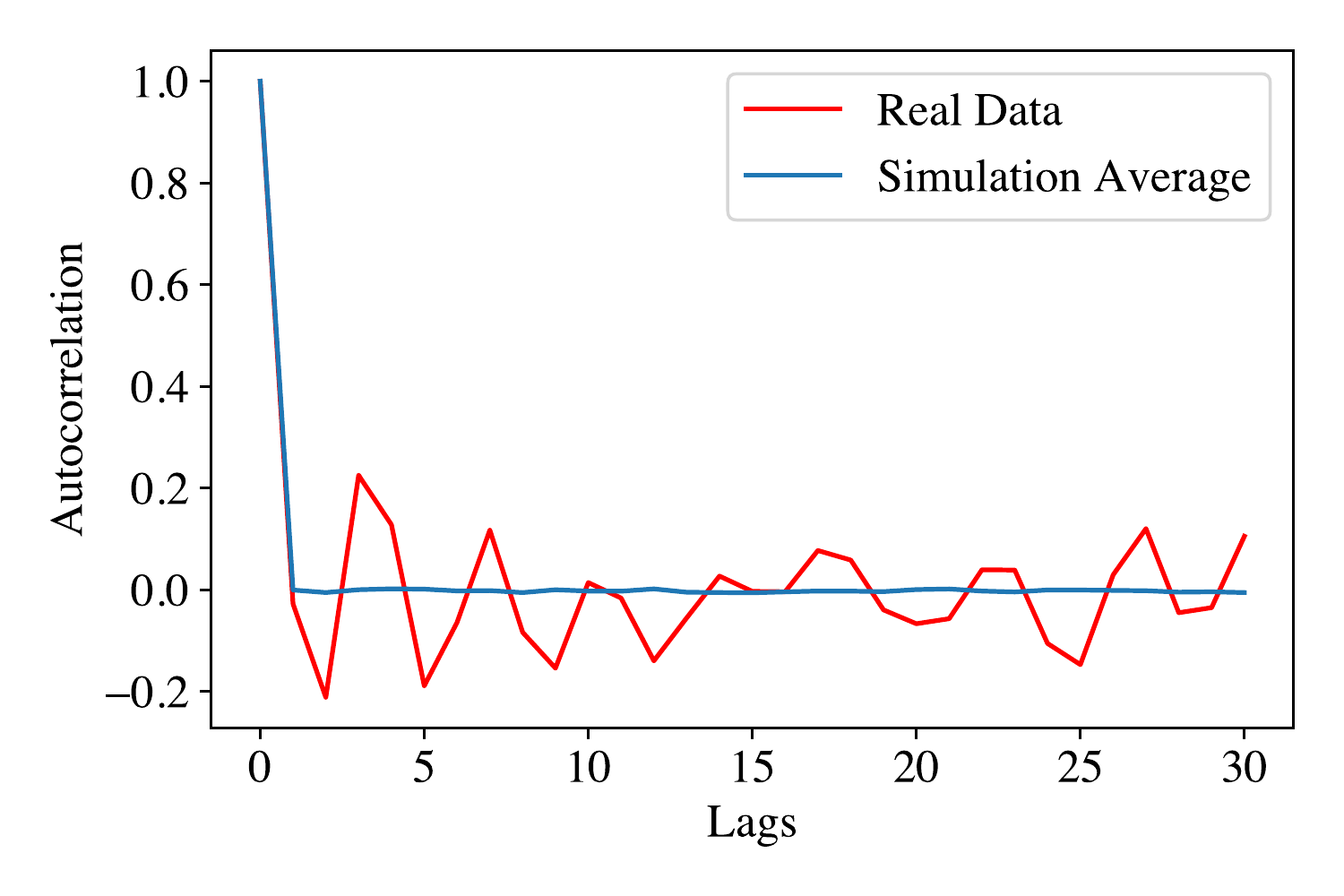}%
}
\subfigure[RBM]{%
\includegraphics[width=0.5\linewidth]{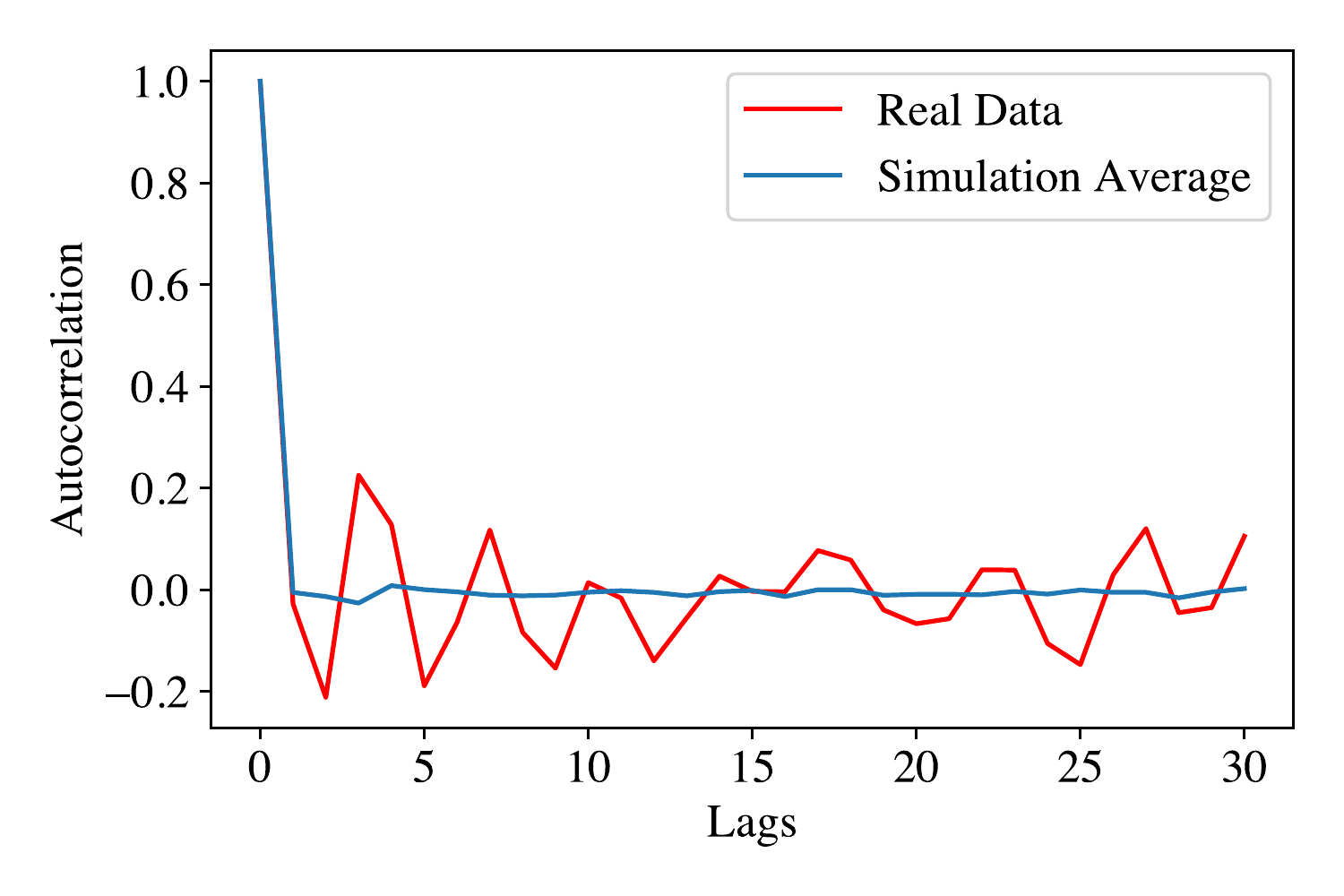}%
}%
\hspace*{\fill}
\subfigure[CVAE]{%
\includegraphics[width=0.5\linewidth]{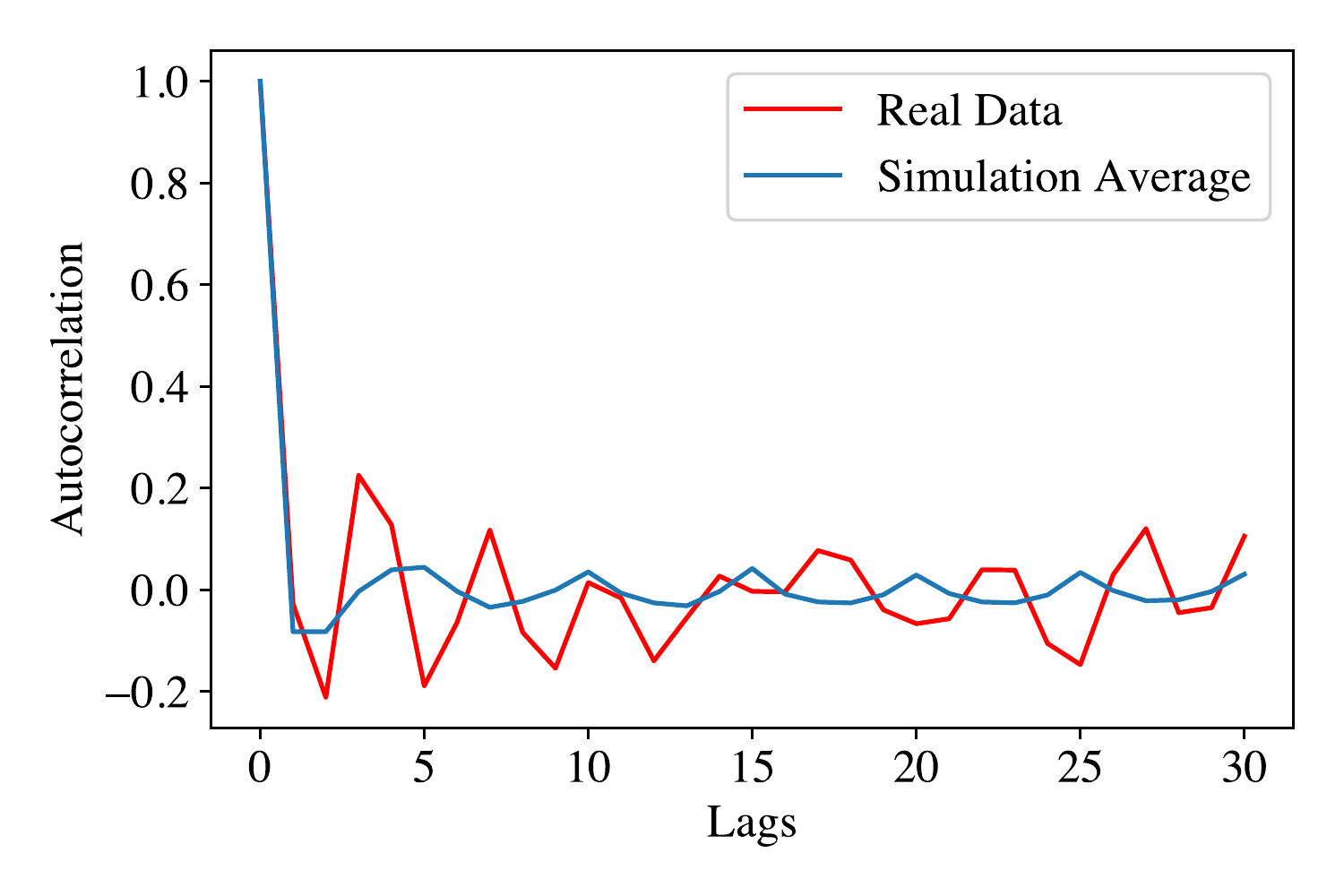}%
}
\caption[Autocorrelation plots for 26 weeks test period]{Average autocorrelation over 26 weeks of generated returns vs. autocorrelation of real S\&P500 Log Returns.}
\subfigure[FHS]{%
\includegraphics[width=0.5\linewidth]{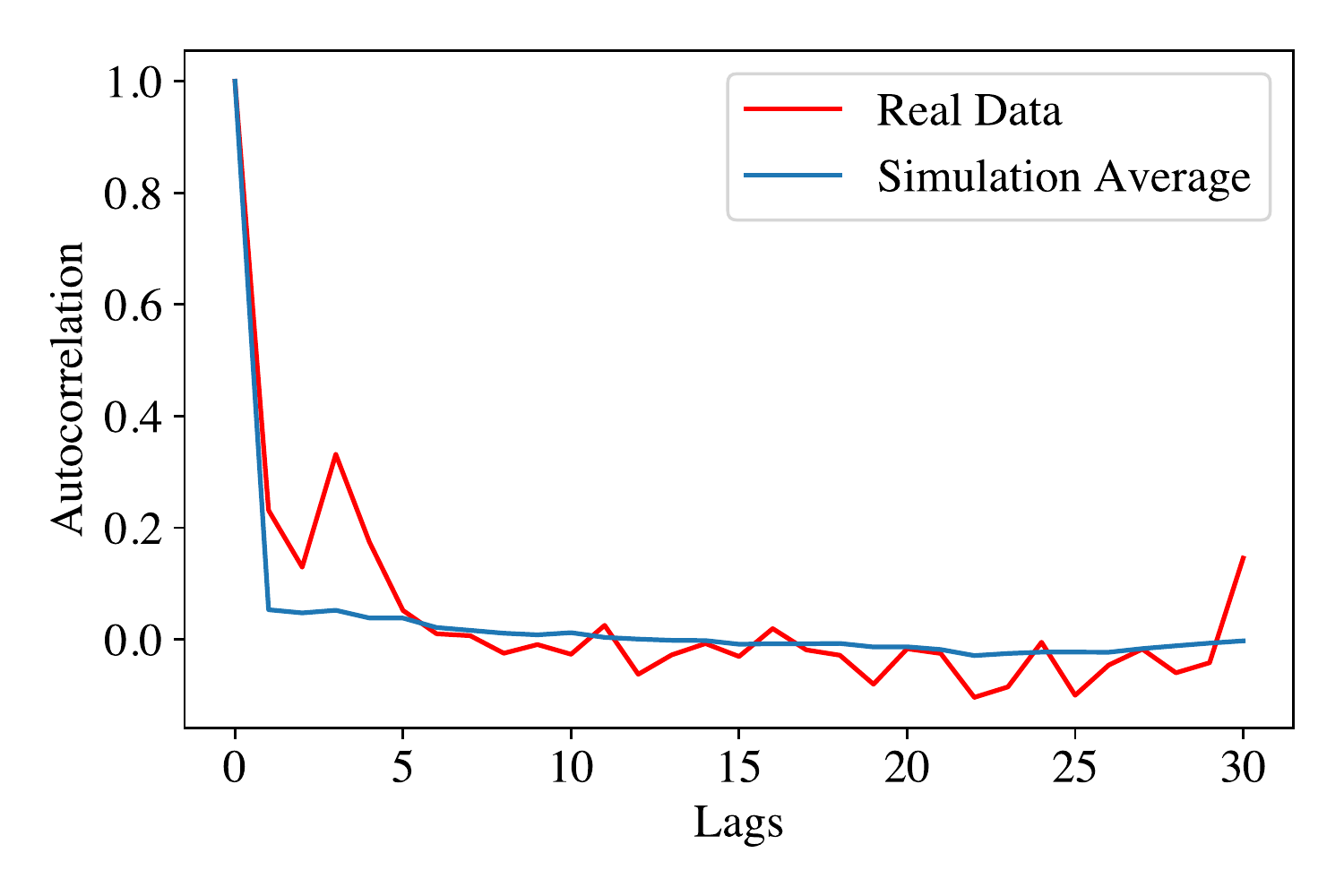}%
}%
\hspace*{\fill}
\subfigure[GARCH]{%
\includegraphics[width=0.5\linewidth]{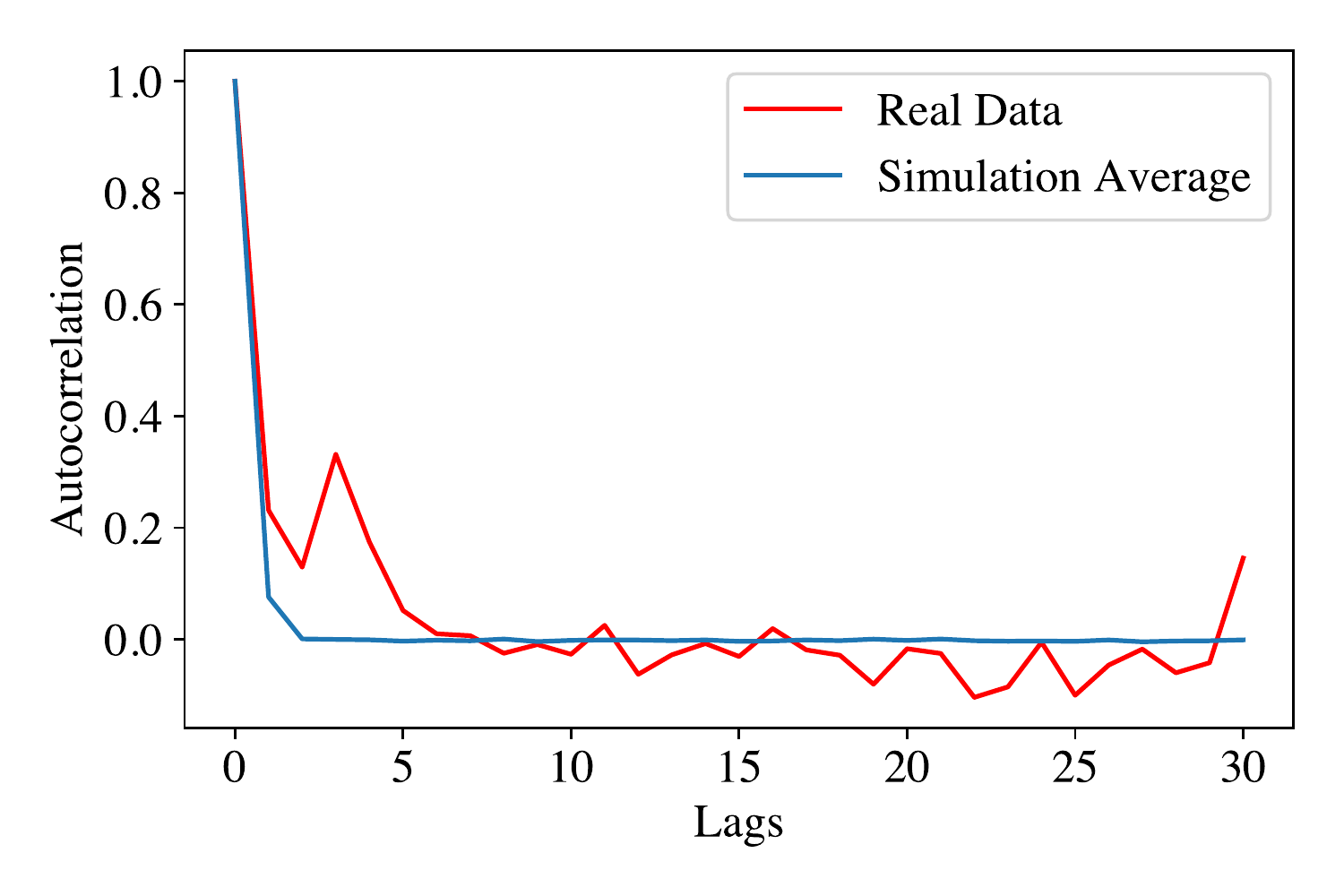}%
}
\subfigure[RBM]{%
\includegraphics[width=0.5\linewidth]{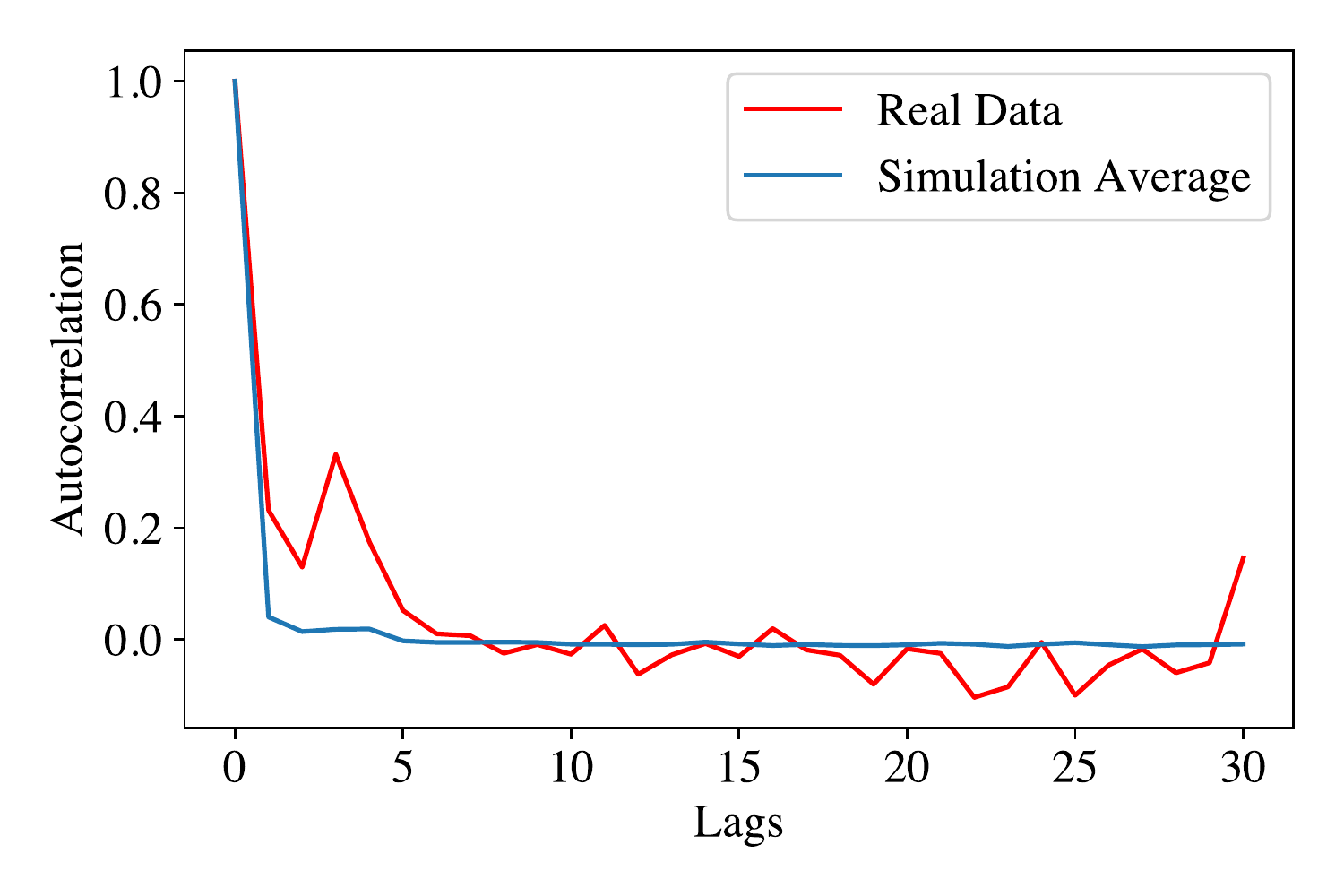}%
}%
\hspace*{\fill}
\subfigure[CVAE]{%
\includegraphics[width=0.5\linewidth]{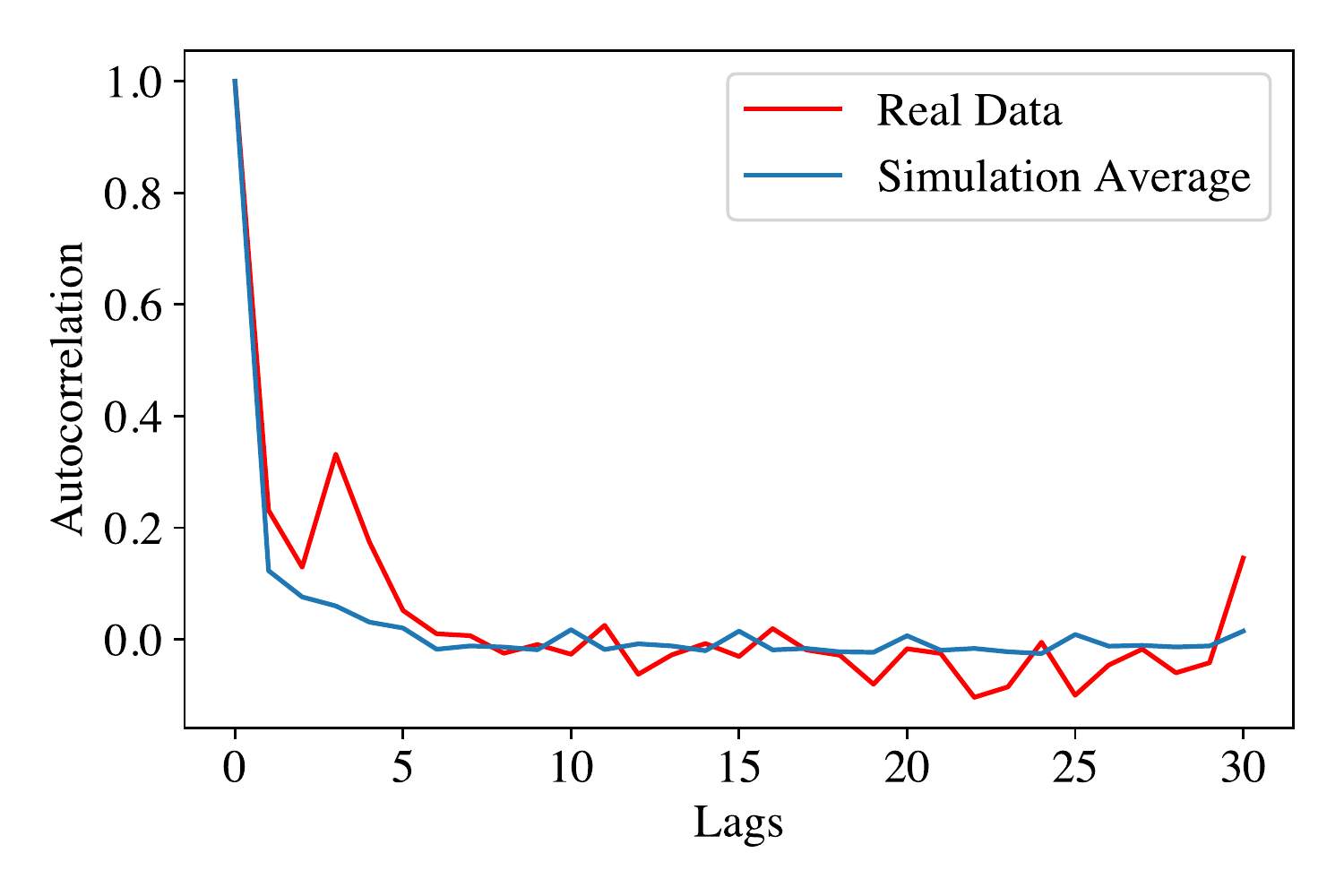}%
}
\caption[Autocorrelation plots of squared samples for 26 weeks test period]{Average autocorrelation over 26 weeks of squared generated returns vs. autocorrelation of squared S\&P500 Log Returns.}
\label{sqacfrealfour}
\end{adjustbox}
\end{figure} 

\begin{figure*}
\centering  
\SetFigLayout{4}{3}
  \subfigure[Data vs. Normal]{%
\includegraphics[width=0.32\linewidth, trim=2cm 0.1cm 3.9cm 1.55cm, clip]{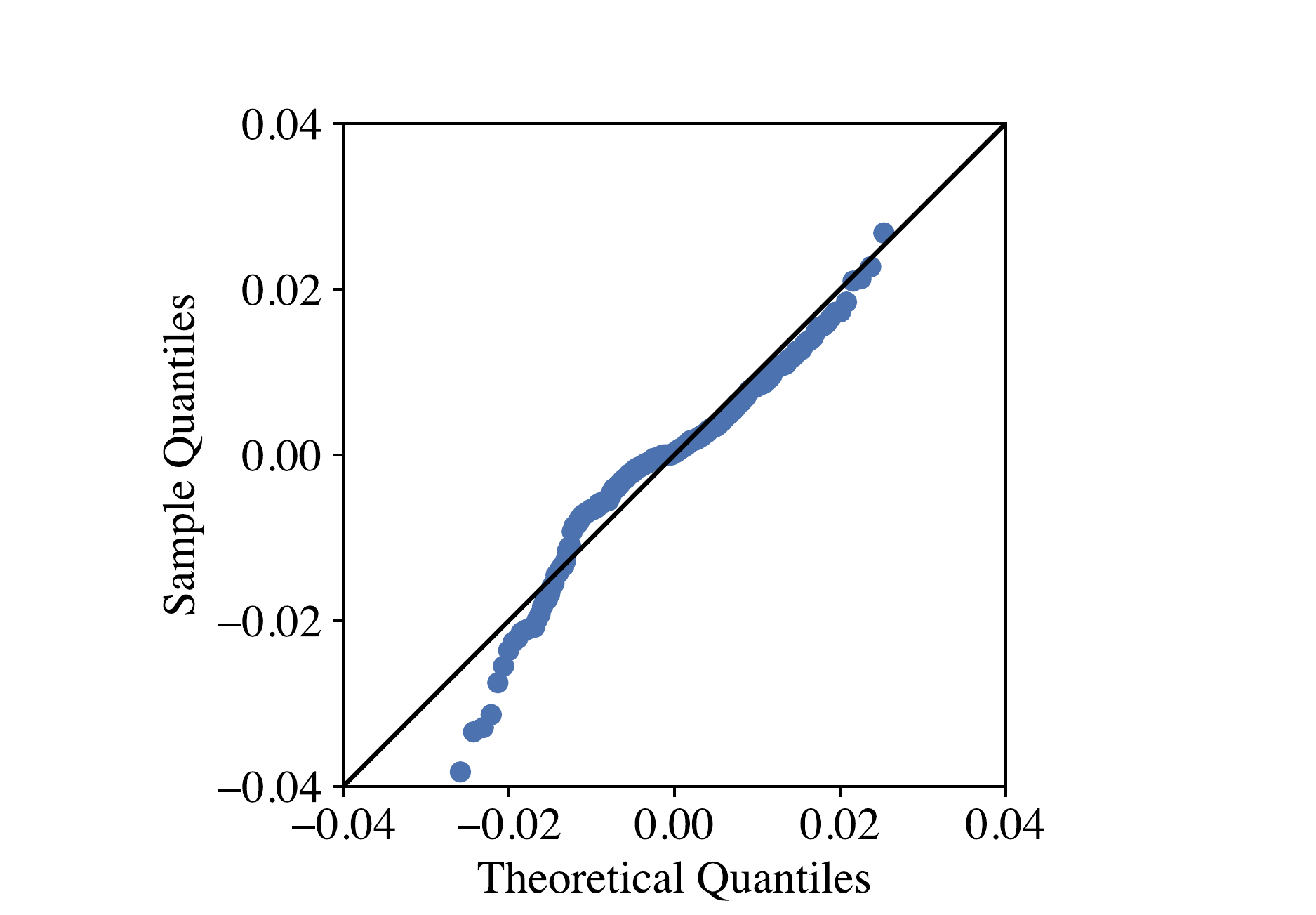}%
}%
\hspace*{\fill}
\subfigure[FHS vs. Normal]{%
\includegraphics[width=0.32\linewidth,trim=2cm 0.1cm 3.9cm 1.55cm, clip]{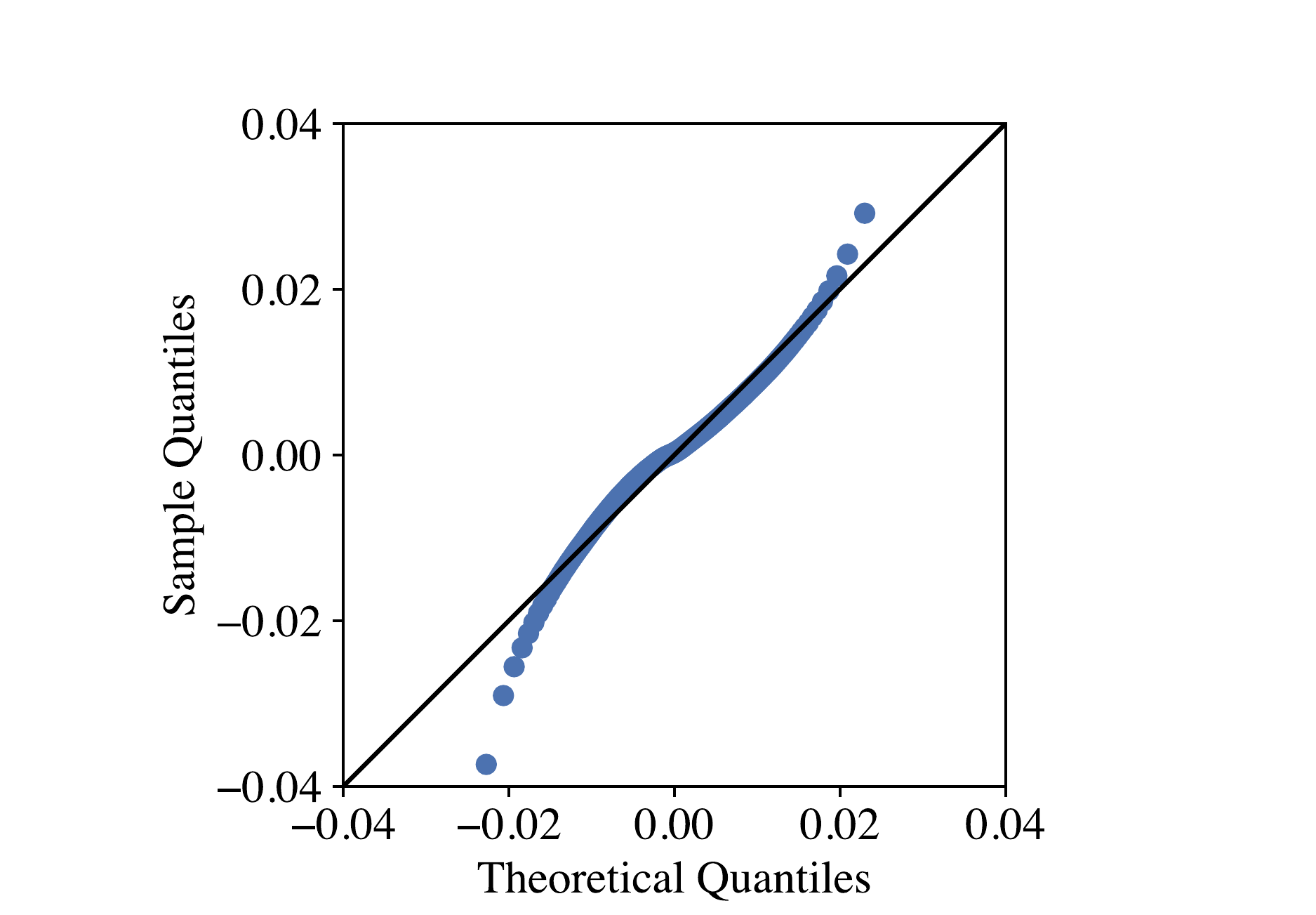}%
}%
\hspace*{\fill}
\subfigure[Data vs. FHS]{%
\includegraphics[width=0.32\linewidth,trim=2cm 0.1cm 3.9cm 1.55cm, clip]{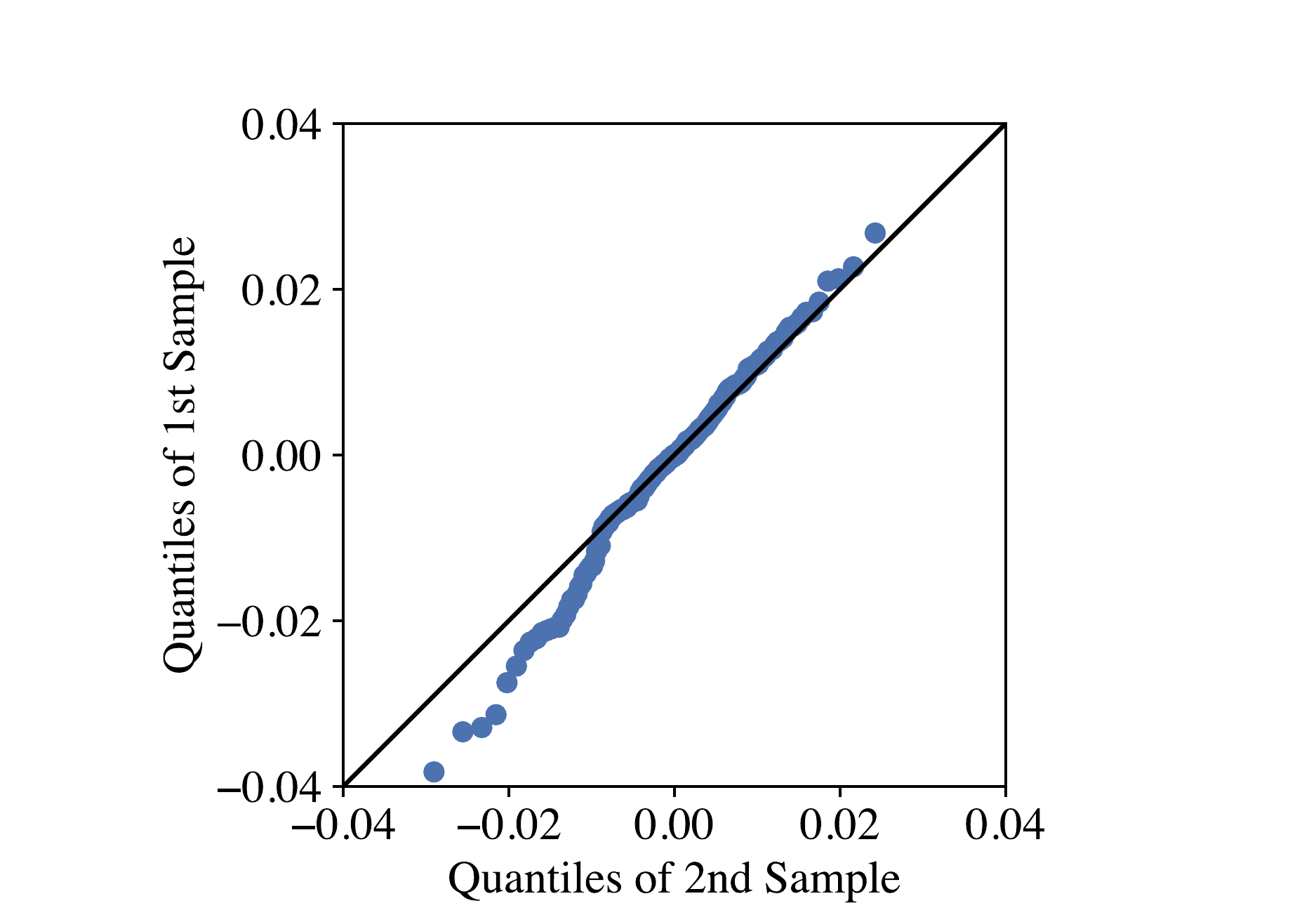}%
}
\subfigure[Data vs. Normal]{%
\includegraphics[width=0.32\linewidth, trim=2cm 0.1cm 3.9cm 1.55cm, clip]{figures/Real_Data/robust/qq/52real.pdf}%
}%
\hspace*{\fill}
\subfigure[GARCH vs. Normal]{%
\includegraphics[width=0.32\linewidth, trim=2cm 0.1cm 3.9cm 1.55cm, clip]{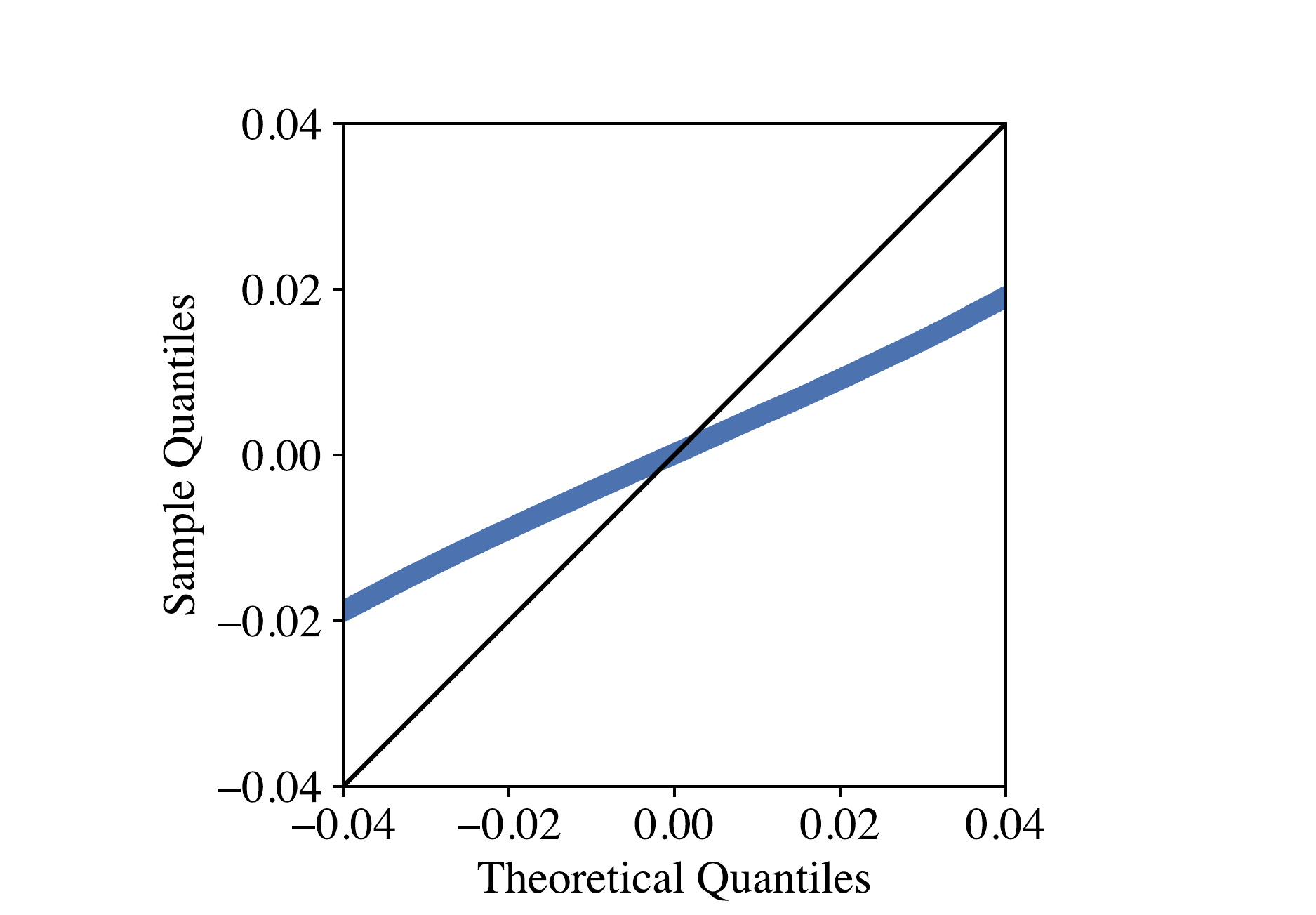}%
}%
\hspace*{\fill}
\subfigure[Data vs. GARCH]{%
\includegraphics[width=0.32\linewidth, trim=2cm 0.1cm 3.9cm 1.55cm, clip]{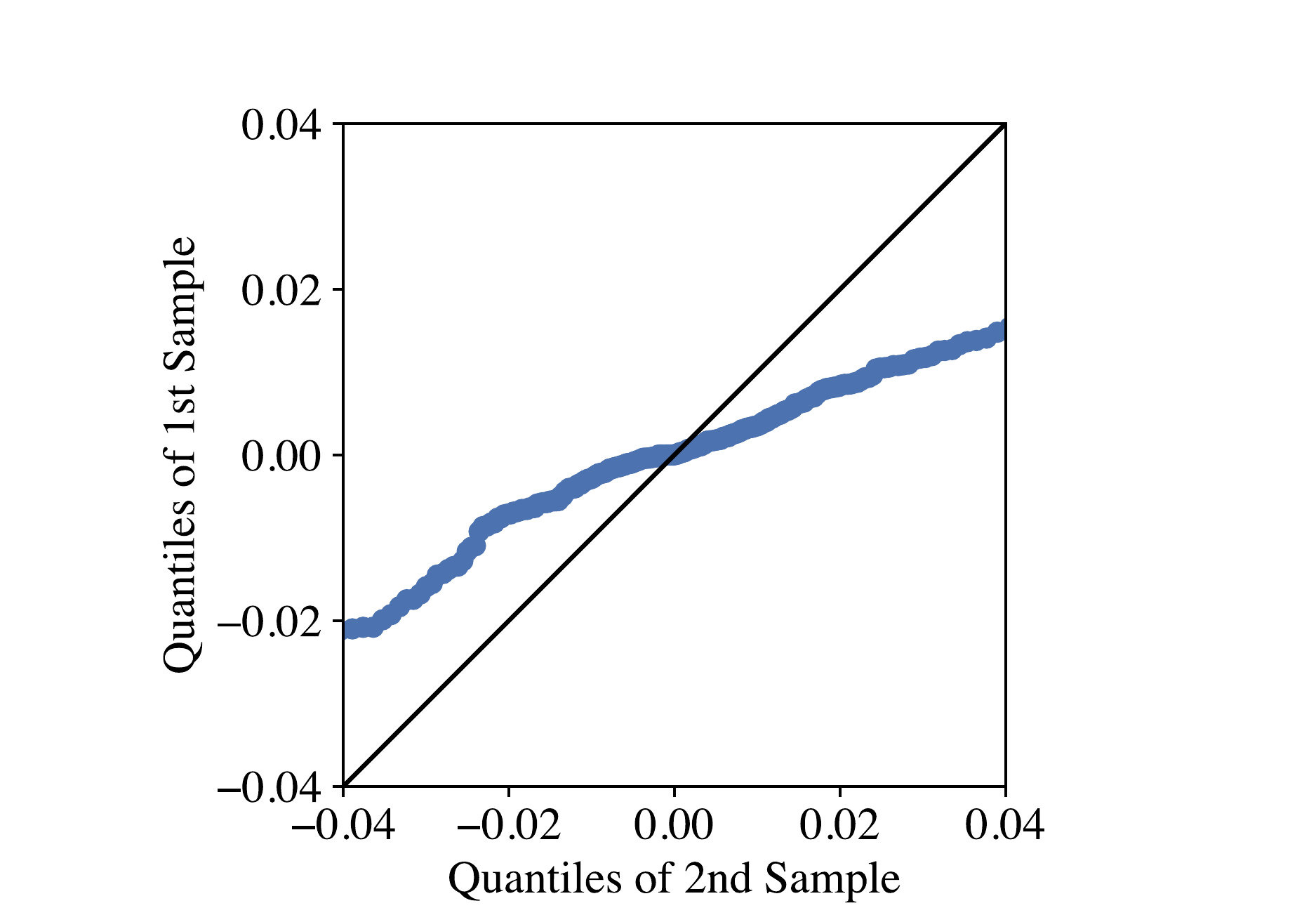}%
}
\subfigure[Data vs. Normal]{%
\includegraphics[width=0.32\linewidth, trim=2cm 0.1cm 3.9cm 1.55cm, clip]{figures/Real_Data/robust/qq/52real.pdf}%
}%
\hspace*{\fill}
\subfigure[RBM vs. Normal]{%
\includegraphics[width=0.32\linewidth, trim=1.8cm 0.1cm 3cm 0.5cm, clip]{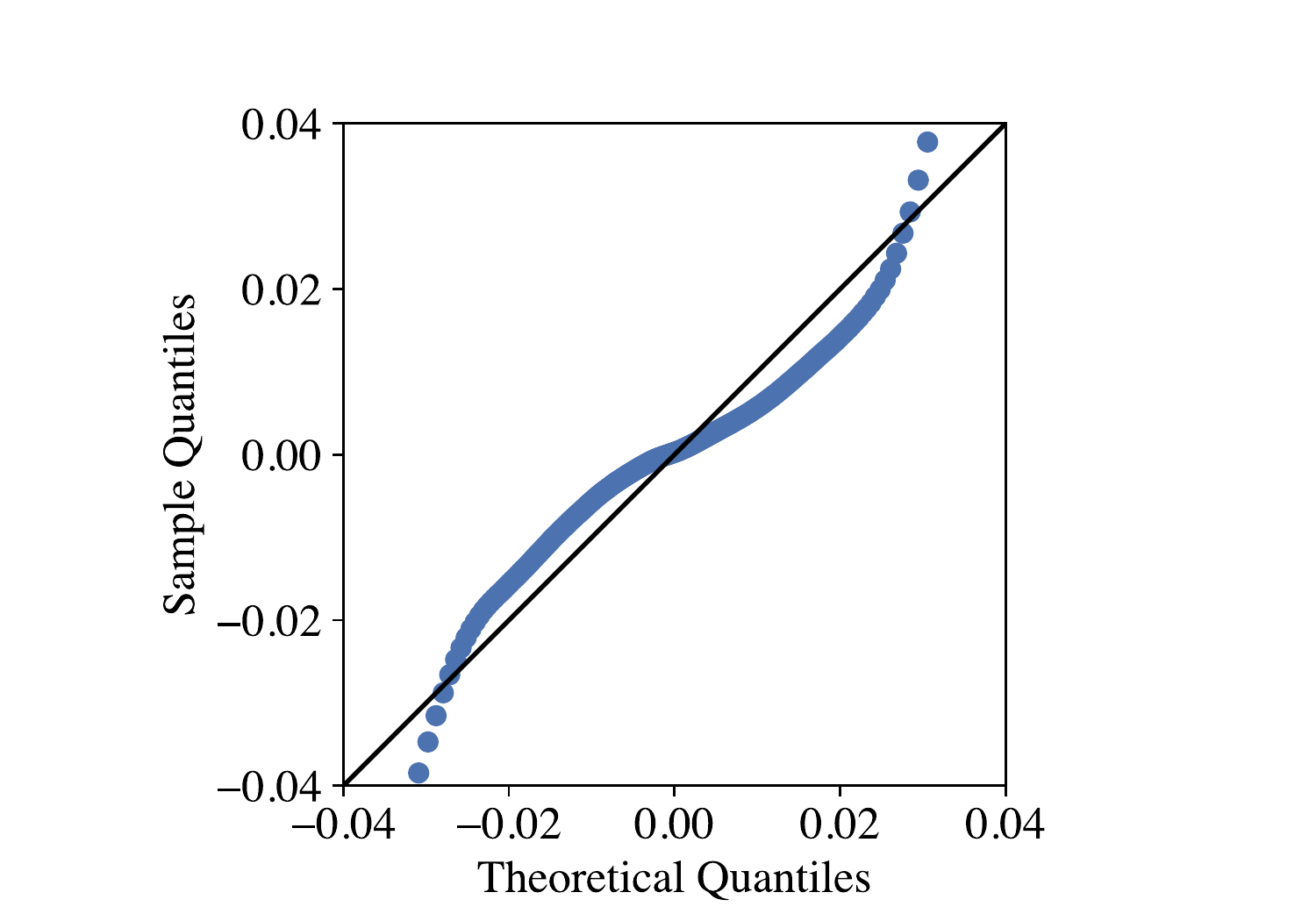}%
}%
\hspace*{\fill}
\subfigure[Data vs. RBM]{%
\includegraphics[width=0.32\linewidth, trim=1.8cm 0.1cm 3cm 0.5cm, clip]{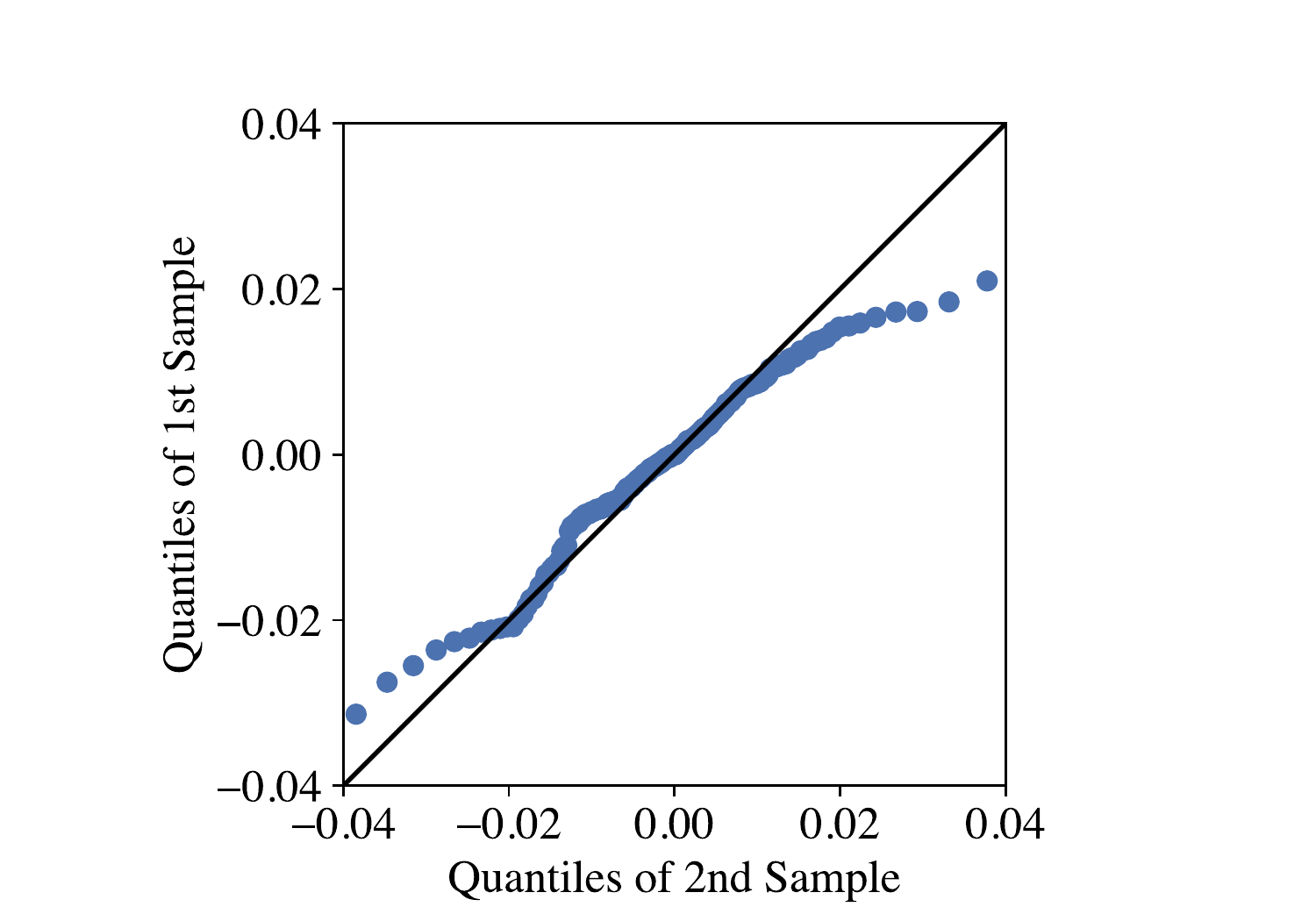}%
}
\subfigure[Data vs. Normal]{%
\includegraphics[width=0.32\linewidth, trim=2cm 0.1cm 3.9cm 1.55cm, clip]{figures/Real_Data/robust/qq/52real.pdf}%
}%
\hspace*{\fill}
\subfigure[CVAE vs. Normal]{%
\includegraphics[width=0.32\linewidth, trim=2cm 0.1cm 3.9cm 1.55cm, clip]{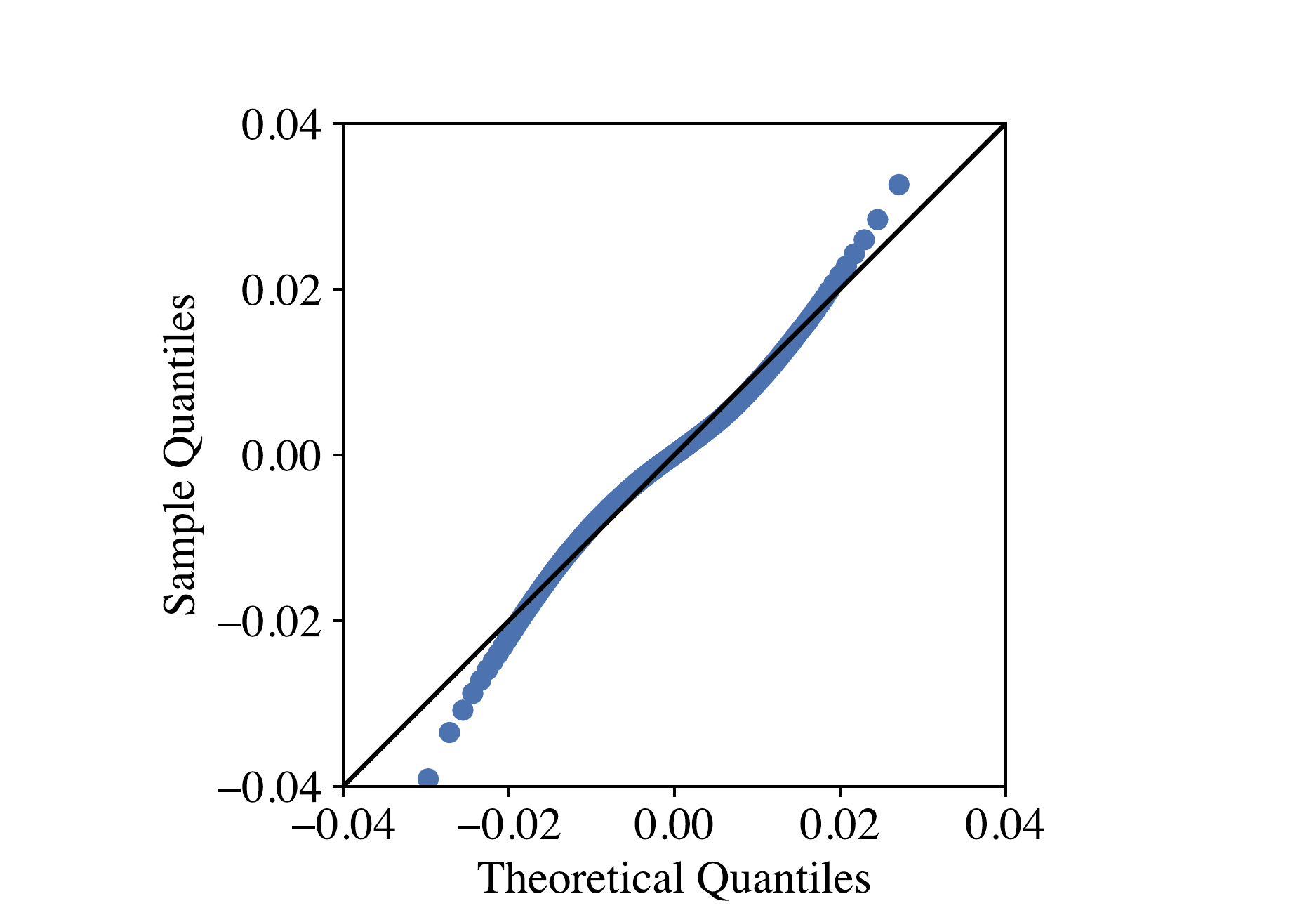}%
}%
\hspace*{\fill}
\subfigure[Data vs. CVAE]{%
\includegraphics[width=0.32\linewidth, trim=2cm 0.1cm 3.9cm 1.55cm, clip]{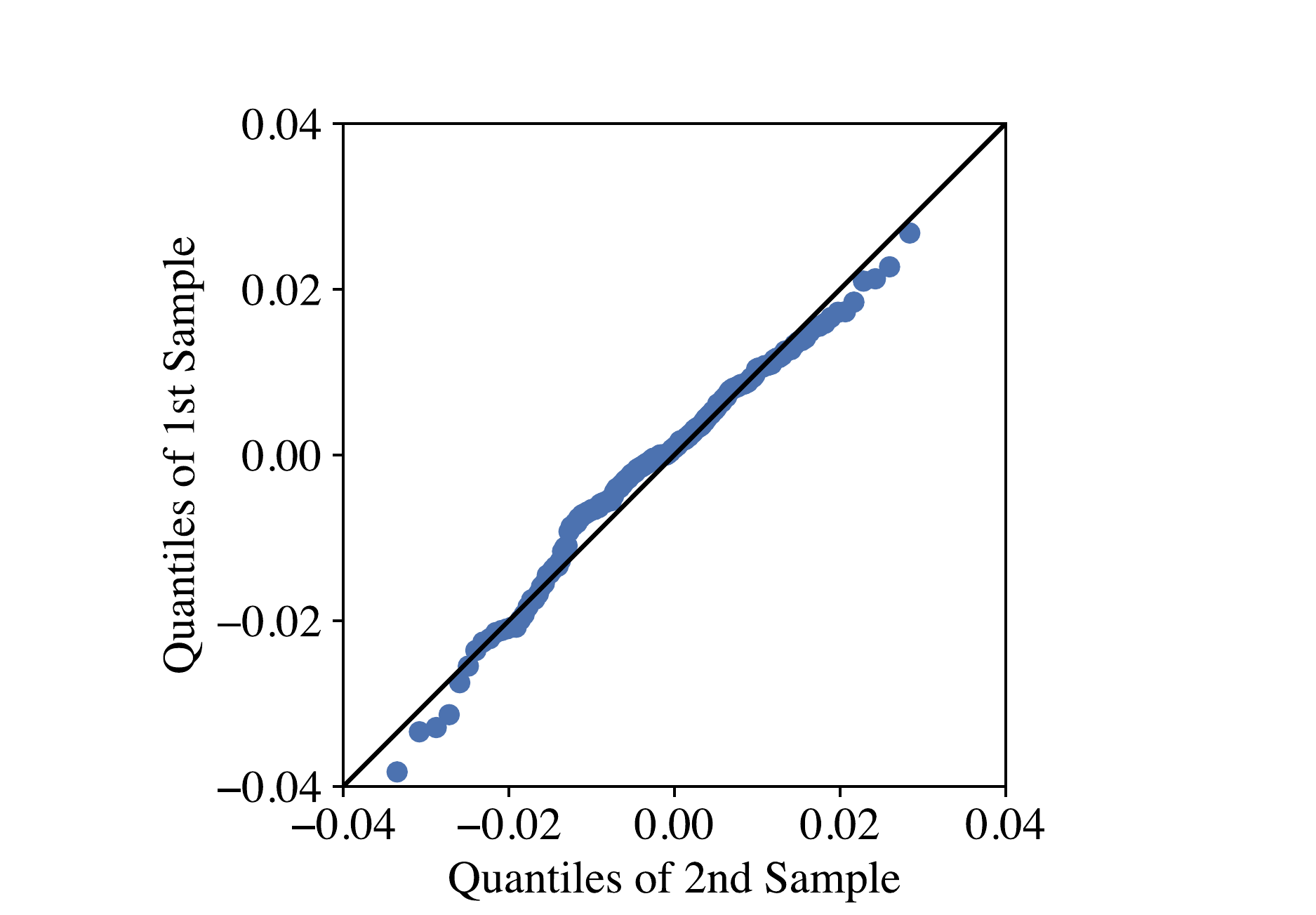}%
}
\caption[QQ Plots for 52 weeks test period]{QQ Plots of real Log Returns and generated samples over 52 weeks period: a)-c): Filtered Historical Simulation; d)-f): GARCH; g)-i): RBM; j)-l): CVAE.}
\label{sqacfreal52}
\end{figure*}

\begin{figure}[tbh]
\centering
\begin{adjustbox}{minipage=\textwidth, scale=0.87}
\subfigure[FHS]{%
\includegraphics[width=0.5\linewidth]{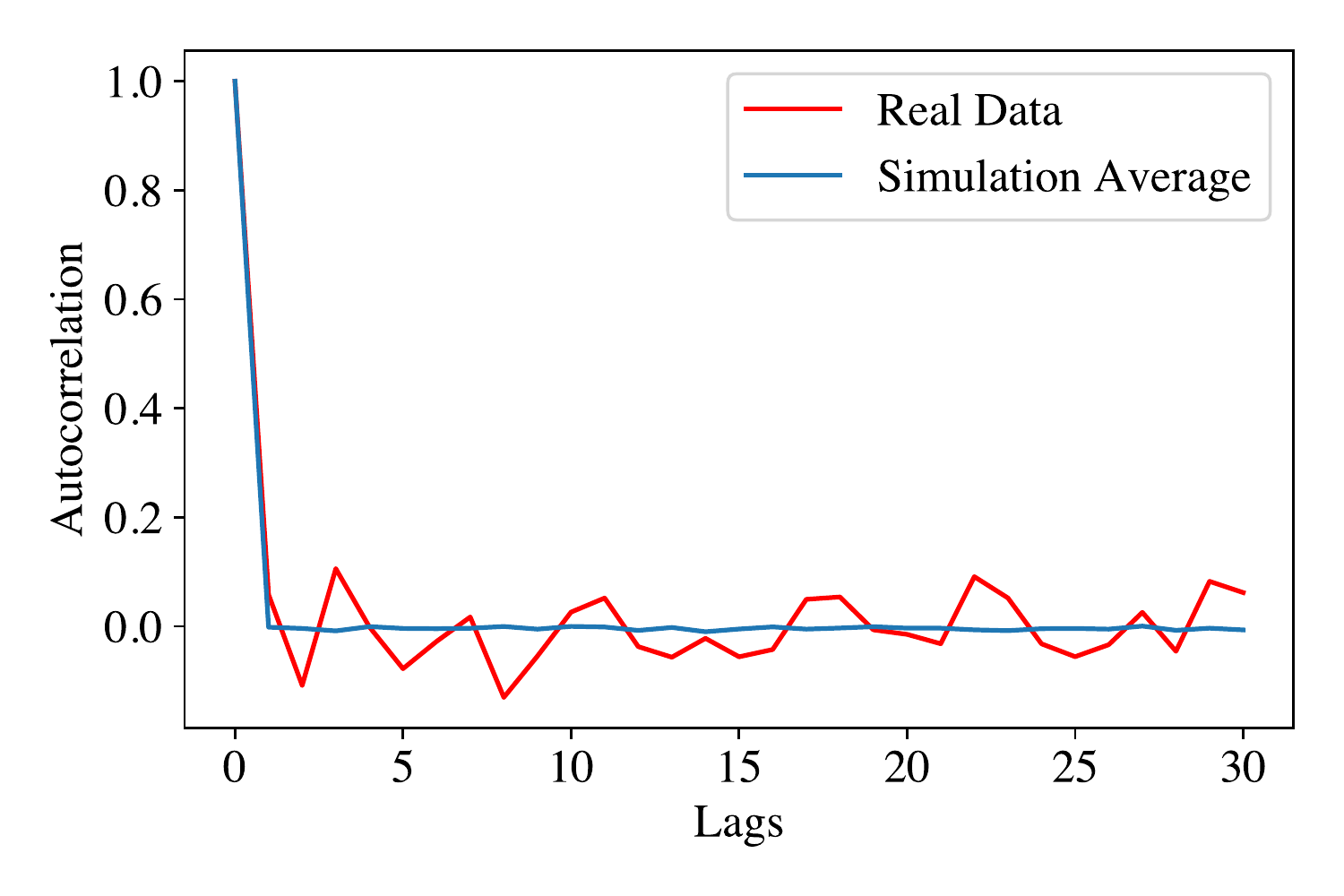}%
}%
\hspace*{\fill}
\subfigure[GARCH]{%
\includegraphics[width=0.5\linewidth]{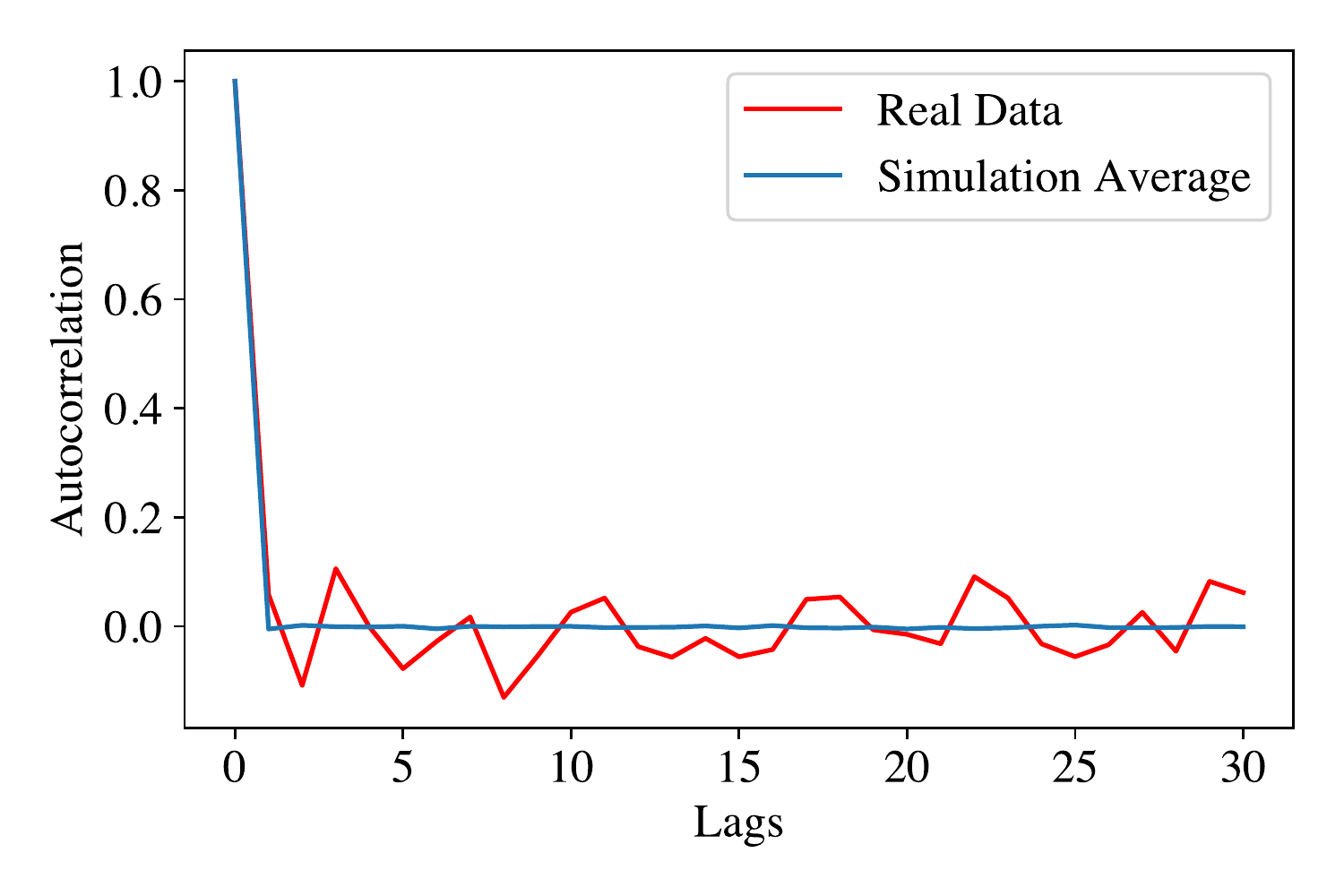}%
}
\subfigure[RBM]{%
\includegraphics[width=0.5\linewidth]{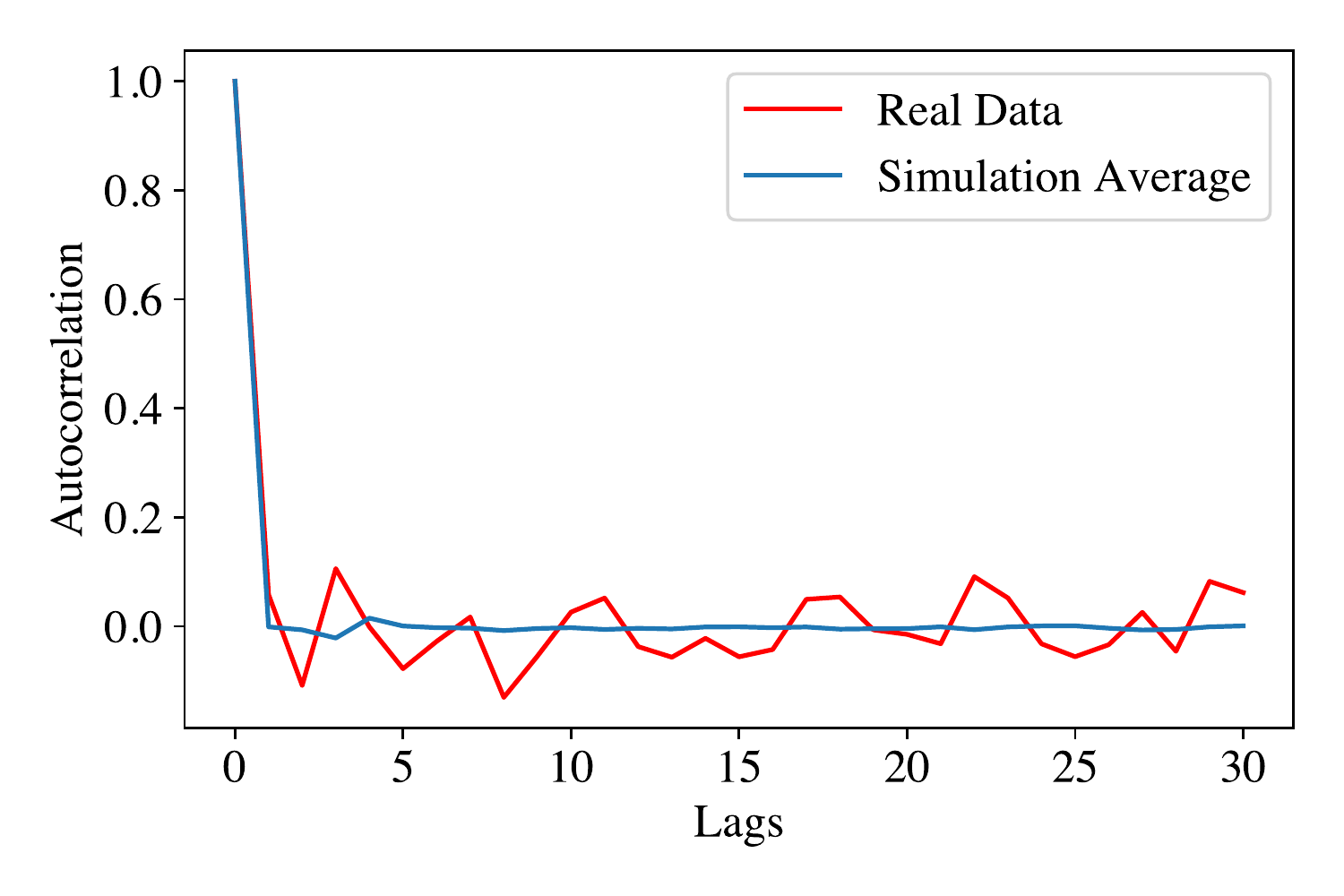}%
}%
\hspace*{\fill}
\subfigure[CVAE]{%
\includegraphics[width=0.5\linewidth]{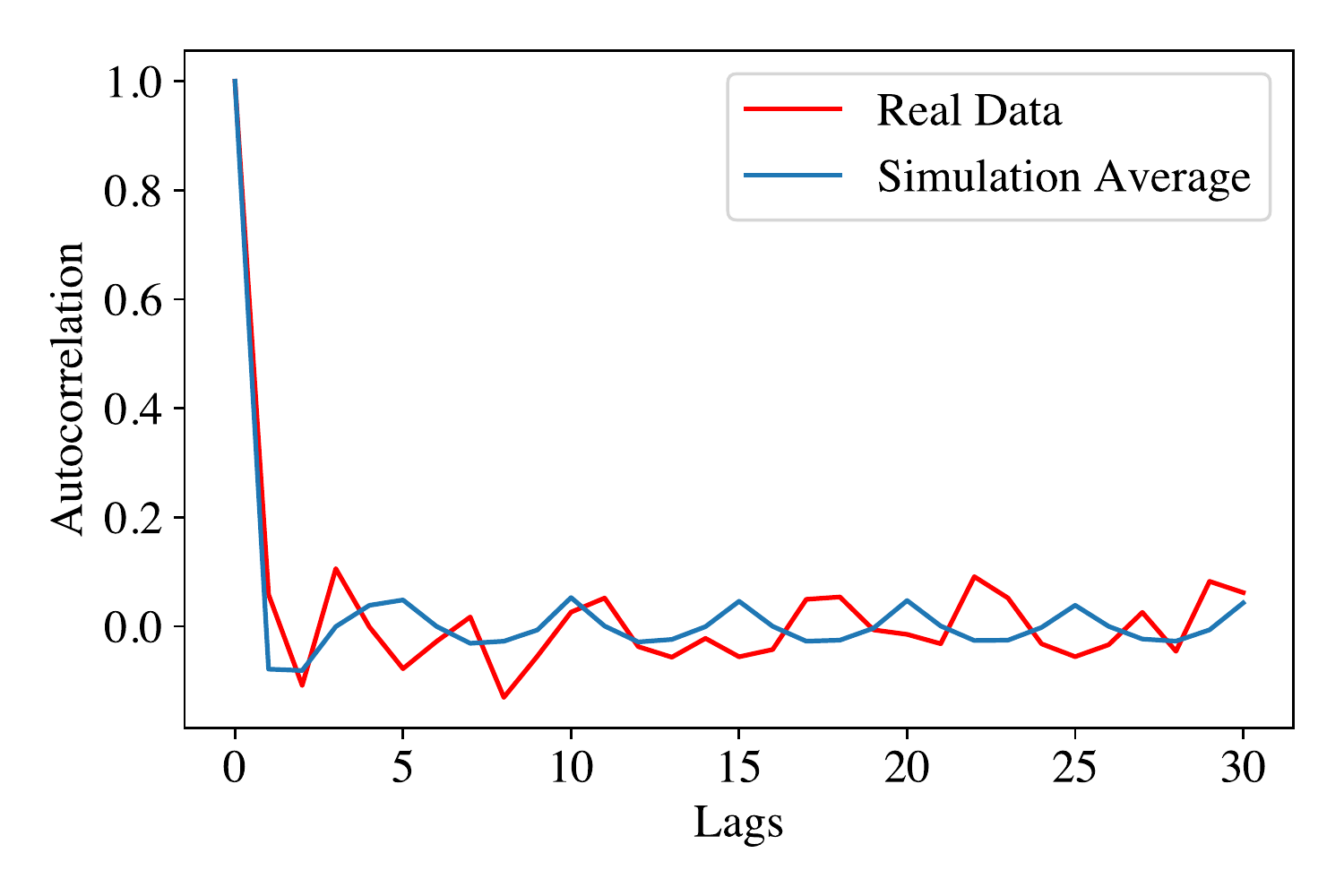}%
}
\caption[Autocorrelation plots for 52 weeks test period]{Average autocorrelation over 52 weeks period of generated returns vs. autocorrelation of real S\&P500 Log Returns.}
\subfigure[FHS]{%
\includegraphics[width=0.5\linewidth]{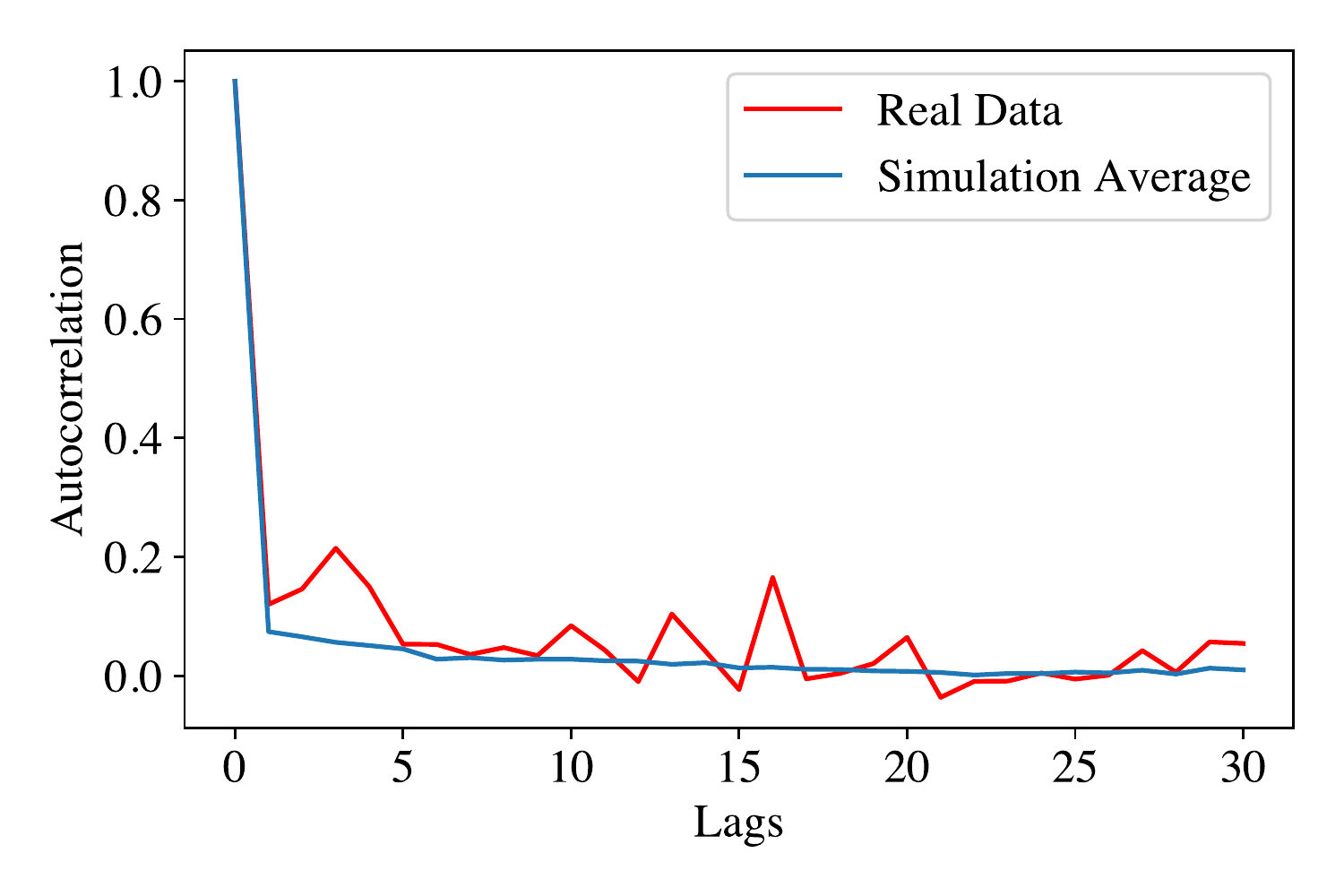}%
}%
\hspace*{\fill}
\subfigure[GARCH]{%
\includegraphics[width=0.5\linewidth]{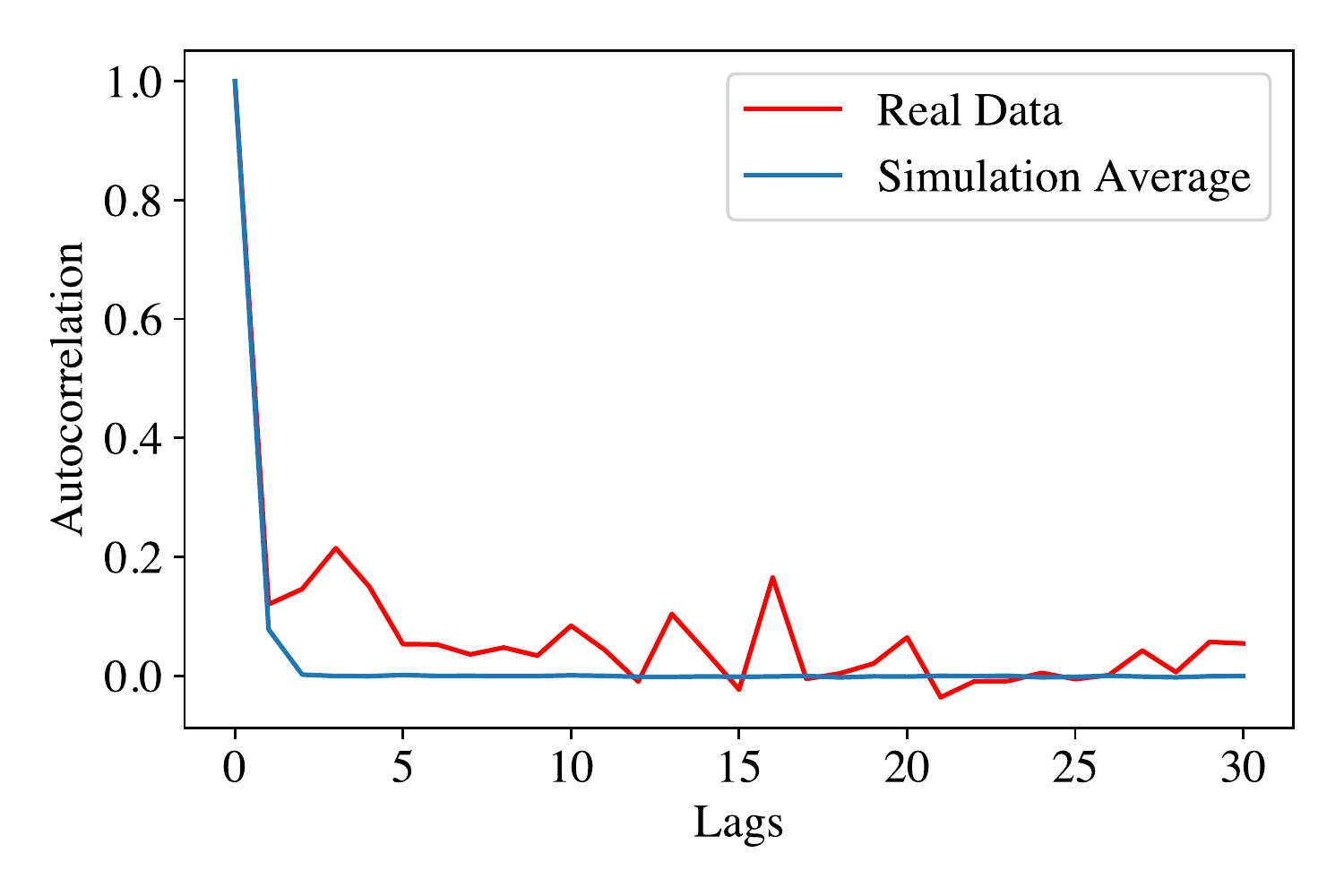}%
}
\subfigure[RBM]{%
\includegraphics[width=0.5\linewidth]{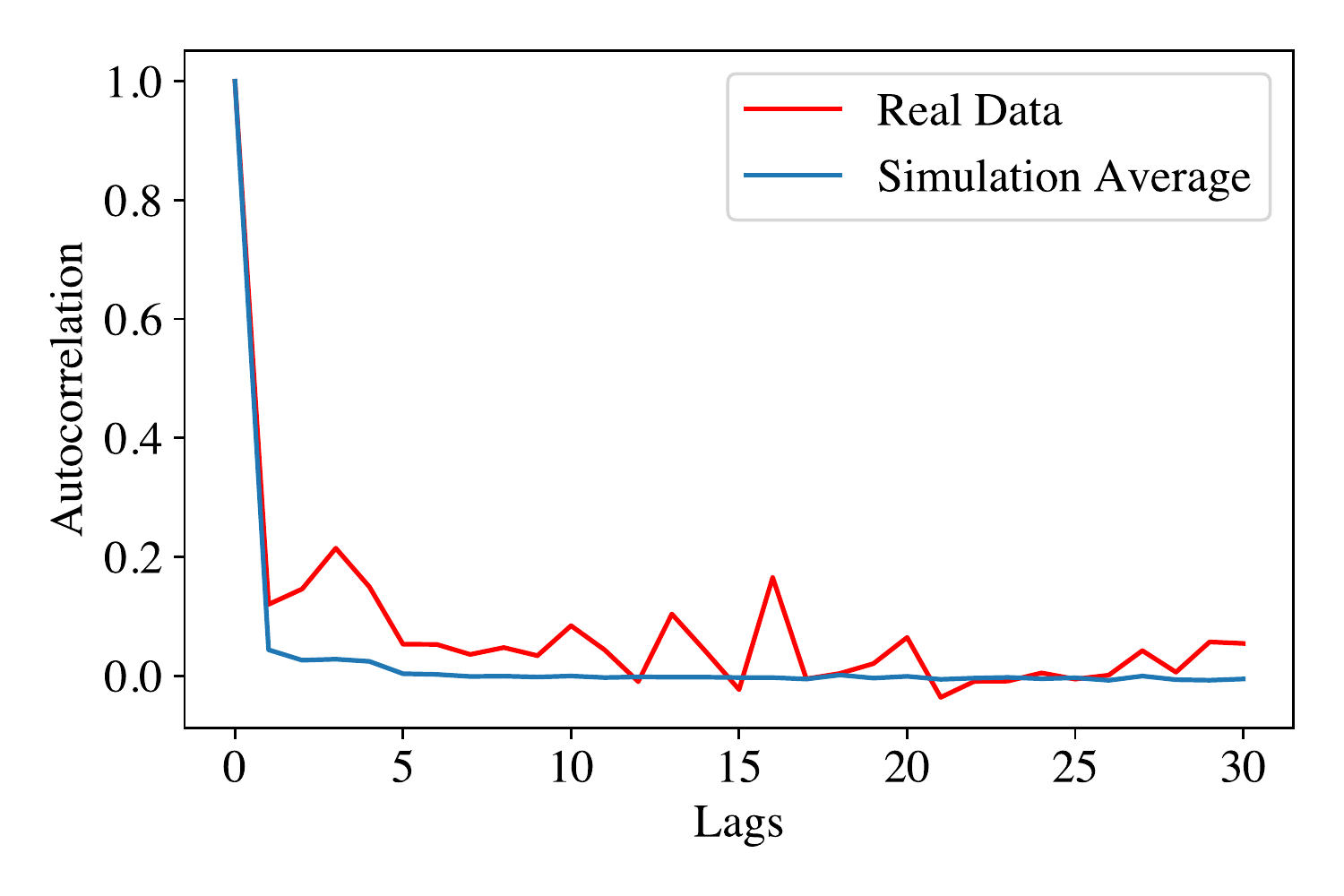}%
}%
\hspace*{\fill}
\subfigure[CVAE]{%
\includegraphics[width=0.5\linewidth]{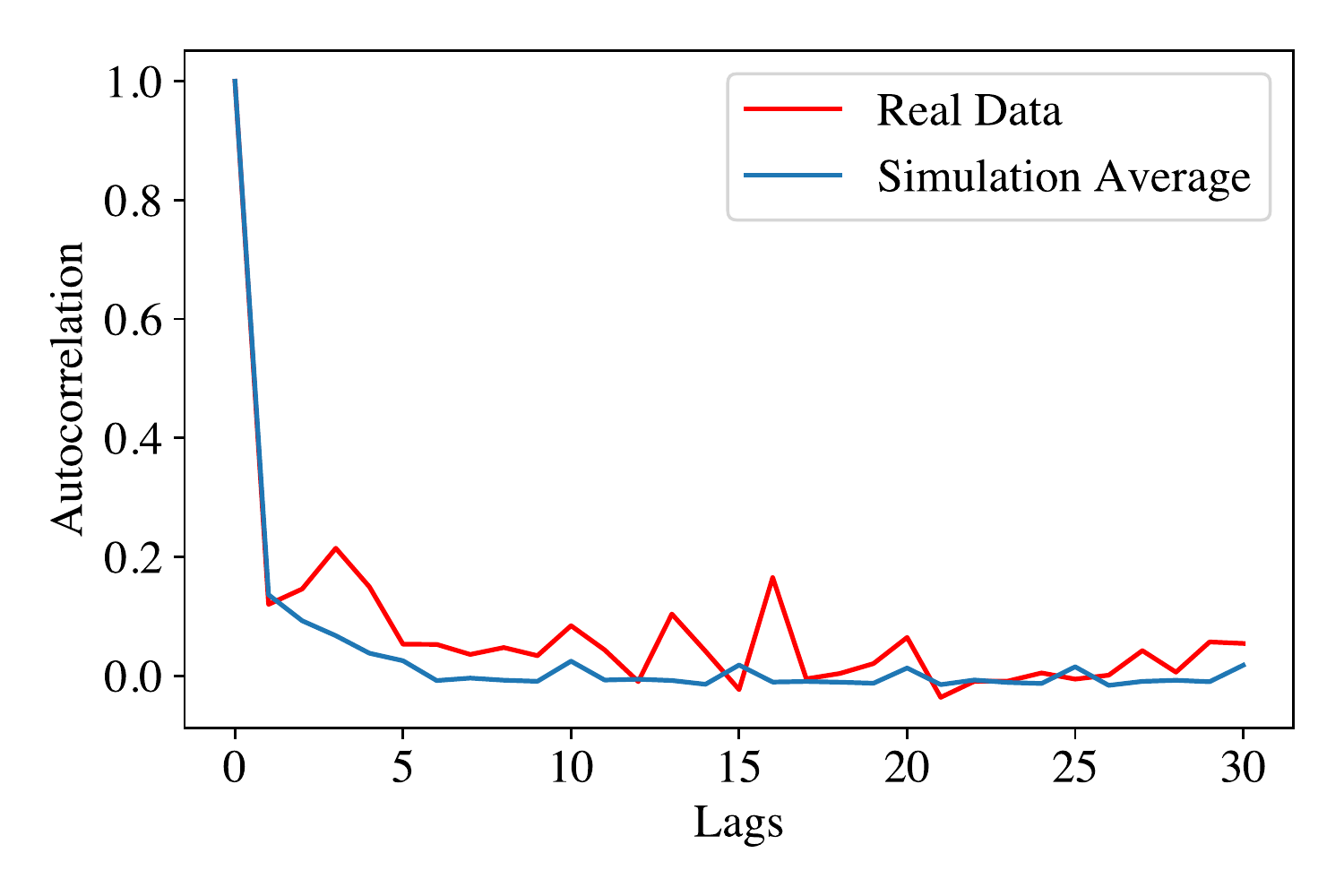}%
}
\caption[Autocorrelation plots of squared samples for 52 weeks test period]{Average autocorrelation over 52 weeks period of squared generated returns vs. autocorrelation of squared S\&P500 Log Returns.}
\label{sqacfrealfour}
\end{adjustbox}
\end{figure} 

\backmatter

\printbibliography[title={References}]

@article{indexreview,
  title={Disentangling diverse measures: a survey of financial stress indexes},
  author={Kevin L. Kliesen and Michael T. Owyang and E. K. Vermann},
  journal={Canadian Parliamentary Review},
  year={2012},
  volume={94},
  pages={369-398}
}

@techreport{CISS,
author = {D\'{a}niel Holl\'{o} and Manfred Kremer and Marco Lo Duca},
number = {1426},
publisher = {European Central Bank (ECB)},
title = {CISS - a composite indicator of systemic stress in the financial system},
type = {ECB Working Paper},
url = {http://hdl.handle.net/10419/153859},
year = {2012}
}

@Book{QRM,
  author={Alexander J. McNeil and Rüdiger Frey and Paul Embrechts},
  title={{Quantitative Risk Management: Concepts, Techniques and Tools Revised edition}},
  publisher={Princeton University Press},
  year=2015,
}

@Misc{greenspan,
  author =       {Alan Greenspan},
  year =         {1995},
  title =        {Remarks at a Research Conference on Risk Measurement and Systemic Risk, Washington, D.C.[Accessed: 2021 01 15] },
  month =        {11},
url ={https://fraser.stlouisfed.org/title/452/item/8552},
}

@Misc{basel2,
  author =       {Basel Committee on
Banking Supervision},
  title =        {Basel II: International Convergence of Capital Measurement and Capital Standards: a Revised Framework - Comprehensive Version},
  year =         {2006},
  month =        {06},
  day=          {30},
}

@Misc{solv2,
  author =       {European Union},
  title =        {Directive 2009/138/EC of the European Parliament and of the Council of 25 November 2009 on the taking-up and pursuit of the business of Insurance and Reinsurance (Solvency II)},
  year =         {2009},
  month =        {11},
  day=          {25},
}

@article{10.2307/41300453,
 ISSN = {13573217, 20440456},
 author = {E. M. Varnell},
 journal = {British Actuarial Journal},
 number = {1},
 pages = {121--179},
 publisher = {Cambridge University Press},
 title = {Economic Scenario Generators and Solvency II},
 volume = {16},
 year = {2011}
}

@article{ESG,
    author =       "Hal Pedersen and Mary Pat Campbell and Stephan L. Christiansen and Samuel H. Cox and Daniel Finn and Ken Griffin and Nigel Hooker and Matthew Lightwood and Stephen M. Sonlin and Chris Suchar",
    title =        "Economic Scenario Generators - A practical guide",
    journal =      "Society of Actuaries",
    year =         "2016",
    
}

@article{usehistory,
    author =       "Gavin Kretzschmar and Alexander J. McNeil and Axel Kirchner",
    title =        "Integrated models of capital adequacy - Why banks are undercapitalised",
    journal = "Journal of Banking \& Finance",
    volume = "34",
    number = "12",
    pages = "2838 - 2850",
    year = "2010"
}

@misc{cali,
    author = {Misha van Beek},
    title = "{Consistent Calibration of Economic Scenario Generators: the Case for Conditional Simulation}",
    archivePrefix = "arXiv", 
    note = {arXiv:2004.09042},
    year = 2020,
}

@article{MDS,
    author =       "Bennett Golub and David Greenberg and Ronald Ratcliffe",
    title =        "Market-Driven Scenarios: An Approach for Plausible Scenario Construction",
    volume =       "44",
    number =        "5",
    pages =        "6-20",
    journal =      "The Journal of Portfolio Management",
    year =         "2018",
}

@article{BLM,
    author =       "Fischer Black and Robert Litterman",
    title =        "Global Portfolio Optimization",
    volume =       "48",
    number=       "5",
    pages =        "28-43",
    journal =      "Financial Analysts Journal",
    year =         "2018",
}

@article{stylizedfacts,
    author =       "Rama Cont",
    title =        "Empirical properties of asset returns: stylized facts and statistical issues",
    volume =       "1",
    pages =        "223–236",
    journal =      "Quantitative Finance",
    year =         "2001",
}

@article{cGANtrade,
author = { Adriano   Koshiyama  and  Nick   Firoozye  and  Philip   Treleaven },
title = {Generative adversarial networks for financial trading strategies fine-tuning and combination},
journal = {Quantitative Finance},
volume = {0},
number = {0},
pages = {1-17},
year  = {2020},
URL = {https://doi.org/10.1080/14697688.2020.1790635},

}

@book{meucci,
      title     = "Risk and Asset Allocation",
      author    = "Meucci, Attilio",
      year      = 2000,
      publisher = "Springer Finance",
    }

@article{combiprob,
    author =       "Robert T. Clemen and Robert L. Winkler",
    title =        "Combining Probability Distributions From Experts in Risk Analysis",
    volume =       "19",
    number=         "2",
    pages =        "187-203",
    journal =      "Risk Analysis",
    year =         "1999",
}

@article{meanvol,
    author =       "Edward Qian and Stephen Gorman",
    title =        "Conditional Distribution in Portfolio Theory",
    volume =       "57",
    number=         "2",
    pages =        "44-53",
    journal =      "Financial Analysts Journal",
    year =         "2001",
}

@incollection{MDSori,
  author      = "Paul H. Kupiec",
  title       = "Stress testing in a value at risk framework",
  editor      = "Michael A. H Dempster",
  booktitle   = "Risk management : value at risk and beyond",
  publisher   = "Cambridge Univ. Press",
  year        = 2002,
  pages       = "76-100",
  chapter     = 3,
}

@article{condfacmod,
    author =       "Martin van der Schans and Hens Steehouwer",
    title =        "Views, Factor Models and Optimal Asset Allocation",
    volume =       "29",
    pages =        "122-134",
    journal =      "Procedia Economics and Finance",
    year =         "2015",
}

@article{timebl,
    author =       "Martinvan der Schans and Hens Steehouwer",
    title =        "Time-Dependent Black–Litterman",
    volume =       "18",
    pages =        "371–387",
    journal =      "Journal of Asset Management volume",
    year =         "2017",
}

@article {Whaley12,
	author = {Whaley, Robert E.},
	title = {The Investor Fear Gauge},
	volume = {26},
	number = {3},
	pages = {12--17},
	year = {2000},
	journal = {The Journal of Portfolio Management}
}

@InProceedings{equitystylizedfacts,
  author     = {Omar Rojas and Carlos Trejo-Pech},
  title      = {Financial Time Series: Stylized Facts for the Mexican Stock Exchange Index Compared to Developed Markets},
  year       = {2013},
  eventtitle = {12th. International Business and Economy Conference},
  eventdate  = {27th to 29th July 2020 },
  venue      = {Vancouver, Canada},
}

@article{voladuration,
  author={Manfred Kremer},
  title={Regime shifts in stock market volatility: a historical perspective on the US market},
  year=2018,
  month="May",
  journal =      "European Central Bank Financial Stability Review",
  volume=1,
}

@article {Giot92,
	author = {Pierre Giot},
	title = {Relationships Between Implied Volatility Indexes and Stock Index Returns},
	volume = {31},
	number = {3},
	pages = {92--100},
	year = {2005},
	journal = {The Journal of Portfolio Management}
}

@article{jrfm12040156,
author = {Bahram Adrangi and Arjun Chatrath and Joseph Macri and Kambiz Raffiee},
title = {Dynamic Responses of Major Equity Markets to the US Fear Index},
journal = {Journal of Risk and Financial Management},
volume = {12},
year = {2019},
number = {4}
}

@article{vixforecast,
author = { Charles J. Corrado and Thomas W. Miller, Jr. },
title = {The forecast quality of CBOE implied volatility indexes},
journal = {Journal of Futures Markets},
volume = {25},
number = {4},
pages = {339-373},
year = {2005}
}

@article{varfed,
  author={Darryll Hendricks},
  title={Evaluation of value-at-risk models using historical data},
  year=1996,
  journal =      "Federal Reserve Bank of New York Economic Policy Review",
  volume= "2",
  number="1"
}

@article{varhs,
  author={Tanya Styblo Beder},
  title={VAR: Seductive but Dangerous},
  year=1995,
  journal =      "Federal Reserve Bank of New York Economic Policy Review",
  volume= "51",
  number="5",
  pages="12-24",
}

@article{bsori,
    author =       "Brad Efron",
    title =        "Bootstrap Methods: Another Look at the Jackknife",
    volume =       "7",
    number =       "1",
    pages =        "1-26",
    journal =      "The Annals of Statistics",
    year =         "1979",
}

@article{fhsori,
    author =       "John Hull and Alan White",
    title =        "Incorporating volatility updating into the historical simulation method for value at risk",
    volume =       "1",
    pages =        "5-19",
    journal =      "Journal of Risk",
    year =         "1998",
}

@article{fhs,
    author =       "Giovanni Barone‐Adesi and Kostas Giannopoulos and Les Vosper",
    title =        "VaR without correlations for portfolios of derivative securities",
    volume =       "19",
    number =       "5",
    pages =        "583-602",
    journal =      "The Journal of Futures Markets",
    year =         "1999",
}

@article{bsvix,
    author =       "Marcus Nossman and Anders Vilhelmsson",
    title =        "Non-parametric Forward Looking Value-at-Risk",
    volume =       "16",
    number =       "4",
    pages =        "103-123",
    journal =      "Journal of Risk",
    year =         "2014",
}

@article{d2020learning,
  title={Learning the Ising model with generative neural networks},
  author={D'Angelo, Francesco and B{\"o}ttcher, Lucas},
  journal={Physical Review Research},
  volume={2},
  number={2},
  pages={023266},
  year={2020},
}

@misc{RBM,
    author    = "Alexei Kondratyev and Christian Schwarz",
    title     = "Learning Curve Dynamics with Artificial Neural Networks",
    url       = "https://papers.ssrn.com/sol3/papers.cfm?abstract_id=3384948",
    year      = 2019
}

@misc{cvae,
    author    = "Hans Bühler and Blanka Horvath and Terry Lyons and Imanol P. Arribas and Ben Wood",
    title     = "A Data-driven Market Simulator for Small Data Environments",
    archivePrefix = "arXiv", 
    note =          {arXiv:2006.14498},
    year      = 2020
}

@misc{genmodels,
    author    = "Pierre Henry-Labordere",
    title     = "Generative Models for Financial Data",
    url       = "https://ssrn.com/abstract=3408007",
    year      = 2019
}

@misc{roncalli,
    author    = "Edmond Lezmi and Jules Roche and Thierry Roncalli and Jiali Xu",
    title     = "Improving the Robustness of Trading Strategy Backtesting with Boltzmann Machines and Generative Adversarial Networks",
    url       = "https://ssrn.com/abstract=3645473",
    year      = 2020
}

@incollection{RBMori,
  author      = "Paul Smolensky",
  title       = "Information processing in dynamical systems: foundations of harmony theory",
  editor      = "David E. Rumelhart and James L McClelland",
  booktitle   = "Parallel distributed processing: explorations in the microstructure of cognition",
  publisher   = "MIT Press",
  year        = 1986,
  pages       = "194–281",
  chapter     = 6,
}

@InProceedings{rbmcond,
author="Asja Fischer and Christian Igel",
editor="Luis Alvarez and Marta Mejail and Luis Gomez and Julio Jacobo",
title="An Introduction to Restricted Boltzmann Machines",
booktitle="Progress in Pattern Recognition, Image Analysis, Computer Vision, and Applications",
year="2012",
publisher="Springer Berlin Heidelberg",
pages="14-36",
}

@inproceedings{Kingma2014,
  author = {Diederik P. Kingma and Max Welling},
  booktitle = {2nd International Conference on Learning Representations, {ICLR} 2014, Banff, AB, Canada, April 14-16, 2014, Conference Track Proceedings},
  title = {{Auto-Encoding Variational Bayes}},
  year = 2014
}

@incollection{KL,
  author      = "Solomon Kullback",
  title       = "Information Theory and Statistics",
  editor      = "L. B. Heilprin",
  booktitle   = "Science",
  publisher   = "Wiley",
  year        = 1960,
  pages       = "917-918",
}

@inproceedings{Ghosh2020From,
title={From Variational to Deterministic Autoencoders},
author={Partha Ghosh and Mehdi S. M. Sajjadi and Antonio Vergari and Michael Black and Bernhard Scholkopf},
booktitle={International Conference on Learning Representations},
year={2020},
url={https://openreview.net/forum?id=S1g7tpEYDS}
}

@techreport{mse,
      author        = "Adrian Alan Pol and Victor Berger and Gianluca Cerminara and Cecile Germain and Maurizio Pierini",
      title         = "{Anomaly Detection With Conditional Variational
                       Autoencoders}",
      number        = "arXiv:2010.05531",
      year          = "2020",
      url           = "http://cds.cern.ch/record/2742923",
      note          = "Presented at ICMLA 2019",
}

@inproceedings{cvaeori, 
author = {Kihyuk Sohn and Xinchen Yan and Honglak Lee}, 
title = {Learning Structured Output Representation Using Deep Conditional Generative Models},
year = {2015}, 
publisher = {MIT Press}, 
booktitle = {Proceedings of the 28th International Conference on Neural Information Processing Systems - Volume 2}, 
pages = {3483–3491}, 
series = {NIPS'15} }

@inproceedings{GAN,
 author = {Goodfellow, Ian and Pouget-Abadie, Jean and Mirza, Mehdi and Xu, Bing and Warde-Farley, David and Ozair, Sherjil and Courville, Aaron and Bengio, Yoshua},
 booktitle = {Advances in Neural Information Processing Systems},
 editor = {Z. Ghahramani and M. Welling and C. Cortes and N. Lawrence and K. Q. Weinberger},
 pages = {2672--2680},
 title = {Generative Adversarial Nets},
 url = {https://proceedings.neurips.cc/paper/2014/file/5ca3e9b122f61f8f06494c97b1afccf3-Paper.pdf},
 volume = {27},
 year = {2014}
}

@article{crbmori, 
author = {Graham W. Taylor and Geoffrey E. Hinton and Sam T. Roweis}, 
title = {Two Distributed-State Models For Generating High-Dimensional Time Series}, 
year = {2011}, 
volume = {12},  
journal = {Journal of Machine Learning Research}, 
month = 7, 
pages = {1025–1068}}

@InProceedings{userbm1,
  author     = {Ruslan Salakhutdinov and Andriy Mnih and Geoffrey Hinton},
  title      = {Restricted Boltzmann Machines for Collaborative Filtering},
  year       = {2007},
  eventtitle = {Proceedings of the Twenty-Fourth International Conference (ICML 2007)},
  eventdate  = {2007-06-20/2007-06-24},
  venue      = {Corvallis, USA},
}

@article{hinton2002training,
  title={Training products of experts by minimizing contrastive divergence},
  author={Hinton, Geoffrey E},
  journal={Neural Comp.},
  volume={14},
  number={8},
  pages={1771--1800},
  year={2002},
  publisher={MIT Press}
}

@article{garch,
    author =       "Tim Bollerslev",
    title =        "Generalized autoregressive conditional heteroskedasticity",
    volume =       "31",
    numer=          "3",
    pages =        "307-327",
    journal =      "Journal of Econometrics",
    year =         "1986",
}

@article{arch,
    author =       "Robert F. Engle",
    title =        "Autoregressive Conditional Heteroscedasticity with Estimates of the Variance of United Kingdom Inflation",
    volume =       "50",
    number =       "4",
    pages =        " 987-1007",
    journal =      "Econometrica",
    year =         "1982",
}

@article{heston,
    author =       "Steve Heston",
    title =        "A Closed-Form Solution for Options with Stochastic Volatility with Applications to Bond and Currency Options",
    volume =       "6",
    numer=          "2",
    pages =        "327-343",
    journal =      "Review of Financial Studies",
    year =         "1993",
}

@article{garchvixformula,
    author =       "Jinji Hao and Jin E. Zhang",
    title =        "GARCH Option Pricing Models, the CBOE VIX, and Variance Risk Premium",
    volume =       "11",
    number =       "3",
    pages =        "556-580",
    journal =      "Journal of Financial Econometrics",
    year =         "2013",
}

@article{garchvixoption,
    author =       "Juho Kanniainen and Binghuan Lin and Hanxue Yang",
    title =        "Estimating and using GARCH models with VIX data for option valuation",
    volume =       "43",
    pages =        "200-211",
    journal =      "Journal of Banking and Finance",
    year =         "2014",
}

@article{garchmulti,
    author =       "Marcos Escobar-Anel and Javad Rastegari and Lars Stentoft",
    title =        "Affine multivariate GARCH models",
    volume =       "118",
    pages =        "200-211",
    journal =      "Journal of Banking and Finance",
    year =         "2020",
}

@misc{SDESG,
    author =       "Juan-Pablo Ortega and Rainer Pullirsch and Josef Teichmann and Julian Wergieluk",
    title =        "A new approach for scenario generation in Risk management",
    archivePrefix = "arXiv", 
    note =          {arXiv:0904.0624},
    year = 2009,
}

@article{garchcomp,
    author =       "Peter R. Hansen and Asger Lunde",
    title =        "A Forecast Comparison of Volatility Models: Does Anything Beat a GARCH (1, 1)?",
    volume =       "20",
    pages =        "873- 889",
    journal =      "Journal of Applied Econometrics",
    year =         "2005",
}

@article{dccgarch,
    author =       "Robert Engle",
    title =        "Dynamic Conditional Correlation: A Simple Class of Multivariate Generalized Autoregressive Conditional Heteroskedasticity Models",
    volume =       "20",
    pages =        " 339-350",
    journal =      "Journal of Business Economic Statistics",
    year =         "2002",
}

@article{rewritevix,
    author =       "Peter Carr and Liuren Wu",
    title =        "A Tale of Two Indices",
    volume =       "13",
    number =        "2",
    pages   =       "13-29",
    journal =      "The Journal of Derivatives",
    year =         "2006",
}

@article{garchQ,
    author =       "Jin‐Chuan Duan",
    title =        "The GARCH option pricing model",
    volume =       "5",
    number =        "1",
    pages   =       "13-32",
    journal =      "Mathematical Finance",
    year =         "1995",
}

@article{automle,
    author =       "Charles M. Beach and James G. MacKinnon",
    title =        "Maximum Likelihood Estimation of Singular Equation Systems with Autoregressive Disturbances",
    volume =       "20",
    number =        "2",
    pages   =       "459-464",
    journal =      "International Economic Review",
    year =         "1979",
}

@article{stdt1,
 author = {Simon R. Hurst and Eckhard Platen},
 journal = {Lecture Notes-Monograph Series},
 pages = {301--314},
 title = {The Marginal Distributions of Returns and Volatility},
 volume = {31},
 year = {1997}
}

@article{stdt2,

 author = {Harry M. Markowitz and Nilufer Usmen},
 journal = {Journal of Risk and Uncertainty},
 number = {3},
 pages = {221--247},
 title = {The Likelihood of Various Stock Market Return Distributions, Part 2: Empirical Results},
 volume = {13},
 year = {1996}
}

@article{blackscholes,
    author =       "Fischer Black and Myron Scholes",
    title =        "The Pricing of Options and Corporate Liabilities",
    volume =       "81",
    number =       "3",
    pages =        "637-654",
    journal =      "The Journal of Political Economy",
    year =         "1979",
}

@article{KAMINSKI2014234,
title = {When do stop-loss rules stop losses?},
journal = {Journal of Financial Markets},
volume = {18},
pages = {234-254},
year = {2014},
author = {Kathryn M. Kaminski and Andrew W. Lo},
}

\end{document}